\documentclass{ieeeaccess}
\usepackage{cite}
\usepackage{amsmath,amssymb,amsfonts}
\usepackage{algorithmic}
\usepackage{graphicx}
\usepackage{textcomp}
\usepackage{multirow}
\usepackage{soul}
\usepackage[hidelinks]{hyperref}
\usepackage[T1]{fontenc}
\usepackage{booktabs} 
\usepackage[caption=false]{subfig}
\usepackage{caption,setspace}
\usepackage{textcomp}
\usepackage[ruled]{algorithm2e} 

\DeclareMathOperator{\arctantwo}{arctan2}
\DeclareMathOperator{\sign}{sign}

\DeclareMathOperator*{\argmin}{argmin}
\SetAlFnt{\small}
\SetAlCapFnt{\small}
\SetAlCapNameFnt{\small}
\SetAlCapHSkip{0pt}
\IncMargin{-\parindent}
\def\BibTeX{{\rm B\kern-.05em{\sc i\kern-.025em b}\kern-.08em
    T\kern-.1667em\lower.7ex\hbox{E}\kern-.125emX}}
\begin{document}
\history{Date of publication xxxx 00, 0000, date of current version xxxx 00, 0000.}
\doi{10.1109/ACCESS.2017.DOI}

\title{Mo\textsuperscript{3}: a Modular Mobility Model for future generation mobile wireless networks}
\author{\uppercase{Luca De Nardis}\authorrefmark{1}, \IEEEmembership{Member, IEEE},
\uppercase{Maria-Gabriella Di Benedetto\authorrefmark{1}},
\IEEEmembership{Fellow, IEEE}}
\address[1]{Department of Information Engineering,  Electronics and Telecommunications,  Sapienza University of Rome,  00184 Rome,  Italy (e-mails: luca.denardis@uniroma1.it, mariagabriella.dibenedetto@uniroma1.it)}

\tfootnote{This work was partially supported by Sapienza University of Rome,  grant nos.  RP11916B88A04AE6,  RP11816433F508D1,  RP11816426A9F174.}
\markboth
{L. De Nardis \headeretal: Mo\textsuperscript{3}: a Modular Mobility Model for future generation mobile wireless networks}
{L. De Nardis \headeretal: Mo\textsuperscript{3}: a Modular Mobility Model for future generation mobile wireless networks}

\corresp{Corresponding author: Luca De Nardis (e-mail: luca.denardis@uniroma1.it).}

\begin{abstract}
Mobility modeling in 5G and beyond 5G must address typical features such as time-varying correlation between mobility patterns of different nodes, and their variation ranging from macro-mobility (kilometer range) to micro-mobility (sub-meter range). Current models have strong limitations in doing so: the widely used reference-based models, such as the Reference Point Group Mobility (RPGM), lack flexibility and accuracy, while the more sophisticated rule-based (i.e.  behavioral) models are complex to set-up and tune.\\
This paper introduces a new rule-based Modular Mobility Model, named Mo\textsuperscript{3}, that provides accuracy and flexibility on par with behavioral models,  while preserving the intuitiveness of the reference-based  approach,  and is based on five rules: 1) Individual Mobility,  2) Correlated Mobility,  3) Collision Avoidance,  4) Obstacle Avoidance and 5) Upper Bounds Enforcement. Mo\textsuperscript{3} avoids introducing acceleration vectors to define rules, as behavioral models do, and this significantly reduces complexity. Rules are mapped one-to-one onto five modules, that can be independently enabled or replaced.\\
Comparison of time-correlation features obtained with Mo\textsuperscript{3} vs. reference-based models, and in particular RPGM, in pure micro-mobility and mixed macro-mobility / micro-mobility scenarios,  shows that  Mo\textsuperscript{3} and RPGM generate mobility patterns with similar topological properties (intra-group and inter-group distances),  but that Mo\textsuperscript{3} preserves a spatial correlation that is lost in RPGM - at no price in terms of complexity - making it suitable for adoption in 5G and beyond 5G.

\end{abstract}


\titlepgskip=-15pt

\maketitle

\section{Introduction}
\label{sec:introduction}
\PARstart{T}{he} design of wireless mobile networks evolved in the last 20 years, from GSM/GPRS to UMTS/HSDPA, from LTE to 5G and the upcoming beyond 5G, along two main trends: increased bandwidth and increased spatial density of wireless devices. Large channel bandwidths require greater physical layer flexibility, so to meet user needs and provide better robustness to channel impairments. One of the physical layer parameters that highlights this trend is the Transmission Time Interval (TTI), defined as the shortest time interval over which link configuration can be adjusted. Figure \ref{fig:TTI} shows TTI across four generations of wireless standards (from 20 ms in GSM/Edge \cite{AxeBjo06} to about 0.15 ms in 5G systems \cite{KelCos15}, \cite{BhuJi17}). Most likely, TTI will further decrease in beyond 5G, to support ultra-Reliable Low Latency Communications (uRLLC) \cite{BojMil21}.
\Figure[t]()[width=0.45\textwidth]{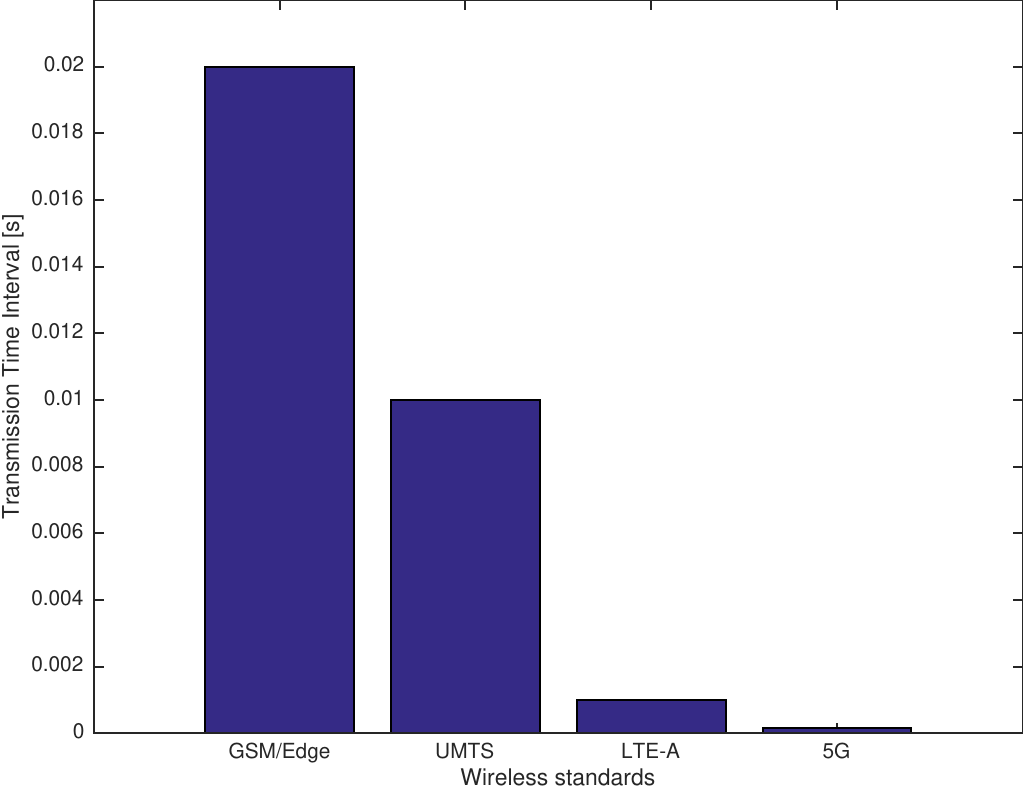}
{Evolution of Transmission Time Interval across generations of cellular networks.\label{fig:TTI}}

Spatial density of wireless devices went from about 1000 devices per square kilometer in GSM, to millions of devices per square kilometer in 5G \cite{Hua13}. The average distance between transmitter and receiver therefore decreased: from hundreds of meters in GSM to a few meters and below in 5G. The steady increase in device density across generations led to major shifts in the design of physical and network layers. At physical layer, signal processing techniques were developed in order to cope with challenging throughput and latency requirements of dense deployments. In particular, beamforming, based on Multiple Input Multiple Output (MIMO), introduced in 3G HSPA and 4G and now fully integrated in 5G, is expected to play a key role in beyond 5G, with the deployment of Massive MIMO \cite{MarCai13}. Steering the beam is possible if the relative position of transmitter vs. receiver is known, and requires swift reactions to position changes \cite{KelCos16}. At network layer, network topology went from a purely centralized configuration with links spanning over thousands vs. hundreds  of meters in 2G vs. 3G, to a mixed nature, with shorter links including both infrastructure to device and Device-To-Device (D2D) connections, as proposed - albeit with limited success - in LTE \cite{LiuKat15}. Cellular networks in 5G and beyond 5G are expected to take full advantage of direct connectivity between devices \cite{HusEls19} and address scenarios - so far restricted to current and legacy Wireless Local Area Networks technologies - in which devices directly exchange data and move in a coordinated manner, although keeping a certain degree of independence in their individual mobility patterns. Examples are:
\begin{itemize}
\item search and rescue in response to emergency calls or disasters. Best practice rules require operators to work in groups of at least two individuals who keep visual or voice contact with one another \cite{OSHA134};
\item tactical and security teams, with on-demand formation, merging, and splitting of groups \cite{TacTeam12};
\item swarms of Unmanned Aerial Vehicles (UAVs) flying in variable formation \cite{MazCap16}, \cite{VigMaz10};
\item cooperative communications in cognitive networks \cite{DeNPer10}.
\end{itemize}
Note that although distances between devices in a same group may be small (i.e. small micro-mobility, in the order of the meter or sub-meter) the same may not be true for distances covered by groups (i.e. large macro-mobility, in the order of the km). Scenarios can be thus characterized as either mixed macro-mobility / micro-mobility, where groups move over distances much larger than distances within a group, or pure micro-mobility where covered distances are small for both groups and within groups. Moreover, when compared against legacy WLANs, 5G and beyond 5G communications are characterized by less favourable propagation features: the combination of higher operation frequencies and possibly lower radiated power due to the massive number of devices will in fact lead to a reduced radio coverage; cooperative and coordinated mechanisms between devices may be thus introduced \cite{ZheHaa17}. A limited radio coverage also amplifies the impact of device mobility on network connectivity, even when movements are within a short range \cite{GeYe16}.\\

\begin{table*}
\caption{Comparison of Mo\textsuperscript{3} with existing mobility models based on the set of design features as proposed in \cite{AunSee15}, complemented with the obstacle avoidance feature. The set of selected mobility models follows \cite{AunSee15}.}
\label{tab:FeatureComparison}
\centering
\begin{tabular}{|p{1.8cm}|p{1.7cm}|p{1.5cm}|p{2.3cm}|p{1.7cm}|p{1.7cm}|p{1.5cm}|p{1.5cm}|}
\hline
  \textbf{Model} & \textbf{Spatial constraints}&\textbf{Stable group structure}&\textbf{Group destination (Random / Predefined)}& \textbf{Group coordination} & \textbf{Group merge/split}&\textbf{Collision avoidance}&\textbf{Obstacle avoidance}\\
\hline
RPGM \cite{HonGer99}&  & $\checkmark$ & R & & & & \\
\hline
RVGM \cite{WanLi02}&  &  &  R& & & & \\
\hline
GFMM \cite{WilHua09} &  &  & R & & & $\checkmark$ & $\checkmark$\\
\hline
MGCM \cite{WuYu06} &  &  & P & $\checkmark$ & & $\checkmark$ & \\
\hline
RRGM \cite{NgZha05} &  &  & P &  & $\checkmark$ & & \\
\hline
VTGM \cite{ZhoXu04} & $\checkmark$ & $\checkmark$ & P &  & $\checkmark$ & & \\
\hline
CMM \cite{MusMas06} / ECMM \cite{VasYan12} &  &  & P &  & $\checkmark$ & & \\
\hline
Mo\textsuperscript{3} & $\checkmark$ & $\checkmark$ & R/P$^{\star}$ & $\checkmark$ & $\checkmark$ & $\checkmark$& $\checkmark$\\
\hline
\multicolumn{8}{p{16cm}}{$^{\star}$ Depending on the selected individual mobility model: see Section \ref{sec:Mo3_individual_model}.}\\
\end{tabular}
\end{table*}
Mobility models for 5G and beyond 5G should therefore accurately address both individual and group mobility, independent of distance. This should be done dynamically to also cover situations in which the correlations of mobility patterns of individual nodes may or may not lead to the emergence of groups. Early models of correlated mobility, referred to as reference-based models, are unsuited for the task, since in those models groups are imposed at start, and not dynamically created; for this reason, they are also called \textit{group} mobility models. Another inherent limitation of these models is the stringent constraints imposed on the mobility patterns of nodes within a group, that limit movements to random variations around either a group reference position (see the seminal Reference Point Group Mobility (RPGM) model \cite{HonGer99}), or a group reference speed (see the Reference Vector Group Mobility (RVGM) model \cite{WanLi02}). Extensions of the above models introduced new features such as group disbanding and merging, and management of geographical constraints, with no change in the underlying reference-based mechanism. Modelling complex mobility patterns of individual nodes was addressed by so-called behavioral models \cite{LegBor06}, \cite{WilHua09}. These models define rules for the behavior of each node that may include interaction with other nodes and with the environment. For example, two or more nodes may be assigned with a mutual attraction rule that keeps them in close proximity. Collision and obstacle avoidance, an absent component in reference-based models, is obtained by repulsion rules preventing collisions between nodes or with obstacles. In behavioral models, each rule is typically implemented as a force; the intensity of the force is determined based on position, speed and direction of the node itself and of other nodes, and on the position of obstacles. Each force in turn defines an acceleration vector associated with the corresponding rule, and the sum of the acceleration vectors determines the acceleration vector of a node, and by integration over time its speed vector. In behavioural models, groups are not defined at start, but rather emerge from nodes sharing a same set of rules, and having thus similar mobility patterns; each node has however its own speed vector, and thus its individual mobility is modeled without any loss of accuracy caused by the introduction of group mobility. Furthermore, by applying different rules to different nodes, behavioural models can describe correlated mobility scenarios that cannot be addressed by reference-based models, thus providing higher flexibility. Tuning behavioral model to obtain specific spatial correlated mobility patterns is, however, particularly complex, and scenarios that are easily modeled in RPGM become unfeasible with behavioural models \cite{TriHan10}. Consider for instance, nodes of a same group that must be within a predefined maximum distance; obtaining this behaviour with behavioural models requires a difficult-to-find balance between mutual attraction vs. repulsion forces, based on the careful selection of several weighting variables \cite{WilHua09}. For this reason, behavioural models - although attractive - are unpractical and found little application in wireless networking, where the simplicity of reference-based models was most often preferred over accuracy. This choice becomes however too simplistic in view of 5G and beyond 5G network scenarios, calling for a model that combines the intuitiveness of reference-based models and the accuracy and flexibility of behavioral models.\\

This paper introduces a new mobility model, called Modular Mobility Model (Mo\textsuperscript{3}), that models, as behavioural models, the mobility of individual agents, but is as straightforward to tune as a reference-based model. In analogy to behavioral models, Mo\textsuperscript{3} determines the mobility of nodes by defining rules that describe their individual mobility and the effect of mutual  interactions as well as with the environment. Mo\textsuperscript{3}, however, does not implement rules as forces. Rather, each rule determines a modification of the speed and direction of movement of a node, without requiring the definition of an acceleration vector and avoiding thus the introduction of weighting variables. The set of rules in Mo\textsuperscript{3} was defined based on the observation that mobility patterns typically result from modifications to planned trajectories, i.e. a target destination and an initial speed and direction. The deviation from a planned trajectory is usually caused by either voluntary interactions with other agents (correlated mobility) or reactions to external events (collision and obstacle avoidance). Consider for example the case of an individual who visits an art exhibition and is part of a group. At a given point in time, the individual may depart the planned trajectory toward an art piece of interest, due to either voluntary behavior (reunite with the group) or external events (avoiding collision with another visitor or with an obstacle). Note that any modification to speed and direction will stay within the physical limits of the individual. Similar behavior can be observed for fleets of vehicles or animal herds, for example.\\
Moving from this observation, Mo\textsuperscript{3} defines five rules: Individual Mobility, Correlated Mobility, Collision Avoidance, Obstacle Avoidance and Upper Bounds Enforcement. Note that the phenomena that determine a given mobility behavior are beyond the scope of this paper; Mo\textsuperscript{3} takes this information as an input, coded in the Individual Mobility and Correlated Mobility rules settings, and generates accurate and artifact-free mobility patterns, independent of the considered grouping behavior and mobility scale. The Individual Mobility rule, in particular, can be implemented by adopting any existing individual mobility model, to be selected according to the desired mobility behavior.\\
Table~\ref{tab:FeatureComparison} shows a comparison of Mo\textsuperscript{3} against existing reference-based and behavioral models, as proposed in \cite{AunSee15}. Table~\ref{tab:FeatureComparison} considers the same models analyzed in \cite{AunSee15} and includes the features identified as relevant in \cite{AunSee15}, with the addition of the obstacle avoidance feature.
Table~\ref{tab:FeatureComparison} shows that Mo\textsuperscript{3} provides all the relevant features, and highlights that existing models only provide a subset of the features, in most cases the ones required to model specific mobility scenarios.\\ 

The paper is organized as follows. Section \ref{sec:MobilityModels} reviews and discusses existing models. Specifically, Section \ref{sec:IndividualModels} reviews individual mobility models in order to identify suitable candidates for the Individual Mobility rule in Mo\textsuperscript{3}, while Section \ref{sec:CorrelatedModels} provides a critical review of the existing correlated mobility models. 
Section \ref{sec:Mo3} includes an exhaustive description of the Mo\textsuperscript{3} model. The section highlights the modular nature of Mo\textsuperscript{3}: each of the rules characterizing  Mo\textsuperscript{3} is paired with a module, that can be replaced without affecting the others. This modularity favours extensions and modifications to Mo\textsuperscript{3} and, to this purpose, access to an open source software implementation of Mo\textsuperscript{3} is provided. Section \ref{sec:Mo3_mimetic_abilities} provides examples of emulation of existing correlated mobility models using Mo\textsuperscript{3}. Section \ref{sec:PerfEvalIntro} compares then Mo\textsuperscript{3} against other existing models in pure micro-mobility and mixed macro-mobility/micro mobility scenarios. Finally, Section \ref{sec:Conclusion} draws conclusions.

\section{Related work on mobility modeling}
\label{sec:MobilityModels}
\subsection{Individual mobility models}
\label{sec:IndividualModels}

\begin{figure}[t]
  \centering
\includegraphics[width=3.2in]{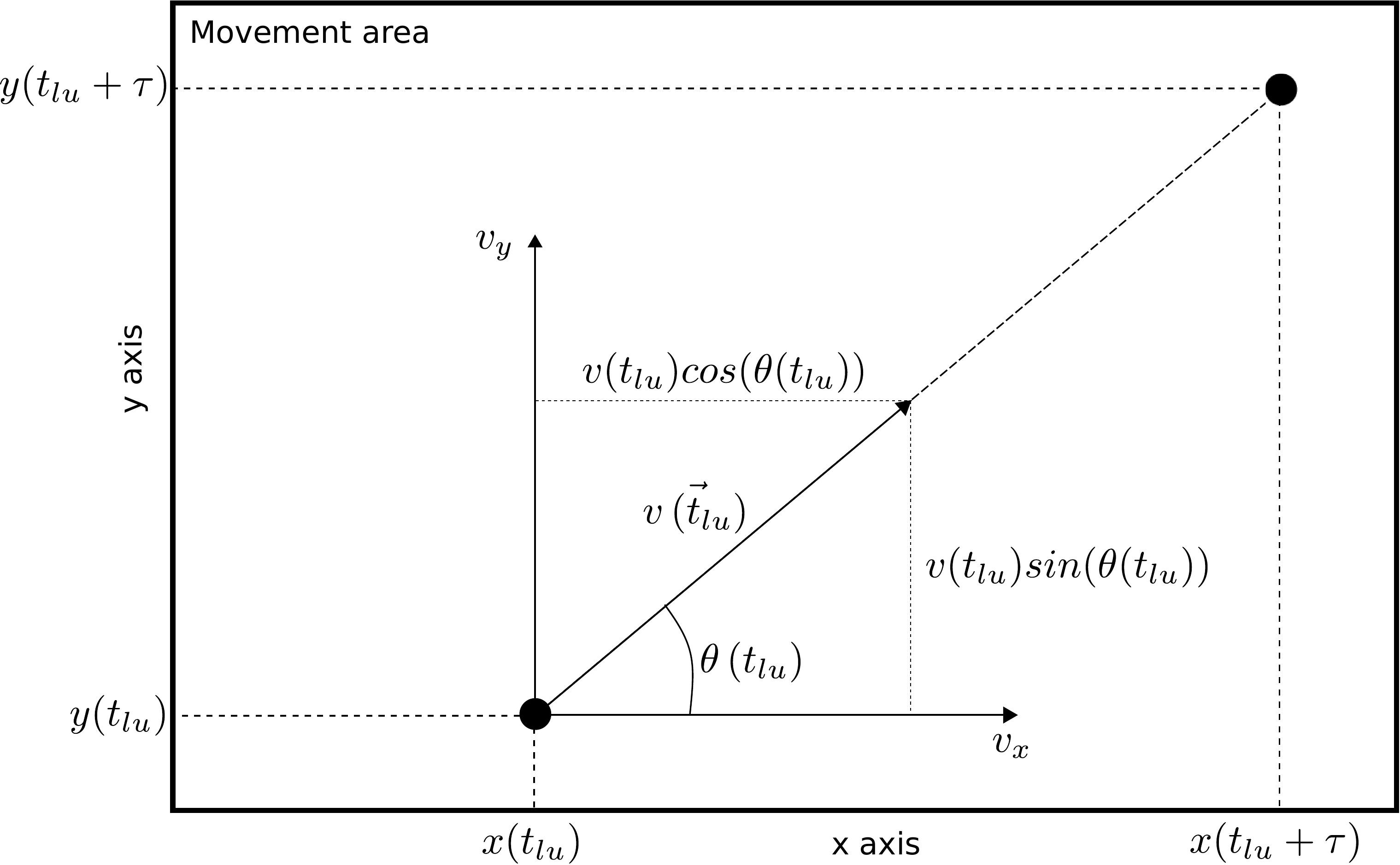}
 \caption{Position at time $t_{lu}+\tau$ as a function of speed vector updated at time $t_{lu}$.\label{fig:IndividualModels_pos}}
\end{figure}

Individual mobility models determine the pattern of each node by varying its speed vector, defined at generic time $t$ as: $\vec{v(t)}=v(t)e^{j\theta(t)}$, where $v\left(t\right)$ is the speed of movement and $\theta\left(t\right)$ is the direction of movement, defined as the angle between the $x$ axis and the speed vector in the selected coordinate system. The position of a node at any time $\tau$ between two updates of $\vec{v}$ is thus:
\begin{equation}
\left\{ \begin{array}{l}
x\left( t_{lu} + \tau \right) = x\left( t_{lu} \right) + v\left( t_{lu} \right)\cos \left( {\theta \left( t_{lu} \right)} \right)\tau \\ 
y\left( t_{lu} + \tau \right) = y\left( t_{lu} \right) + v\left( t_{lu} \right)\sin \left( {\theta \left( t_{lu} \right)} \right)\tau, \\ 
 \end{array} \right.
 \label{eq:position_inter_update}
\end{equation}
where $t_{lu}$ is the time instant of last speed vector update (see Figure \ref{fig:IndividualModels_pos}).
Models are commonly classified as \emph{memoryless} vs. \emph{memory-based}, based on how $v$ and $\theta$ are updated.
\subsubsection{Memoryless mobility models}
In memoryless models, updates to values of speed and direction are independent of current and previous values. Well-known memoryless models are:
\begin{itemize}
\item the \emph{Random Walk} model, also referred to as Brownian model, widely used to determine the impact of mobility on cellular network performance \cite{ZoDa97}. The model has the desirable property of leading to a uniform spatial distribution of the position of a node in the movement area. Statistical models of average cell crossing time, average channel holding time, and average number of handovers, were developed using a Random Walk user mobility model \cite{ZoDa97}. The model was also used in the analysis of mobile ad-hoc routing protocols in presence of node mobility \cite{HuJohn00}.\\
In the Random Walk model, the selection of a new speed and direction is triggered by either of the following events:
\begin{itemize}
\item A periodic timer, set to a predefined update period $T$, expires \cite{BarKes94};
\item The node covers a predefined distance $D$ \cite{CaBo02}.
\end{itemize}

\item the model proposed by Ko and Vaidya \cite{KoVai98a} as a simple way to introduce mobility in the performance evaluation of the Location Aided Routing protocol, with no claim for specific advantages over other mobility models.
According to the model, a node selects a random direction $\theta$ at simulation start time. The node also selects its speed $v$ according to a uniform distribution within a predefined interval $[v_{min},v_{max}]$, and a distance $d$ to be covered at speed $v$, according to an exponential distribution. After covering the distance, new values are selected for $v$, $\theta$ and $d$. When the node hits a boundary of the simulation area, it is perfectly reflected within the area.

\item the \emph{Random Waypoint} model, originally proposed in \cite{JohnMal96}; this model is similar to the Random Walk model, but the trajectory of a node is determined here by a sequence of destination points to be reached. When a node reaches a destination point it pauses for a random time, and then moves towards the next destination point with a new random speed.
It has been observed that the Random Waypoint models leads to an uneven spatial distribution of nodes \cite{RoMel01}, which causes large variations in the average number of neighboring nodes (i.e. nodes within a given distance), especially on the short term \cite{CaBo02}. Variations of this model have been proposed to address this issue, such as the Random Direction model, where the node selects a speed vector rather than a destination, and moves according to the selected speed vector until it reaches a boundary of the simulation area when, after a predefined pause time, a new speed vector is selected \cite{RoMel01}.
\end{itemize}

All memoryless models share the issue of potentially causing sharp turns and steep variations in speed when a new speed vector is selected, making it impossible to meet upper bounds on linear acceleration and angular speed, and possibly leading to unrealistic mobility patterns. 
%

\subsubsection{Memory-based models}
\label{sec:memory_based_models}
Memory-based models lead to more realistic patterns by introducing memory in the selection of speed and direction. Well-known models that adopt this approach are:
\begin{itemize}
\item the \emph{Inertia} mobility model \cite{BasChl99}, in which new values for $v$ and $\theta$ are selected at each position update with probability $\rho$, while the current set is kept with probability $1-\rho$. Parameter $\rho$ models an inertia that tends to keep the node on the current trajectory: the higher the value of $\rho$, the lower the probability of selecting a new speed vector. For $\rho=0$ the Inertia model coincides with the Random Walk model.
The Inertia model provides a straightforward mechanism for introducing memory in the selection of the speed vector. New values of $v$ and $\theta$ have, however, no correlation with respective previous values. This leads to unrealistic patterns characterized by abrupt turns and speed variations that make it difficult to meet requirements on maximum linear and angular speeds;
\item the \emph{Gauss-Markov} model\cite{LiaHaa03}, in which the component $v_i$ of the speed vector along direction $i$ (with $i \in [x,y]$ in a two-dimensional space) at time $t$ is the outcome of a Gauss-Markov random process $v_i(t)$, that is a stationary Gaussian process characterized by the following autocorrelation function:
\begin{equation}
\phi _{v _i} \left( \tau  \right) = E\left[ {v_i \left( t \right)v_i \left( {t + \tau } \right)} \right] = \sigma _i^2 e^{ - \beta \left| \tau  \right|}  + \mu _i^2,
\label{eq:GaussMarkov_autocorr}
\end{equation}
where $\sigma _i^2$ and $\mu_i$ are the variance and the mean of $v_i(t)$, respectively, and parameter $\beta \ge 0$ introduces a memory effect. The mobility patterns generated by this model are governed by properly setting the $\beta$ parameter.
The Gauss-Markov model provides smoother patterns than Inertia, but still does not provide a straightforward way to meet constraints on maximum speed and rotation in the generation of a mobility pattern. The adoption of a Gaussian probability density function, in particular, may lead to unrealistic values for node speed.
\item the \emph{Boundless} mobility model \cite{Haa97}, named after the idea of mapping a bidimensional simulation area on the surface of a three-dimensional torus: a node that reaches an edge of the area disappears, and reappears instantaneously on a point on the opposite edge, while keeping the same speed vector. 
The algorithm for speed and direction update proposed in \cite{Haa97} can be, however, adopted within a traditional bounded movement area as well. In the Boundless model, the speed vector is updated every $T$ seconds according to the following rules:
\begin{equation}
\left\{ \begin{array}{l}
 v\left( {t + T} \right) = \min \left( {\max \left( {v\left( t \right) + \Delta v,0} \right),v_{max} } \right) \\ 
 \theta \left( {t + T} \right) = \theta \left( t \right) + \Delta \theta, \\ 
 \end{array} \right.
\label{eq:Boundless_speed_vector}
\end{equation}
where:
\begin{itemize}
\item $v_{max}$ is the maximum speed;
\item $\Delta v$ is the speed variation, uniformly selected at every update time in the interval $[-a_{max}T, a_{max}T]$, where $a_{max}$ is the maximum linear acceleration allowed for a node, measured in $m/s^2$;
\item $\Delta \theta$ is the direction variation, uniformly selected at every update time in the interval $[-\gamma_{max}T, \gamma_{max}T]$, where $\gamma_{max}$ is the maximum rotation speed allowed for a node, measured in rad/s.
\end{itemize}
The Boundless model shares with the Gauss-Markov model the capability of producing realistic movement patterns. The model has, however, the advantage of allowing the introduction of limits on speed, acceleration and rotation speed of nodes, making it easier to achieve realistic mobility patterns that meet predefined upper bounds.
\end{itemize}

\begin{figure*}[!t]
    \centering
    \subfloat[$\Delta t=1\,s$]{
        \includegraphics[scale=0.3]{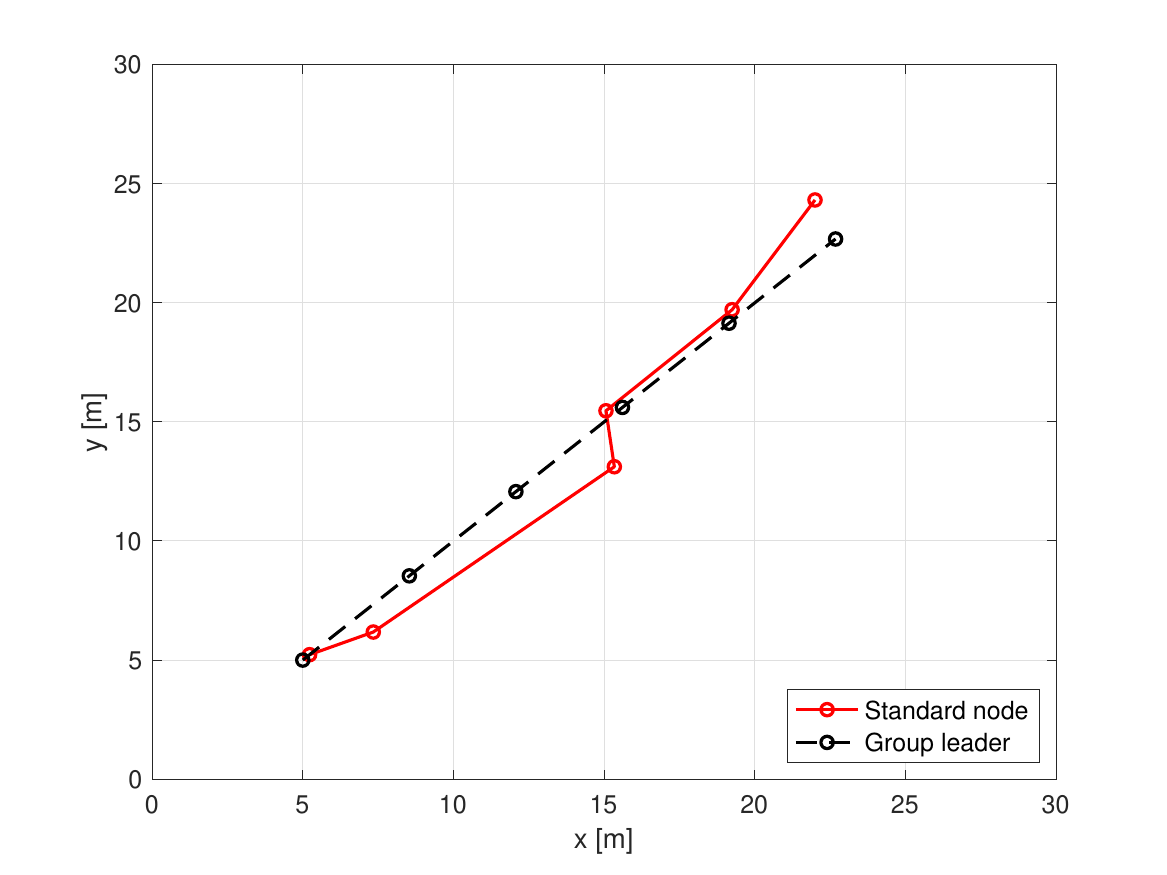}
        \label{fig:dt_1}
    }
    \subfloat[$\Delta t=0.1\,s$]{
        \hspace*{-.1in}
        \includegraphics[scale=0.3]{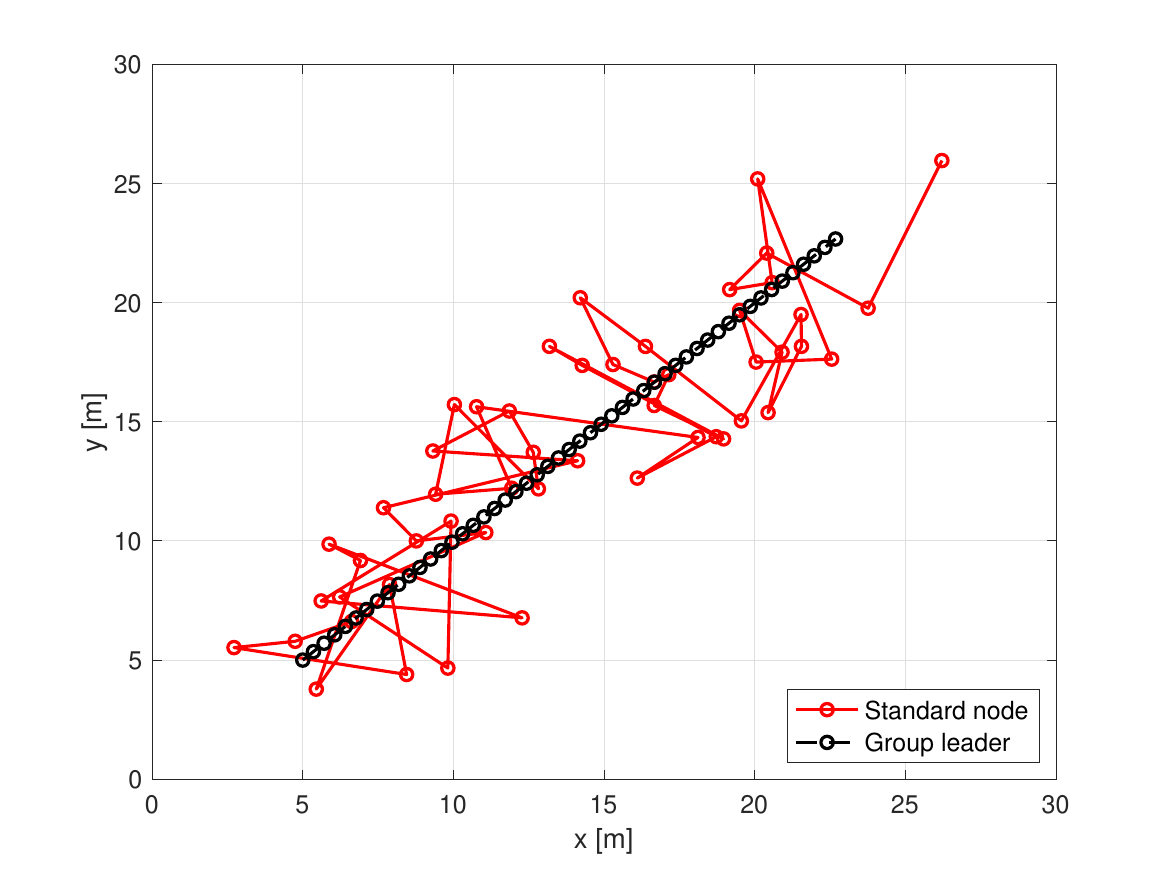}
        \label{fig:dt_01}
    }
    \subfloat[$\Delta t=0.01\,s$]{
        \hspace*{-.1in}
        \includegraphics[scale=0.3]{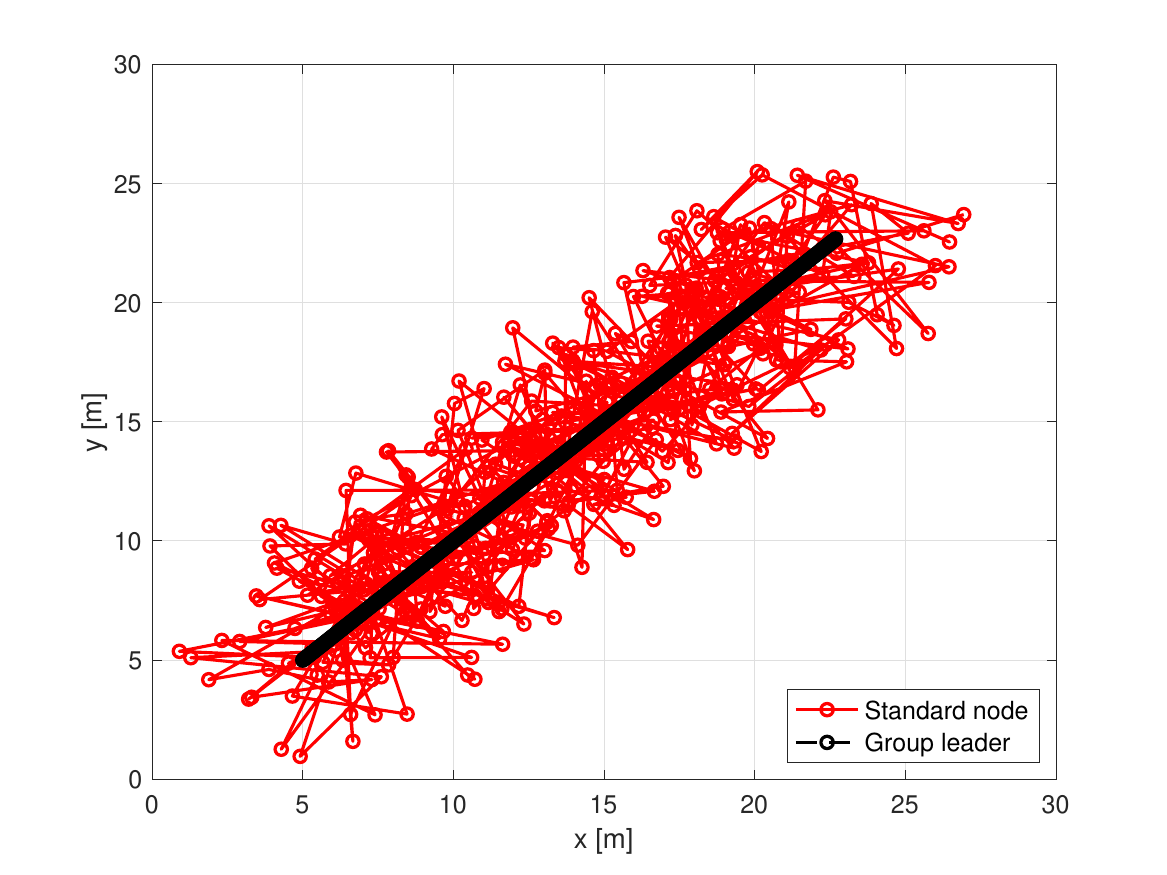}
        \label{fig:dt_001}
    }
    \hfill
    \caption{Patterns obtained for three different $\Delta t$ values with the RPGM model for a group leader moving at $v=5\, m/s$ and a standard node in the same group,  with $d_{max}=5\, m$.}
      \label{fig:dt_effect_RPGM}
\end{figure*}

\Figure[t]()[width=0.48\textwidth]{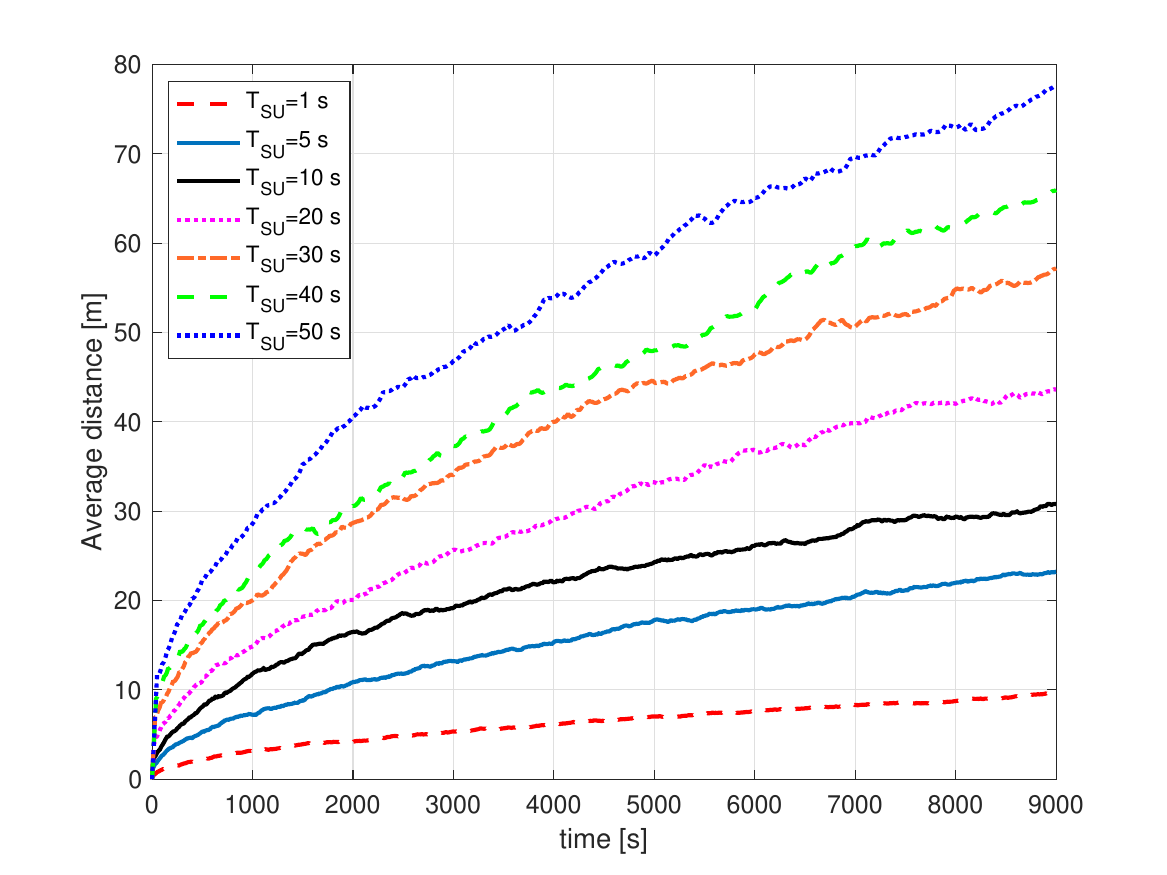}
{Average distance between nodes in a group of $M=5$ nodes as a function of time in the RVGM model for  different values of the update period of the speed vector of standard nodes $T_{SU}$, assuming a fixed reference speed vector with $v=1\, m/s$ and $\theta=\pi/4\, rads$. At each update, the speed vector of each node is determined by randomly extracting a speed in $\left[v-0.1,v+0.1\right]\, m/s$ and a direction in $\left[\theta-\pi/12,\theta+\pi/12\right]\, rads$; all nodes are placed at the same position at $t=0\, s$.\label{fig:RVGM_T_SU}}

\subsection{Correlated mobility models}
\label{sec:CorrelatedModels}
Correlated mobility models can be divided into three families: reference-based models, behavioral models, and models based on social network theory \cite{LiuSic17}. In reference-based models, correlation typically implies the presence of groups; the positions of nodes belonging to a same group is determined as a random deviation from a common reference, defined either as a reference position or as a reference speed vector (see Section \ref{sec:ReferenceBasedModels}). In behavioral models, nodes select their speed and direction according to predefined rules; in these models, groups naturally emerge from more nodes sharing same rules (Section \ref{sec:Behavioral_models}). Social network theory models match the mobility patterns to those of people in a community (Section \ref{sec:Social_network_models}). Section \ref{sec:BenchmarkSelection} compares the different models and identifies benchmarks for performance evaluation.
\subsubsection{Reference-based models}
\label{sec:ReferenceBasedModels}

The Exponential Correlated Random (ECR) mobility model, proposed in \cite{Ber96}, was one of the first models addressing correlated mobility. ECR models the mobility of a group, but not of individual nodes.\\
The Reference Point Group Mobility (RPGM) model was designed in order to overcome the limitations of ECR, by allowing the description of group as well as individual node mobility within a group \cite{HonGer99}. RPGM defines a logical reference point for each group, that often coincides with the position of one of the nodes in the group (group leader), whose movement is followed by all nodes in the group (standard nodes). The path of the group leader is typically generated according to an individual mobility model; for example, the Random Waypoint model (see Section \ref{sec:IndividualModels}) was used in \cite{HonGer99}. The position of standard nodes is generated randomly, according to a uniform distribution for both angle and distance from the reference position of the leader (within $d_{max}$), and is refreshed every $\Delta t$ seconds. The Structured Group Mobility Model (SGMM) \cite{BlaLow04} extended RPGM by introducing different statistical distributions for the position of different standard nodes.\\
In RPGM-based models standard nodes are not associated with a speed vector: their position is in fact only known at each position update at time $t=k\Delta t$. In 5G and beyond 5G scenarios, the adoption of a short $\Delta t$ to match the TTI parameter, leads to erratic mobility patterns for standard nodes, as shown in Figure \ref{fig:dt_effect_RPGM}, presenting the patterns generated for a group leader and a standard node in the same group for three different $\Delta t$ values. Figure \ref{fig:dt_effect_RPGM} highlights that, although on average the standard node follows the group leader for all $\Delta t$ values, its pattern shows an increasing variability as $\Delta t$ decreases. Furthermore, as shown in Section \ref{sec:perfEval}, the maximum speed $v$ for a standard node, for small $\Delta t$, is about $v=2d_{max}/\Delta t$, which leads to unreasonable speeds.\\
The Reference Velocity Group Mobility (RVGM) model \cite{WanLi02} proposed the use of a reference speed vector, rather than a reference position, so that each node has its own speed vector. Here, a random deviation with respect to reference is introduced on the speed vector rather than on the position of standard nodes. The period $T_{SU}$ with which the speed vector of a standard node is updated is an independent factor, that is tuned by the requirements of the mobility scenario. At any instant in time, the position of a node can be determined by a simple calculation based on the current speed vector; this solves the problem of erratic behaviour of standard nodes observed in RPGM, when frequent updates are required. However, RVGM does not provide any mechanism to preserve physical proximity within a group, or to restore it by forcing a standard node to rejoin its group, and this autonomy allows standard nodes to drift away. This eventually leads to a loss of cohesion within groups, as shown in Figure \ref{fig:RVGM_T_SU}. Figure \ref{fig:RVGM_T_SU} shows the average distance within a group of $M=5$ nodes, as a function of time, for different values of $T_{SU}$. All nodes are placed in a same position in $t=0\, s$ and, at each update, relatively small deviations from the reference vector are allowed (see caption of Figure \ref{fig:RVGM_T_SU} for details). Results show that the average distance steadily increases over time for all $T_{SU}$ values, highlighting the loss of physical proximity within the group.\\
A common trait of reference-based models discussed so far is a static definition of groups, that is no mechanisms are defined to change the composition of groups. 
A reference-based model that allows group merging and splitting is the Reference Region Group Mobility (RRGM) model \cite{NgZha05}. RRGM defines a set of target destinations for groups, and corresponding reference regions surrounding the destinations. The model manages group merging by assigning the same destinations/reference regions to two groups, and group splitting by assigning a new reference region to part of the nodes in a group. The model places however several restrictions on when group merging and splitting can happen in time, and on which groups will be merged or split, as a function of their positions and of the position of the new target destination. Furthermore, the complexity in setting up the model is higher than in RPGM and RVGM, losing the simplicity that is a key advantage of reference-based models.
\subsubsection{Behavioral models}
\label{sec:Behavioral_models}
The concept of behavioral mobility modeling was first adopted in the Behavioral Mobility Model (BMM) \cite{LegBor06}. In behavioral models, the mobility pattern of a node is governed by a set of rules, expressed as forces on the node, that is, acceleration vectors. For each force $j$ one has an acceleration vector $\vec{a_j}$; the combination of acceleration vectors leads to a global acceleration vector $\vec{a}$, that in turn determines the speed vector, and eventually the trajectory of the node. The set of rules proposed in \cite{LegBor06} determines the behavior of a node with respect to a) node destination, b) surrounding environment, and c) presence of other nodes. The main "desired destination" rule is the \emph{Path Following} rule, that introduces an acceleration vector towards a preset destination, under a node maximum speed constraint. Environmental rules include \emph{Wall Avoidance} and \emph{Obstacle Avoidance}, that generate repulsive forces. Finally, rules that determine the behavior of a node in the presence of other nodes include \emph{Mutual Avoidance}, avoiding collisions between nodes, and \emph{Group Centering} and \emph{Velocity Matching}, forcing nodes to stay close to one another in the space vs. speed domains, and are the behavioral equivalents of RPGM vs. RVGM. An extension of  \cite{LegBor06}, referred to as Behavioral Mobility Model with Geographic Constraints (BMM-GC) \cite{BeuMiw13}, provides a more accurate modeling of the interaction with obstacles. Another model adopting a behavioral approach for the interaction between nodes quite similar to \cite{LegBor06} is the Group Force Mobility Model (GFMM)\cite{WilHua09}.\\
Behavioral mobility modeling was also investigated in \cite{MedGur10}, where a modular approach is proposed, in which basic rules are combined in order to generate complex behaviors such as group mobility. Basic rules include \emph{Seek}, \emph{Flee}, and \emph{Arrive} for individual mobility, and \emph{Pursuit}, \emph{Evade} and \emph{Interpose} for modeling the interaction between nodes.\\
Finally, in \cite{BorLeg09} the rules generating acceleration vectors are based on a sociological analysis of the impact of social ties between individuals on mobility patterns.\\
A major drawback of behavioral mobility models is their complexity at set-up, due to the need of selecting normalizing and scaling factors to determine the strength of the forces. Furthermore, mobility patterns that are easy to describe in reference-based models, e.g. a maximum distance between nodes in a group, are extremely difficult to describe with rules,  as observed in \cite{TriHan10}.

\subsubsection{Social network theory models}
\label{sec:Social_network_models}
Studies on the interactions among members of a community inspired a third family of correlated mobility models, based on social network theory \cite{LiuSic17}. The Community Based Mobility Model (CMM) \cite{MusMas06} introduces the concept of Interaction Matrix, with elements indicating the degree of social interaction between any two nodes, and the corresponding inclination to spend time together with a value between 0 and 1. Although  CMM and its enhanced version Enhanced CMM \cite{VasYan12} are designed for the specific scenario of social human interactions, the concept of the Interaction Matrix can be applied to different mobility scenarios. A similar approach is adopted in Mo\textsuperscript{3} as part of the Correlated Mobility rule, as explained in Section \ref{sec:Mo3_correlated_mobility}.

\subsubsection{Benchmark selection }
\label{sec:BenchmarkSelection}
The comparison in terms of features presented in Table \ref{tab:FeatureComparison} and the review carried out in this section highlight that a large number of correlated/group mobility models have been proposed through the years; a fair question is thus how to select which models to consider as a benchmark when proposing and evaluating a new model. The proposed selection was based on two factors: the suitability of models to address the same wide range of mobility scenarios targeted by Mo\textsuperscript{3}, and the intuitiveness in describing such scenarios. Based on the review of this Section, one may conclude that models based on social network theory are hardly adaptable to the scenarios identified in Section {\ref{sec:introduction}}. Behavioral models, on the other hand, are potentially capable of addressing any scenario; however, the difficulty in setting them up, preventing their widespread adoption, does not favor their selection as benchmarks. For this reason, the selected benchmarks belong to the reference-based class, as further discussed in Section {\ref{sec:PerfEvalIntro}}: RPGM, by far the most popular and preferred choice for group mobility modeling \cite{TabSal19}, and RVGM, another general purpose reference-based model. The validity of this selection is also reflected by the impact the selected models on the research community, as measured by the citation indexes of related scientific literature (see Appendix {\ref{sec:CitationAnalysis}}).
\Figure[t]()[width=\textwidth]{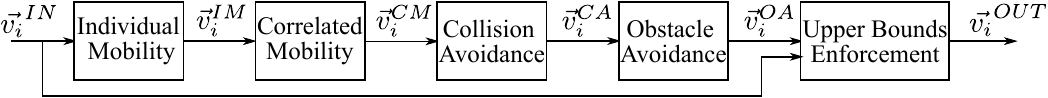}
{The Mo\textsuperscript{3} model. The Mo\textsuperscript{3} rules are applied sequentially on the speed vector, starting with Individual Mobility, followed by Correlated Mobility, Collision Avoidance, Obstacle Avoidance and finally Upper Bounds Enforcement. Note that the Upper Bounds Enforcement rule takes in input both the output speed vector resulting from the application of the Obstacle Avoidance rule and the current speed vector, in order to ensure that the cumulative modifications introduced by the first four rules do not exceed the maximum allowed variations for linear and angular speeds.\label{fig:Mo3_procedure}}

\section{The Mo\textsuperscript{3} model}
\label{sec:Mo3}
Mo\textsuperscript{3} models the mobility of each node by applying five rules: Individual Mobility, Correlated Mobility, Collision Avoidance, Obstacle Avoidance and Upper Bounds Enforcement. The peculiar and novel aspect of Mo\textsuperscript{3} consists in its modularity.
Each rule is implemented in a dedicated module, and each module can be replaced without affecting the other modules, providing thus ample possibility to expand Mo\textsuperscript{3} and tailor it for new mobility scenarios that may emerge in the future. Furthermore, depending on the considered mobility scenario, modules can be independently turned on and off, and can operate with different update periods.\\
The modular nature of Mo\textsuperscript{3} also allows to introduce  mobility in the third dimension for selected rules; Individual Mobility, Correlated Mobility and Upper Bounds Enforcement rules, in particular, support tridimensional mobility, while the introduction of this feature in Collision Avoidance and Obstacle Avoidance rules is left for future work. When tridimensional mobility is selected, the modules corresponding to rules that do not support this feature can be selectively disabled.\\
Mo\textsuperscript{3} supports tridimensional mobility by adopting a spherical coordinate system, where the speed vector $\vec{v(t)}$ associated to a node is represented as a triplet $\left\{v(t),\theta(t),\varphi(t)\right\}$: $v(t)$ and $\theta(t)$ have the same meaning as in the bidimensional speed vector introduced in Section {\ref{sec:IndividualModels}}, while $\phi(t)$ indicates the elevation angle, defined as the angle between the speed vector and the $\left\{x,y\right\}$ plane, as shown in Figure {\ref{fig:3Dvector}}, where the dependence on time is omitted to simplify notation.
\Figure[t]()[width=0.45\textwidth]{./Figures/3Dvector.eps}
{Definition of the speed vector $\vec{v}$ in three dimensions.\label{fig:3Dvector}}
Correspondingly, {\eqref{eq:position_inter_update}} becomes:
\begin{equation}
\left\{ \begin{array}{l}
x\left( t_{lu} + \tau \right) = x\left( t_{lu} \right) + v\left( t_{lu} \right)\cos \left( {\theta \left( t_{lu} \right)} \right)\cos \left( {\varphi \left( t_{lu} \right)} \right)\tau \\ 
y\left( t_{lu} + \tau \right) = y\left( t_{lu} \right) + v\left( t_{lu} \right)\sin \left( {\theta \left( t_{lu} \right)} \right)\cos \left( {\varphi \left( t_{lu} \right)} \right)\tau \\ 
z\left( t_{lu} + \tau \right) = z\left( t_{lu} \right) + v\left( t_{lu} \right)\sin \left( {\varphi \left( t_{lu} \right)} \right)\tau, \\ 
 \end{array} \right.
 \label{eq:position_inter_update_3D}
\end{equation}
where $t_{lu}$ and $\tau$ are defined as in {\eqref{eq:position_inter_update}}, 
 $\theta \in \left[-\pi, \pi\right]$, while $\varphi \in \left[-\pi/2, \pi/2\right]$.\\
A bidimensional mobility scenario can be modeled by setting $\varphi(t)=0\,\forall t$; in this case {\eqref{eq:position_inter_update_3D}} coincides with {\eqref{eq:position_inter_update}} for any arbitrary choice of $z(0)$. This scenario is considered throughout the paper, in order to simplify the graphical representation of patterns and allow for comparison with the bidimensional RPGM and RVGM models adopted as benchmarks; examples of patterns in a tridimensional mobility scenario are however provided in Sections {\ref{sec:Mo3_individual_model}} and {\ref{sec:Mo3_correlated_mobility}}.\\
In Mo\textsuperscript{3}, at each mobility update the speed vector $\vec{v_i}^{IN}$ for a generic node $i$ is modified by sequentially applying each of the five rules. The first rule to be applied takes thus as an input the current speed vector $\vec{v_i}^{IN}$; the resulting speed vector is transferred to the second rule, and so on; when one module is not active or skips a given mobility update, the speed vector is transparently moved from its input to its output, without modifications. The output of the last rule is the new speed vector $\vec{v_i}^{OUT}$.\\ 
Note that the order of application defines a hierarchy between the rules: rules applied later prevail on those applied earlier. The order selected in this work is the following: 1) Individual Mobility; 2) Correlated Mobility; 3) Collision Avoidance; 4) Obstacle Avoidance; 5) Upper Bounds Enforcement. The order was determined as a reasonable model of the behavior described in Section \ref{sec:introduction} for a group of people, a herd of animals, or a fleet of vehicles: the need to meet constraints caused by correlated mobility supersedes the node's individual mobility model, and in turn the need to avoid collisions with other nodes and with obstacles takes precedence on e.g. keeping up with a group. Finally, the enforcement of bounds on linear and angular speeds prevails on everything else, even if this might result in a failure to avoid a collision. The resulting model is presented in Figure \ref{fig:Mo3_procedure}. Figure \ref{fig:Mo3_procedure} highlights the special role of the Upper Bounds Enforcement rule, that compares the speed vector $\vec{v_i}^{OA}$, resulting from the application of the first four rules, with the current speed vector $\vec{v_i}^{IN}$, and ensures that the cumulative modifications applied to $\vec{v_i}^{IN}$ do not exceed the maximum allowed variations for linear and angular speeds. As a final note on the selected order, one might argue that establishing a hierarchy between collision avoidance and obstacle avoidance is somewhat arbitrary; however, as it will explained in Sections \ref{sec:Mo3_CollisionAvoidance} and \ref{sec:Mo3_ObstacleAvoidance}, the Collision Avoidance and Obstacle Avoidance rules are designed to operate on mutually different mobility parameters, so to mitigate the risk of conflicts between the corresponding corrections.\\
The five Mo\textsuperscript{3} rules and corresponding modules are described in sections \ref{sec:Mo3_individual_model} to \ref{sec:Mo3_UpperBounds}; section \ref{sec:Mo3_setup} describes the setup procedure and summarizes the model input parameters, divided by module. Finally, Section \ref{sec:Mo3_availability} provides information on how to access an open-source software that implements the model.\\
Note that the output speed vector of each module is labelled with a corresponding superscript, as shown in Figure \ref{fig:Mo3_procedure}. In the following, superscripts will be however dropped when possible, in order to simplify the notation, in which case the input speed vector is indicated as $\vec{v_i}$, and the corresponding output as $\vec{v_i}'$.
\subsection{Individual Mobility}
\label{sec:Mo3_individual_model}
\begin{figure}[t]
 \centering
\includegraphics[width=0.5\textwidth]{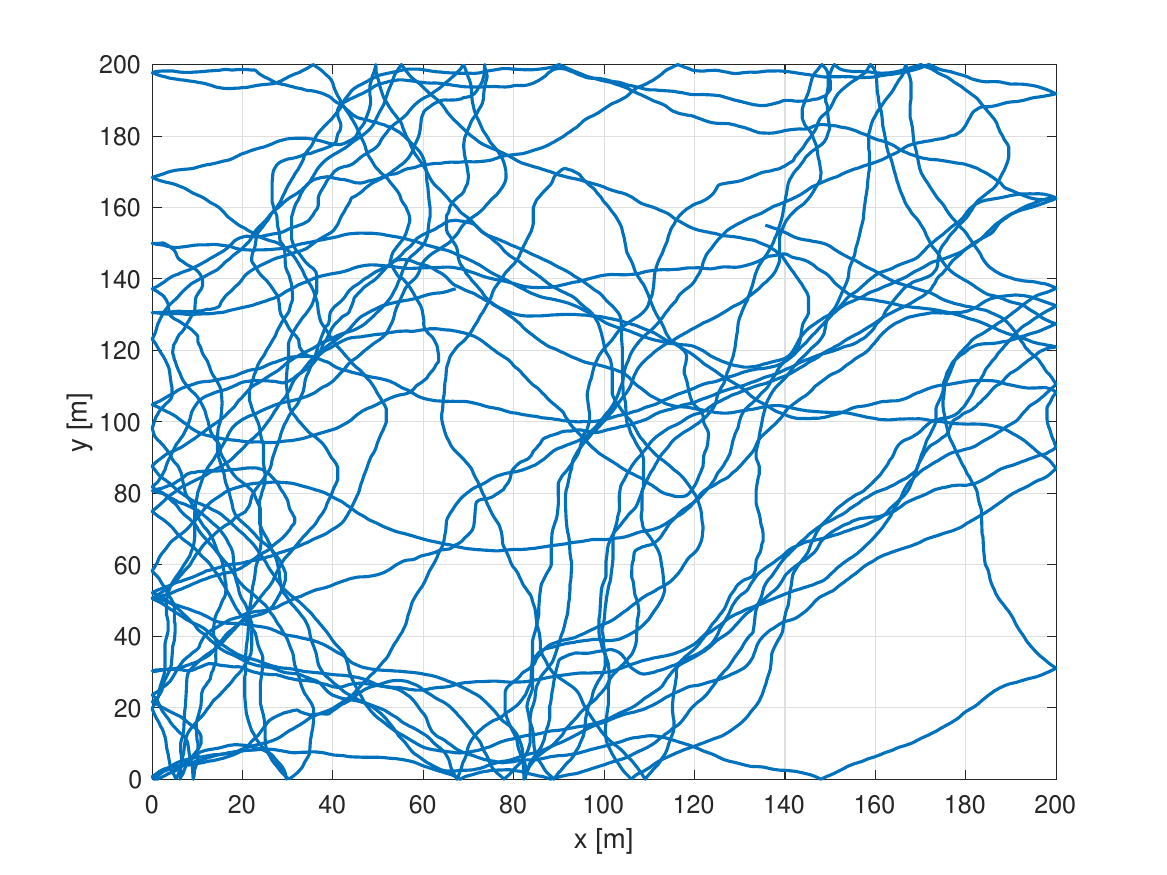}
 \caption{Movement pattern of a node following the Boundless mobility model with $T=0.05\, s$, $\gamma_{max}=\pi / 2\, rad/s$, $a_{max}=5\, m/s^2$ and $v_{max}=5\, m/s$.}
\label{fig:Boundless}
\end{figure}

Any individual mobility model can be adopted in Mo\textsuperscript{3} to implement the Individual Mobility rule: the choice only depends on the specific mobility scenario under consideration, that defines the desired behavior for the nodes in the network.\\ 
The model adopted in this works was selected based on the review carried out in Section \ref{sec:IndividualModels}. The review highlighted that memoryless models typically lead to unrealistic mobility patterns due to the sudden changes in speed and direction. Among the memory-based models, the Boundless model emerged as a good compromise between accuracy and flexibility in providing realistic mobility patterns, as shown in Figure \ref{fig:Boundless}, presenting a node mobility pattern obtained with the Boundless mobility model in an area of $200x200\, m^2$, with $T=0.05\, s$, $\gamma_{max}=\pi / 2\, rad/s$, $a_{max}=5\, m/s^2$ and $v_{max}=5\, m/s$. An individual mobility model inspired by the Boundless model was thus adopted throughout this work, and in particular in the performance evaluation carried out in Section \ref{sec:PerfEvalIntro}.
The model proposed in this work differs from the Boundless model under two aspects: first, the minimum speed, that is always set to 0 in {\eqref{eq:Boundless_speed_vector}} is explicitly defined as $v_{min}$ in the following; second, the model supports tridimensional mobility by means of the following speed vector update rules, that extend {\eqref{eq:Boundless_speed_vector}}:
\begin{equation}
\left\{ \begin{array}{l}
 v\left( {t + T} \right) = \min \left( {\max \left( {v\left( t \right) + \Delta v,v_{min}} \right),v_{max} } \right) \\ 
 \theta \left( {t + T} \right) = \theta \left( t \right) + \Delta \theta, \\ 
 \varphi \left( {t + T} \right) = \varphi \left( t \right) + \Delta \varphi, \\
 \end{array} \right.
\label{eq:Boundless_speed_vector_3D}
\end{equation}
where $v_{max}$, $\Delta v$ and $\Delta \theta$ are defined as in {\eqref{eq:Boundless_speed_vector}}; $\Delta \varphi$ is the elevation angle variation, randomly selected at every update according to a uniform distribution defined on the interval $[-\delta_{max}T, \delta_{max}T]$, where $\delta_{max}$ is the maximum elevation variation speed allowed for a node in rad/s. An example of tridimensional patterns achievable with this rule is presented in Figure {\ref{fig:3DpatternsIM}}.\\
\Figure[t]()[width=0.45\textwidth]{./Figures/3DpatternsFree.eps}
{Example of patterns in three dimensions achievable with the Individual Mobility rule with $T_{IM}=0.05\, s$, $\gamma_{max}=\pi / 2\, rad/s$, $\delta_{max}=\pi / 2\, rad/s$ $a_{max}=5\, m/s^2$, $v_{min}=0.1\, m/s$ and $v_{max}=5\, m/s$.\label{fig:3DpatternsIM}}
It is worth reiterating that all the other rules defined in Mo\textsuperscript{3} and described in the following subsections operate independently from the selected individual mobility model.\\
The Individual Mobility rule is applied with a period $T \equiv T_{IM}$.

\subsection{Correlated Mobility}
\label{sec:Mo3_correlated_mobility}
The Correlated Mobility rule relies on three key concepts: binding, binding condition and grouping condition. The concepts are defined as follows, for a generic node $i$:
\begin{itemize}
    \item A \textbf{binding} is a relation between the mobility patterns of two nodes. If a binding exists between the mobility patterns of $i$ and of a second node $j$, $j$ is referred to as \textit{mate} of node $i$. The mates of nodes $i$ form its so-called binding set $B_i$; $i$ is by definition a member of $B_i$. The size of $B_i$ is indicated with $N_i$.
    \item For each node $j \in B_i$, the following \textbf{binding condition}, inherited from the model proposed in \cite{DeNDiB10}, is defined:
\begin{equation}
d_{ij}\leq D_c,
 \label{eq:Mo3_dist}
\end{equation}
where $d_{ij}=d_{ji}$ is the distance between nodes $i$ and $j$, and the distance $D_c$ is a tunable threshold. If the binding condition in \eqref{eq:Mo3_dist} is satisfied, $i$ is said to be \textit{connected} to $j$. Note that $i$ will always be connected to itself, since $d_{ii}=0$, and the corresponding binding condition is trivial. The set of mates $i$ is connected to is referred to as its \textit{connected set} $C_i$, of size $N_i^c$.
    \item A grouping factor $\rho_i$ is defined as: 
\begin{equation}
\rho_i  = \frac{{N_i^c -1}}{{N_i - 1}},
 \label{eq:Mo3_rho}
\end{equation}
and the following \textbf{grouping condition} is defined on $\rho_i$: 
\begin{equation}
\rho_i \ge \rho _{\min },
 \label{eq:Mo3_condition}
\end{equation}
 where the threshold $\rho _{\min }$ is also a tunable parameter\footnote{For a node $i$ that moves independently of any other node one has $N_i=1$ and $N_i^c=1$ at all times; in this case $\rho_i\equiv1$ by definition, and the grouping condition in {\eqref{eq:Mo3_condition}} is always satisfied.}.
\end{itemize}
The Correlated Mobility rule is applied to each node $i$ with period $T_{CM}$. The application of the rule consists of the following algorithm:
\begin{enumerate}
    \item determine the binding set $B_i$ based on the existing bindings;
    \item determine the connected set $C_i$, by checking the binding condition for each $j\in B_i$;
    \item evaluate $\rho_i$ and determine whether the grouping condition is satisfied. If this is the case, set the node $i$ in a \textbf{Free} state and set $\vec{v_i}'\equiv \vec{v_i}$; if not, set the node $i$ in a \textbf{Forced} state, and take a corrective action by choosing $\vec{v_i}'$ so that the grouping condition can be satisfied in the shortest possible time.
\end{enumerate}
More details on the definition of bindings, on the corrective action associated with the Forced state, and on the definition of connectivity are provided in the three following subsections.
\subsubsection{Binding definition}
\label{sec_Mo3_binding_matrix}
The bindings between nodes are defined by providing a binding matrix $BM$ of size $MxM$, where $M$ is the total number of nodes in the network, defined as follows:
\begin{equation}
 BM = 
\begin{bmatrix} 
b_{1,1}(t)&b_{1,2}(t)&\cdots&b_{1,M}(t)\\
b_{2,1}(t)&b_{2,2}(t)&\cdots&b_{2,M}(t)\\
\vdots &\cdots &\cdots\;&\vdots\\
b_{M,1}(t)\;&b_{M,2}(t)&\cdots\;&b_{M,M}(t)
\end{bmatrix},
\label{eq:Mo3_binding_matrix}
\end{equation}
where 
\begin{equation}
b_{i,j}(t)=
\begin{cases}
1,  & \text{if at time $t$, node $i$ is bound to node $j$}\\
0,  & \text{otherwise}.
\end{cases}
\label{eq:Mo3_binding_eq}
\end{equation}
By definition, $b_{i,i}(t)=1\; \forall t,\,\forall i$. On the other hand, bindings are not necessarily symmetric, so one can have $b_{i,j}(t)\neq b_{j,i}(t)$ for $i\neq j$.\\
The dependence on time $t$ of the elements of $BM$ allows to define dynamic bindings, a feature lacking in most models reviewed in Section \ref{sec:CorrelatedModels}. The impact of dynamic bindings as provided by Mo\textsuperscript{3} will be analyzed in Section \ref{sec:PerfEvalMixed}; out of simplicity, in the remainder of this Section a static case where bindings do not vary in time will be considered, and the dependence on $t$ will be dropped.\\
\begin{figure*}[!ht]
    \centering
    \subfloat[Individual mobility]{
        $
\begin{bmatrix}
        1&0&0&0&0&0&0&0\\
        0&1&0&0&0&0&0&0\\
        0&0&1&0&0&0&0&0\\
        0&0&0&1&0&0&0&0\\
        0&0&0&0&1&0&0&0\\
        0&0&0&0&0&1&0&0\\
        0&0&0&0&0&0&1&0\\
        0&0&0&0&0&0&0&1
\end{bmatrix}
        $
        \label{fig:IndMobilityEx}
    }\hspace{0.5in}
    \subfloat[1 group of 8 nodes]{
        $
        \begin{bmatrix} 
        1&1&1&1&1&1&1&1\\
        1&1&1&1&1&1&1&1\\
        1&1&1&1&1&1&1&1\\
        1&1&1&1&1&1&1&1\\
        1&1&1&1&1&1&1&1\\
        1&1&1&1&1&1&1&1\\
        1&1&1&1&1&1&1&1\\
        1&1&1&1&1&1&1&1
        \end{bmatrix}
$\label{fig:GrpMobilityEx1}
    }\hspace{0.5in}    
    \subfloat[2 groups of 4 nodes each]{
        $
        \begin{bmatrix} 
        1&1&1&1&0&0&0&0\\
        1&1&1&1&0&0&0&0\\
        1&1&1&1&0&0&0&0\\
        1&1&1&1&0&0&0&0\\
        0&0&0&0&1&1&1&1\\
        0&0&0&0&1&1&1&1\\
        0&0&0&0&1&1&1&1\\
        0&0&0&0&1&1&1&1
        \end{bmatrix}
$\label{fig:GrpMobilityEx2}
    }    
    
    \subfloat[2 groups of 3 and 5 nodes, respectively]{
       $
        \begin{bmatrix} 
        1&1&1&0&0&0&0&0\\
        1&1&1&0&0&0&0&0\\
        1&1&1&0&0&0&0&0\\
        0&0&0&1&1&1&1&1\\
        0&0&0&1&1&1&1&1\\
        0&0&0&1&1&1&1&1\\
        0&0&0&1&1&1&1&1\\
        0&0&0&1&1&1&1&1
        \end{bmatrix}
$
        \label{fig:GrpMobilityEx3}
    }\hspace{0.5in}
        \subfloat[1 group of 7 nodes and 1 node moving individually]{
        $
        \begin{bmatrix} 
        1&1&1&1&1&1&1&0\\
        1&1&1&1&1&1&1&0\\
        1&1&1&1&1&1&1&0\\
        1&1&1&1&1&1&1&0\\
        1&1&1&1&1&1&1&0\\
        1&1&1&1&1&1&1&0\\
        1&1&1&1&1&1&1&0\\
        0&0&0&0&0&0&0&1
        \end{bmatrix}
$
        \label{fig:GrpMobilityEx4}
    }\hspace{0.5in}
    \subfloat[2 groups of 3 nodes and 1 group of two nodes, with bindings between groups]{
        $
        \begin{bmatrix} 
        1&1&1&1&0&0&1&0\\
        1&1&1&0&0&0&0&0\\
        1&1&1&0&0&0&0&0\\
        1&0&0&1&1&1&1&0\\
        0&0&0&1&1&1&0&0\\
        0&0&0&1&1&1&0&0\\
        1&0&0&1&0&0&1&1\\
        0&0&0&0&0&0&1&1
        \end{bmatrix}
$
        \label{fig:GrpMobilityEx5}
    }
    \caption{Examples of binding matrices for a network of $M$=8 nodes defining: (a) individual mobility patterns; (b)-(d) group mobility patterns; (e) mixed individual/group mobility patterns; (f) group mobility patterns with group coordination. 
    \label{fig:BM_examples}}
\end{figure*}
The binding matrix provides an intuitive representation of the correlation between mobility patterns: for a given node $i$, the size and the composition of its binding set $B_i$ can be immediately identified by inspecting the $i$-th row of $BM$. A few notable configurations for $BM$ can be identified:
\begin{itemize}
    \item $BM\equiv\mathbb{I}$ -- each node is only bound to itself; this corresponds to a complete absence of correlation. Each mode will move according to its own individual mobility model;
    \item $BM$ is a diagonal block matrix with the generic $LxL$ diagonal block equal to $J_L$\footnote{$J_L$ is defined in linear algebra as an $LxL$ matrix of all ones.} -- each block defines a group as defined in the reference-based models reviewed in Section \ref{sec:CorrelatedModels}.
\end{itemize}

Figure~\ref{fig:BM_examples} presents a few examples of binding matrices falling into one of the two above categories (see Figures \ref{fig:IndMobilityEx}-\ref{fig:GrpMobilityEx4}). Note that mixed scenarios, where some nodes adopt a group mobility behavior while the remaining ones move according to an individual mobility model, can be described by simply defining diagonal blocks of size 1, as shown in Figure~\ref{fig:GrpMobilityEx4}. The flexibility provided by the binding mechanism allows, however, to model any configuration, including those not belonging neither to individual mobility nor to group/mixed mobility as described above. This provides Mo\textsuperscript{3} with the capability of mimicking most of the correlated mobility models introduced in the literature. An example of this feature is shown in Figure \ref{fig:GrpMobilityEx5}, presenting a binding matrix that defines three groups, formed by a) nodes $1$, $2$, $3$, b) nodes $4$, $5$, $6$ and c) nodes $7$, $8$. The binding matrix however also introduces inter-group bindings between nodes $1$, $4$ and $7$, emulating the feature of Group Coordination, as defined in \cite{AunSee15}. The capability of Mo\textsuperscript{3} to emulate other models is further discussed in Section \ref{sec:Mo3_mimetic_abilities}.
 \subsubsection{Corrective action for a node in Forced state}
\label{sec_Mo3_corrective_action}
\begin{figure}[t]
  \centering
\includegraphics[width=3.2in]{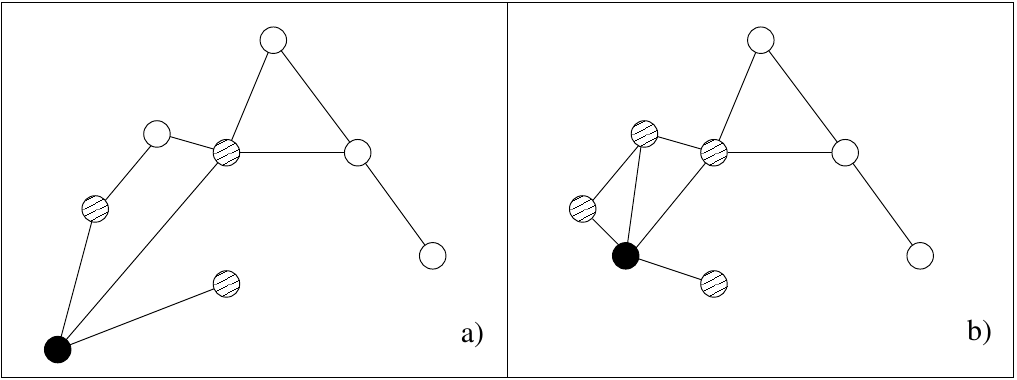}
 \caption{Application of the Correlated Mobility rule for a node with a binding set of size 8, with $\rho_{min}=0.5$. In a) the size of the connected set of the considered node (black circle) is 4 (the node itself and three connected mates, represented as striped circles) leading to a grouping factor $\rho=3/7$, lower than the required $\rho_{min}$. As a consequence, the node enters in \emph{Forced} state and moves towards the closest mate out of its connected set. The node stays in \emph{Forced} state until the grouping condition is satisfied, eventually reaching the configuration in b), where the size of the connected set is increased to 4, corresponding to $\rho\geq \rho_{min}$.}
 \label{fig:Mo3_example}
\end{figure}

When $i$ is in Forced state, the following corrective action is taken:
\begin{enumerate}
    \item select the closest mate not part of $C_i$, defined as:
    \begin{equation}
k = \argmin_{j\in B_i, j \notin C_i} \left\{d_{ij}\right\};
\label{eq:Mo3_closest_mate}
\end{equation}

\item select $v_i'$, $\theta_i'$  and $\varphi_i'$ as follows:
\begin{equation}
v_i'  = v_{max},
\label{eq:Mo3_forced_v}
\end{equation}
\begin{equation}
\theta_i' = \theta_{ki}=\arctantwo\left(y_k-y_i,x_k-x_i\right),
\label{eq:Mo3_forced_theta}
\end{equation}
\begin{equation}
\varphi_i' = \varphi_{ki}=\arcsin\left(\frac{z_k-z_i}{d_{ik}}\right),
\label{eq:Mo3_forced_phi}
\end{equation}
where the $\arctantwo\left(x,y\right)$ operator returns the principal value of $\arctan\left(y/x\right)$ in $\left[-\pi, \pi\right]$. 
\end{enumerate}
Equations \eqref{eq:Mo3_forced_v}, \eqref{eq:Mo3_forced_theta} and \eqref{eq:Mo3_forced_phi} ensure that node $i$ adopts the speed vector that will reach the current position of the selected mate $k$ in the shortest possible time. The pair $\left\{\theta_{ki},\varphi_{ki}\right\}$ identifies in fact the direction of the vector centered in the current position of node $i$ and pointing to the current position of node $k$. The equations also address the case of bidimensional mobility, since in this case $z_i=z_k$ and thus $\varphi_{ki}=0$; $\theta_{ki}$ is therefore the direction from the current position of node $i$, $(x_i,y_i)$, to the current position of node $k$, $(x_k,y_k)$. 
Figure \ref{fig:Mo3_example} shows an example of application of the  Correlated Mobility rule in the case of a node (black circle) with a binding set of size $N=8$, with $\rho_{min}=0.5$; arcs between nodes indicate connectivity as defined in \eqref{eq:Mo3_dist}. In Figure \ref{fig:Mo3_example}a the size $N^c=3$ of the connected set (striped circles) for the node leads to a grouping factor $\rho=0.43$. The grouping condition is thus not satisfied, and the node moves toward the closest mate among those it is not connected to (white circles), until the condition is satisfied (Figure \ref{fig:Mo3_example}b, where $N^c=4$ and $\rho=0.57$).\\%
An example of the movement patterns obtained in a bidimensional mobility scenario by applying the Correlated Mobility rule for a single group of 4 nodes,  corresponding to the binding matrix in Figure \ref{fig:BM_examples}b with $M=4$, is presented in Figure \ref{fig:Mo3_pattern}, that also highlights the state of each node (Free vs. Forced). An example of movement patterns obtained in a tridimensional mobility scenario is shown in Figure {\ref{fig:3DpatternsCM}}. \\ 

\Figure[t]()[width=0.45\textwidth]{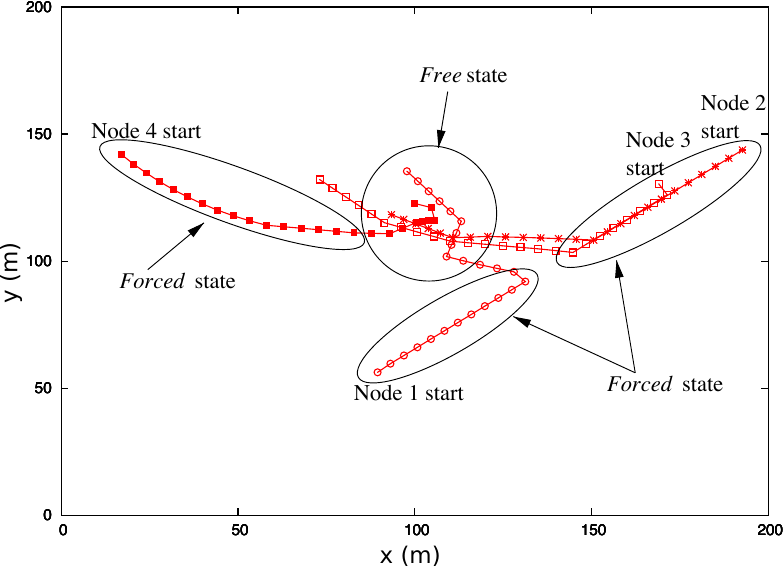}
{Mobility patterns obtained by applying the Correlated Mobility rule for a single group of 4 nodes.  As nodes start from random positions, the grouping factor $\rho$ is below the threshold for all of them. Nodes thus start moving in \emph{Forced} state; as soon they achieve $\rho>\rho_{min}$ they switch to \emph{Free} state. The patterns were obtained using the Individual Mobility rule proposed in Section \ref{sec:Mo3_individual_model} and the following settings: area size $200x200\, m^2$,  $v_{max}=5\, m/s$, $v_{min}=0.1\, m/s$, $a_{max}= 5\, m/(s^2)$, $\gamma_{max}=\pi /2\, rad/s$, $T_{IM}=0.05\, s$, $\rho_{min}=0.7$, $D_c= 30\, m$, $T_{CM} =1\, s$. \label{fig:Mo3_pattern}}

\Figure[t]()[width=0.45\textwidth]{./Figures/3DpatternsCorr.eps}
{Mobility patterns in three dimensions obtained by applying the Correlation Mobility rule for a single group of 4 nodes and the following settings: area size $200x200\, m^2$,  $v_{max}=5\, m/s$, $v_{min}=0.1\, m/s$, $a_{max}= 5\, m/(s^2)$, $\gamma_{max}=\pi /2\, rad/s$,$\delta_{max}=\pi /2\, rad/s$, $T_{IM}=0.05\, s$, $\rho_{min}=0.7$, $D_c= 10\, m$, $T_{CM} =0.01\, s$.\label{fig:3DpatternsCM}}

\subsubsection{Definition of connectivity and meaning of $D_c$}
\label{sec_Mo3_connectivity}
The definition of connectivity, and the corresponding meaning of the threshold $D_c$, depends on the mobility scenario; the binding condition adopted in Mo\textsuperscript{3} is general enough to address a wide range of scenarios. Two possible examples are the following:
\begin{itemize}
\item connectivity related to radio communications - in this case two mates will be considered as connected if they can communicate through a direct radio link (physical layer connectivity), and $D_c$ will be set depending on the radio coverage;
\item connectivity based on a radio-independent parameter - for example, if a group corresponds to a security team, connectivity may correspond to physical visibility: a team member will be connected to another if they are in line of sight.
\end{itemize}

\subsection{Collision Avoidance}
\label{sec:Mo3_CollisionAvoidance}
\begin{figure}[t]
 \centering
\includegraphics[width=3.2in]{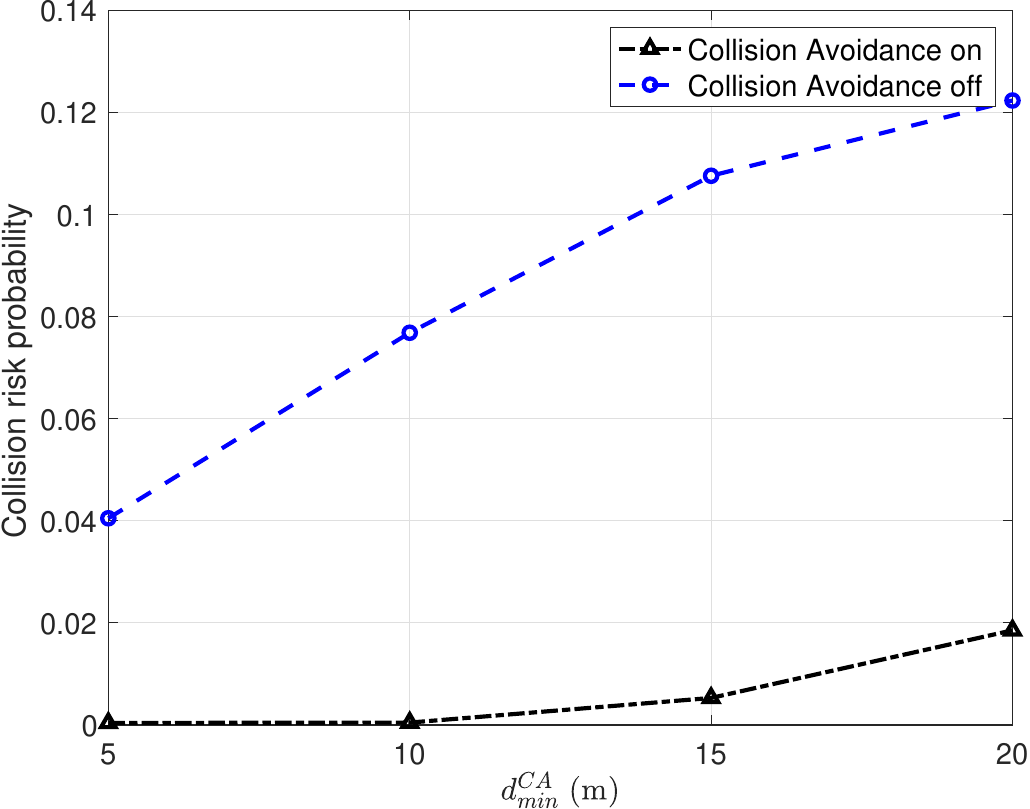}
 \caption{Collision risk probability as a function of $d_{min}^{CA}$ observed using the Collision Avoidance module in Mo\textsuperscript{3} vs. no collision avoidance, in a scenario with $M=5$ nodes moving in an area of $50x50\, m^2$.  Mo\textsuperscript{3} was configured as follows: $T_{IM}=0.1\, s$, $d^{CA}_{trigger}=20\, m$, $T_{CA}=0.01\, s$, $\gamma_{max}=\pi / 2\, rad/s$, $a_{max}=2\, m/s^2$, $v_{max}=0.5\, m/s$ and $v_{min}=0.001\, m/s$; the Correlated Mobility module was disabled.}
\label{fig:CA_effect}
\end{figure}

Collision avoidance in Mo\textsuperscript{3} aims at predicting potential collisions, based on the positions and speed vectors of nodes, and taking a corrective action before they happen. A corrective action could in general include a change in both speed and direction of a node; in Mo\textsuperscript{3}, however, the Collision Avoidance rule introduces modifications to speed only, unless a change of direction is absolutely necessary to avoid a frontal collision \footnote{The probability of a frontal collision is negligible when speeds and directions are selected randomly. For this reason, this issue is addressed in Appendix \ref{sec:CAdetails}, focusing in this Section on the more common case of collisions that are not frontal.}. This choice has two justifications: first, it allows to avoid collisions with other nodes without changing course, which is a reasonable model of what happens in real world mobility; secondly, it minimizes the conflicts between the corrections introduced by Collision Avoidance vs. Obstacle Avoidance rules. As it will be detailed in Section \ref{sec:Mo3_ObstacleAvoidance}, in fact, the Obstacle Avoidance rule only operates on direction of movement and not on speed.\\
The core idea in the Collision Avoidance rule proposed in  Mo\textsuperscript{3} is to identify collision risks based on the current trajectories of nodes rather than on their mutual distance; for example, two nodes that move on parallel lines will never trigger a collision risk alert in Mo\textsuperscript{3}, regardless of their distance. A collision risk is in fact identified for a node only if its current speed vector will place it within less than $d_{min}^{CA}$ meters from any other node when either of them reaches the location where their paths would cross (\textit{crossing point}). An example of a scenario potentially causing a collision risk is presented in Figure \ref{fig:CA_example}. Figure \ref{fig:CA_example} highlights for either node the positions that would trigger a collision risk alert when the other node is at the crossing point, corresponding to the segment of its planned path that falls within a circle of radius $d_{min}^{CA}$ centered on the crossing point.\\The Collision Avoidance rule is applied for each node $j$ with period $T_{CA}$.%
\Figure[t]()[width=0.45\textwidth]{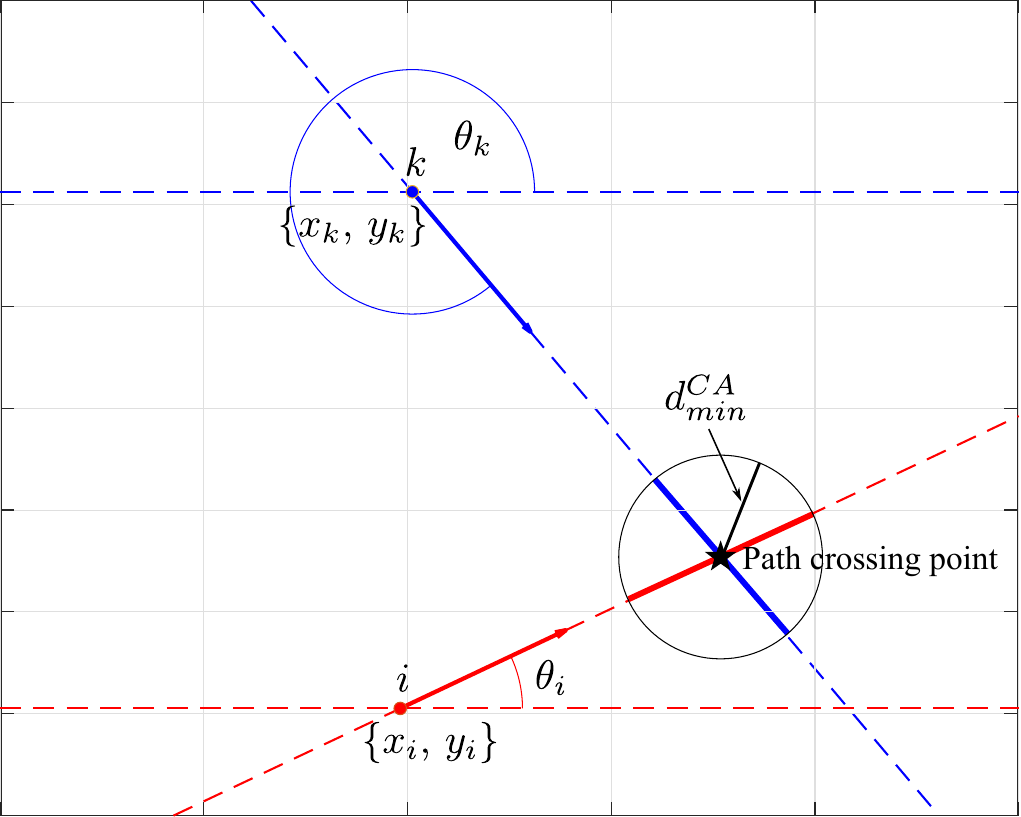}%
{Example of possible collision risk considering two nodes $i$ and $k$ with crossing paths. A collision risk is detected if the current speeds of the two nodes would lead either node to occupy any location in the segment of its own path highlighted in bold when the other node is at the path crossing point; the length of each segment is equal to $2*d_{min}^{CA}$.\label{fig:CA_example}}
A brief description of the algorithm implementing the rule is provided in the following; details on the computations carried out at each step are provided in Appendix \ref{sec:CAdetails}.\\
The output speed $v_i'$ for node $i$ is determined as follows:
\begin{itemize}
\item \textit{Path crossing identification} - trajectories of all nodes within a distance $d_{trigger}^{CA}$ from $i$ are analyzed, and nodes that are on trajectories crossing the current trajectory of $i$ are added to the set of Path Crossing Nodes $\left\{PCN\right\}$;
\item \textit{Speed bounds determination} - for each node $k \in \left\{PCN\right\}$, the range $[v_L^k,\; v_U^k]$ is determined for $v_i'$, that guarantees that $d_{ik} \geq d_{min}^{CA}$ when either node reaches the crossing point; a corresponding constraint is defined, expressed by the inequality $v_L^k \leq v_i' \leq v_U^k$;
\item \textit{Collision avoidance} - the set of constraints on the new speed $v_i'$ determined at the previous step is analyzed in order to determine whether the corresponding system of inequalities can be solved. As shown in Appendix~\ref{sec:CAdetails}, the analysis will lead to one of three possible outcomes:
\begin{enumerate}
    \item no collision risk is identified, and no action is taken, leading to $v_i'\equiv v_i$;
    \item a collision risk involving one or more nodes in $\left\{PCN\right\}$ is identified, and can be addressed by choosing a $v_i'\neq v_i$;
    \item a collision risk involving one or more nodes in $\left\{PCN\right\}$ is identified and cannot be addressed: the $v_i'\neq v_i$ that best mitigates the risk is selected, and a further attempt to fully address it will take place at the next application of the Collision Avoidance rule.
\end{enumerate} 
\end{itemize}
Figure \ref{fig:CA_effect} shows the probability of a collision risk for a node, estimated as the average frequency of observed collision risks in generated mobility patterns, as a function of $d_{min}^{CA}$ in a scenario considering $M=5$ nodes moving independently in an area $50x50\, m^2$. Figure \ref{fig:CA_effect} highlights that the Collision Avoidance rule is effective in mitigating the risk of collisions. As the required $d_{min}^{CA}$ increases, the probability of a collision risk increases as well, since the condition to be met is harder and harder to satisfy; nevertheless the application of the Collision Avoidance rule reduces in all cases the probability of a collision risk by about one order of magnitude.
\subsection{Obstacle Avoidance}
\label{sec:Mo3_ObstacleAvoidance}
Several approaches to obstacle avoidance were proposed in the past. The Obstacle Model \cite{JarBel03} was designed to model the movement of people walking around and through buildings; it defined obstacles shaped as polygons, and restricted nodes to move on a set of paths equidistant from obstacles, determined by partitioning the movement area according to a Voronoi diagram, with the corners of obstacles as location points for the diagram. An evolution of this model was proposed in \cite{Hua05}, in which obstacles were defined as rectangles of arbitrary size and orientation, but nodes' movement patterns were still restricted to a set of paths between obstacles, albeit richer than the one obtained in \cite{JarBel03}. GFMM \cite{WilHua09} adopted instead a behavioral approach, that allowed nodes to occupy any position in areas not covered by obstacles, by defining a repulsion force from obstacles, but without providing a mechanism to define obstacles of arbitrary size and shape.\\
Mo\textsuperscript{3} introduces an Obstacle Avoidance rule that shares with \cite{JarBel03}, \cite{Hua05} a geometrical approach to the definition of obstacles, but provides complete freedom of movement for nodes around and between obstacles as in \cite{WilHua09}. Section \ref{sec:ObstacleDefinition} introduces the basic obstacle shapes available in Mo\textsuperscript{3} and how they are defined, while Section \ref{sec:ObstacleAvoidanceAlgorithm} describes how obstacles are detected and avoided by nodes in Mo\textsuperscript{3}.
\subsubsection{Obstacle definition}
\label{sec:ObstacleDefinition}
Obstacles in Mo\textsuperscript{3} can take the basic shapes of either a rectangle or an ellipse\footnote{Both ellipses and rectangles can only be defined with one axis/side parallel to the horizontal axis of the movement area. The extension to shapes with arbitrary rotation is left for future work.}. The obstacles are defined by providing the geometric information needed to place them in the movement area, that is: 1) the coordinates of the center and the length of horizontal semi axis and vertical semi axis for an ellipse, and  2) the coordinates of the center and the length of horizontal and vertical sides for a rectangle. Obstacles with more elaborate shapes can be obtained by combining (possibly overlapping) basic shapes.\\
The choice of defining obstacles over a set of basic shapes allowed to define an Obstacle Avoidance rule based on a geometric approach, as explained in the following subsection.
\subsubsection{Obstacle avoidance rule}
\label{sec:ObstacleAvoidanceAlgorithm}
\begin{figure*}[!t]
    \centering
    \subfloat[2 obstacles, $M=1$ node]{
        \includegraphics[scale=0.45]{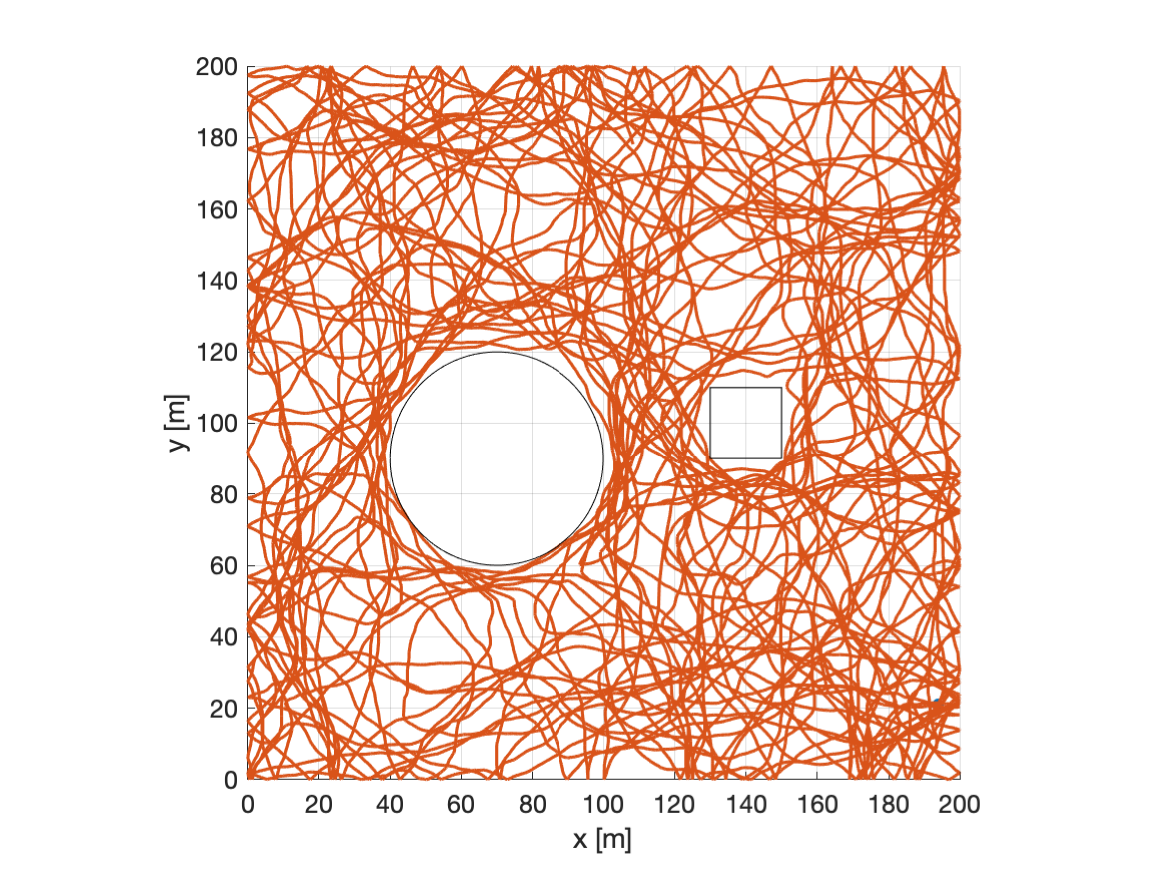}
        \label{fig:OA_two_obstacles_1_node}
    }
    \subfloat[2 obstacles, $M=5$ nodes]{
        \includegraphics[scale=0.45]{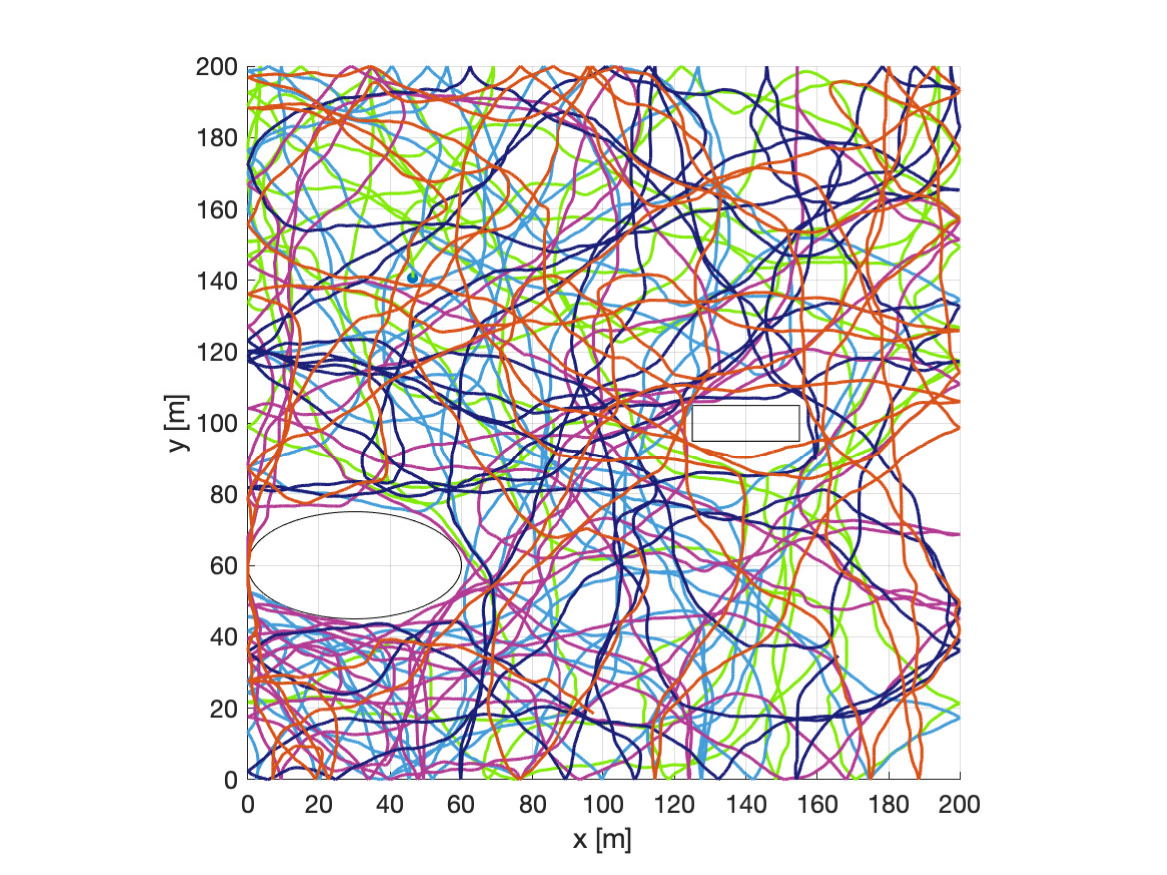}
        \label{fig:OA_two_obstacles_5_nodes}
    }
    
    \subfloat[1 obstacle, $M=1$ node, $d_{trigger}^{OA}=10\,m$]{
        \includegraphics[scale=0.45]{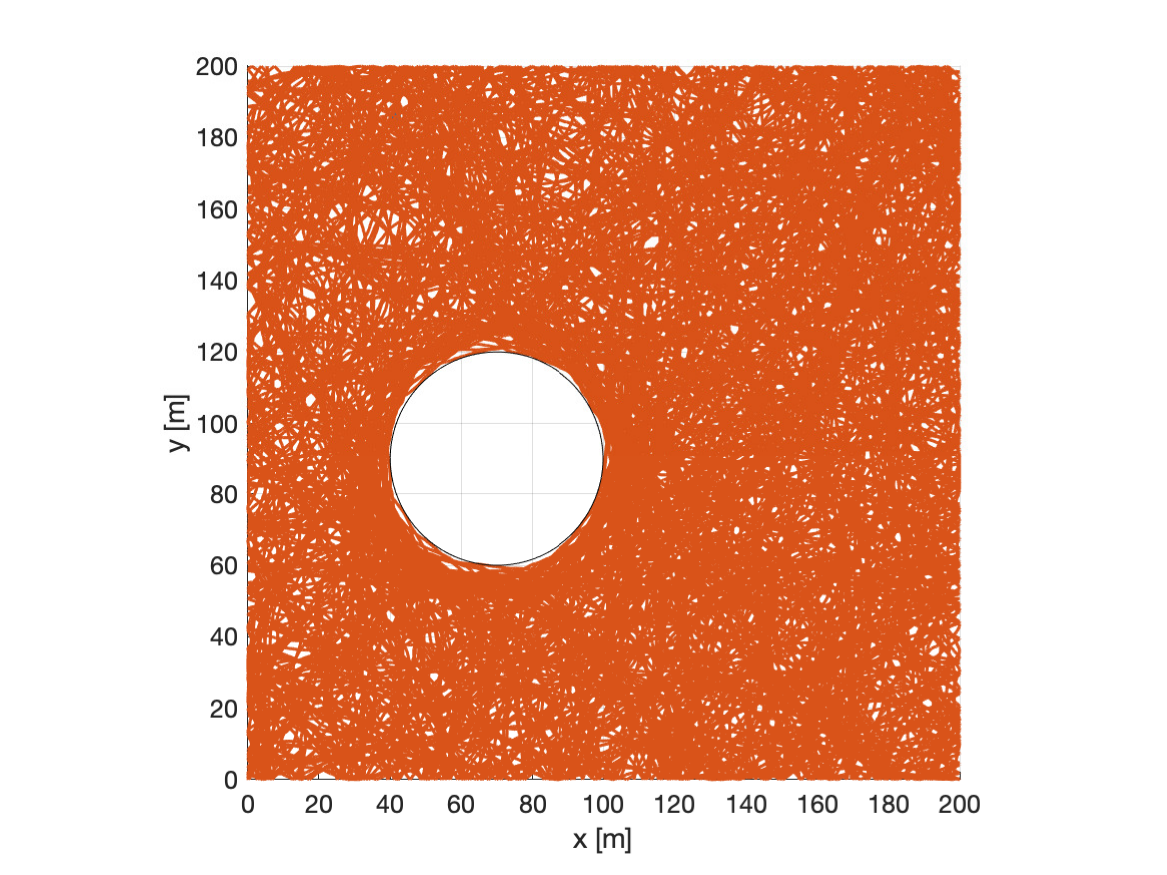}
        \label{fig:OA_one_obstacles_d_10}
    }
        \subfloat[1 obstacle, $M=1$ node, $d_{trigger}^{OA}=40\,m$]{
        \includegraphics[scale=0.45]{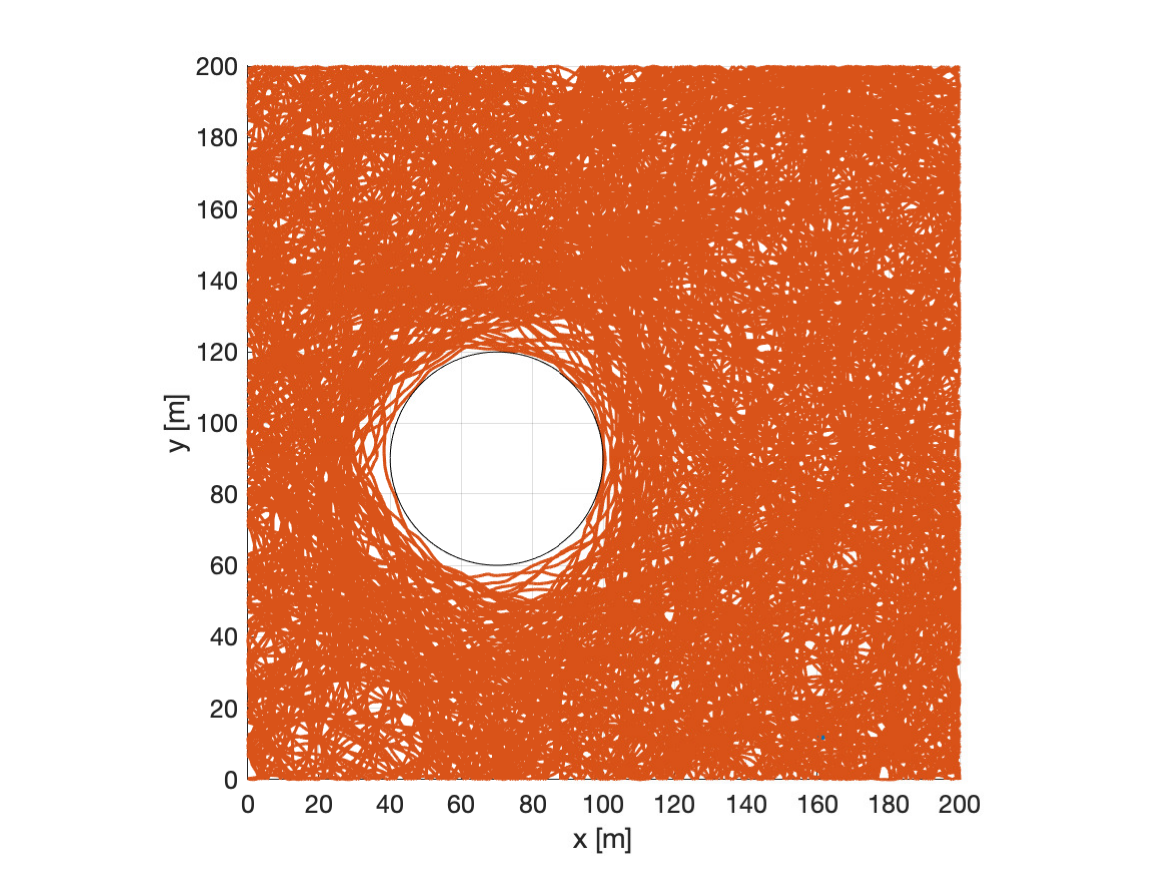}
        \label{fig:OA_one_obstacles_d_40}
    }
    
    \caption{Mobility patterns obtained adopting the Collision Avoidance algorithm: (a) 1 node, two obstacles; (b) 5 nodes,  two obstacles; (c) 1 node, one obstacle, $d_{trigger}^{OA}=10\,m$; (d) 1 node, one obstacle, $d_{trigger}^{OA}=40\,m$. All remaining settings are as follows: $T_{OA}=0.05\, s$, $T_{IM}=0.05\, s$, $\gamma_{max}=\pi / 2\, rad/s$, $a_{max}=5\, m/s^2$, $v_{max}=5\, m/s$ and $\theta^{OA}=\pi/10$.
    \label{fig:OA_mobility_patterns}}
\end{figure*}

\begin{figure*}[!t]
    \centering
    \subfloat[Perfect reflection]{
        \includegraphics[scale=0.6]{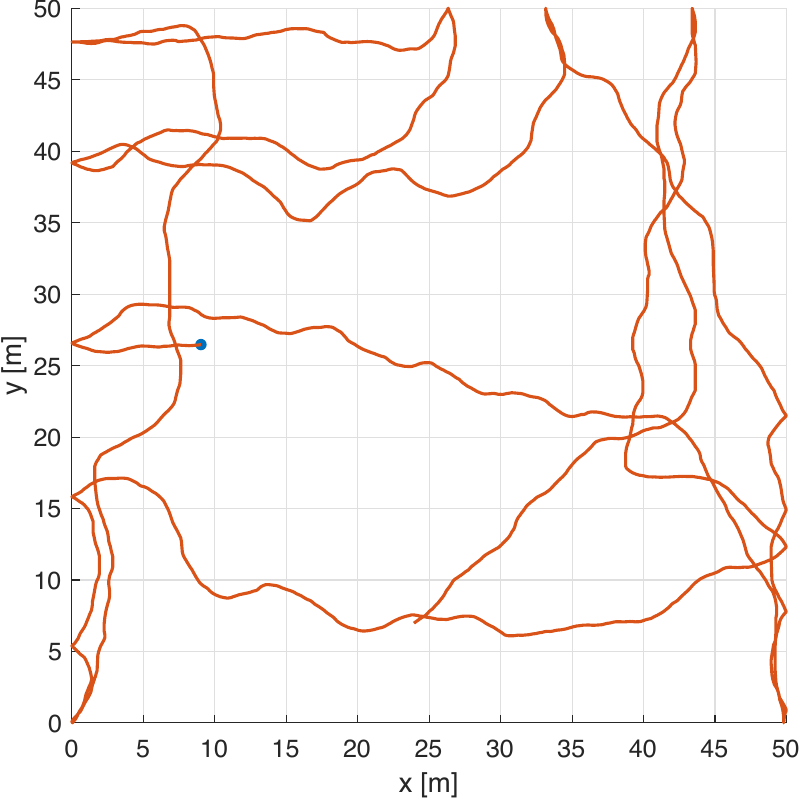}
        \label{fig:reflection}
    }
    \hfill
    \subfloat[Bounding box defined using the Obstacle Avoidance rule in Mo\textsuperscript{3}]{
        \hspace*{-.1in}
        \includegraphics[scale=0.6]{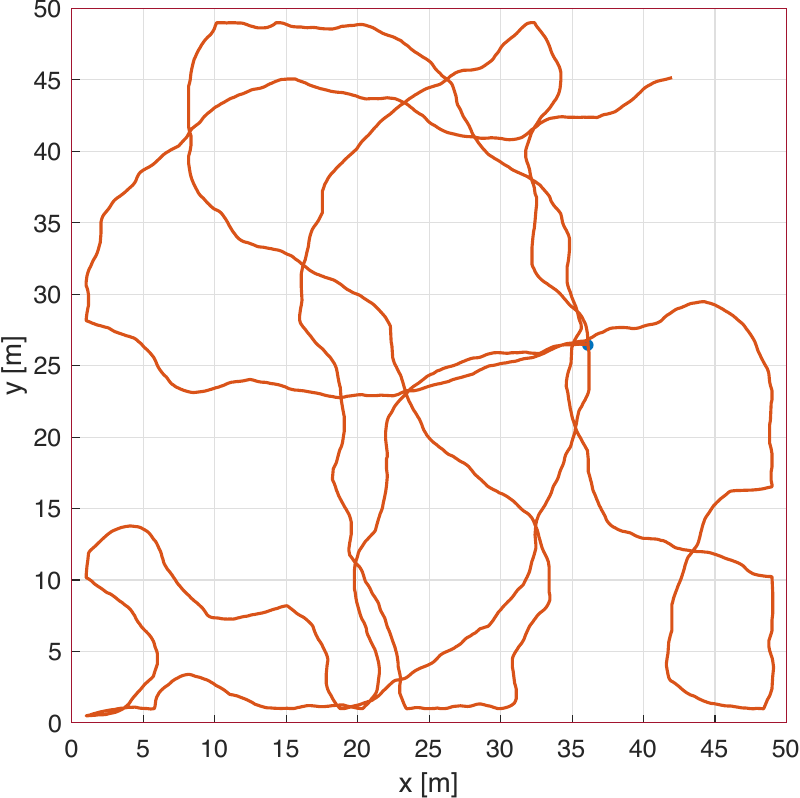}
        \label{fig:smooth}
    }
    \hfill
    \caption{Patterns obtained using: a) a perfect reflection law when hitting a boundary of the movement area vs. b) the obstacle avoidance feature in Mo\textsuperscript{3} to introduce a bounding box by means of four narrow rectangles. Mo\textsuperscript{3} was configured as follows: $T_{OA}=0.01\, s$, $d^{OA}_{trigger}=1\, m$, $T_{IM}=0.1\, s$, $\gamma_{max}=\pi / 2\, rad/s$, $a_{max}=2\, m/s^2$, $v_{max}=0.5\, m/s$ and $\theta^{OA}=\pi/10$.}
      \label{fig:smoothborders}
\end{figure*}

Since a change in speed would not avoid a collision with an obstacle, the Obstacle Avoidance only modifies the direction of movement. As anticipated, this choice has also the advantage of decoupling the corrections applied by this rule from those introduced by the Collision Avoidance rule, mitigating the risk of conflicts.\\
The Obstacle Avoidance rule is applied for each node with period $T_{OA}$, taking into account all obstacles within a distance $d_{trigger}^{OA}$ from the node. The algorithm that determines the new direction $\theta_i'$ is as follows:
\begin{itemize}
    \item For each obstacle $k$, determine whether the minimum distance from $i$ to any point in $k$ is lower than $d_{trigger}^{OA}$. If this is not the case, ignore it, otherwise find the two angles $\theta_{i,k}^{min}$ and $\theta_{i,k}^{max}$ that determine the range of directions that would lead to a collision, based on the geometry of the obstacle and the current position of $i$ (see Appendix \ref{sec:OAdetails} for details), and label such range as forbidden\footnote{The concept of forbidden range of directions is similar to the one of \emph{obstruction cone} introduced in \cite{JarBel03}. However, in \cite{JarBel03} the concept was adopted only to model radio blockage, and only for rectangular obstacles.}.
    \item Determine the range of allowed directions $A_\theta$ as the complement to $\left[-\pi,\pi\right]$ of the union of all the forbidden ranges determined at the previous step, and compare it with the current direction $\theta_i$. The analysis can lead to either of the following outcomes\footnote{The possible outcomes are defined assuming that the range of allowed directions is not empty. The case where no allowed direction exists corresponds to a scenario where a node is surrounded in all directions by obstacles at distance shorter than $d_{trigger}^{OA}$, that is extremely unlikely in common mobility scenarios.}:
    \begin{enumerate}
        \item $\theta_i$ is within the allowed range: no collision risk is identified, and no action is taken, leading to $\theta_i'\equiv \theta_i$;
        \item $\theta_i$ is not within the allowed range: a collision risk involving one or more obstacles is identified, and is addressed by selecting a $\theta_i'\in A_\theta$ defined as follows:
            \begin{equation}
\theta_i' = \argmin_{\theta_i'\in A_\theta} \left\{\left|\theta_i' - \theta_i\right|\right\}+\theta^{OA}\cdot\sign\left(\theta_i' - \theta_i\right).
\label{eq:Mo3_OA_best_angle}
\end{equation}
    \end{enumerate}
The angle identified by \eqref{eq:Mo3_OA_best_angle} is the one minimizing the correction introduced on the input direction, shifted by a small margin $\theta^{OA}$ that ensures that $i$ will not touch the obstacle at the tangent point.
\end{itemize}
An example of the application of the Obstacle Avoidance rule for a generic node $i$ is presented in Figure~\ref{fig:OA_example}, showing a scenario with three obstacles: a rectangle, labeled with $1$, an ellipse ($2$), and a circle ($3$). Obstacle $3$ is excluded from the computation of the forbidden ranges, since it is not within a distance $d_{trigger}^{OA}$ from $i$. The allowed range $A_\theta$, obtained as the complement to $\left[-\pi,\pi\right]$ of the union of the forbidden ranges determined by obstacles $1$ and $2$, is identified by the green shaded area. A collision risk is identified, since $\theta_i\notin A_\theta$, and in the considered scenario $\theta_i'= \theta_{i,1}^{max}+\theta^{OA}$ is selected to address the risk.\\
\Figure[t]()[width=0.45\textwidth]{./Figures/Figure_OA_example.eps}
{Application of the  Obstacle Avoidance rule for a node $i$ in presence of three obstacles: a rectangle, labeled with $1$, an ellipse ($2$), and a circle ($3$) at distance larger than $d_{trigger}^{OA}$; the allowed range of directions $A_\theta$ is highlighted with a shade of green. A collision risk is detected with the rectangle, and $\theta_i'=\theta_{i,1}^{max}+\theta^{OA}$ is adopted to address it.\label{fig:OA_example}}
Several examples of mobility patterns obtained using the Obstacle Avoidance rule are presented in Figure~\ref{fig:OA_mobility_patterns}. Figure \ref{fig:OA_two_obstacles_1_node} and Figure \ref{fig:OA_two_obstacles_5_nodes} consider two scenarios with 2 obstacles, respectively a circle and a square with $M=1$ vs. a rectangle and an ellipse with $M=5$. Figure \ref{fig:OA_one_obstacles_d_10} and \ref{fig:OA_one_obstacles_d_40} highlight the impact of $d_{trigger}^{OA}$, by showing the patterns obtained in the same scenario with a single obstacle and $M=1$ for $d_{trigger}^{OA}=10\,m$ vs. $d_{trigger}^{OA}=40\,m$ over a long observation time, that allows for an almost full coverage of the movement area by the node. A comparison between Figure \ref{fig:OA_one_obstacles_d_10} and \ref{fig:OA_one_obstacles_d_40} shows that collisions are prevented in both cases, but a larger $d_{trigger}^{OA}$ allows an early application of the Collision Avoidance rule, resulting in trajectories farther away from the obstacle compared to those obtained for a small $d_{trigger}^{OA}$, as highlighted by a lower density of trajectories in close vicinity to the obstacle in Figure  \ref{fig:OA_one_obstacles_d_40}.\\
The Obstacle Avoidance rule also allows to define pathways and corridors, by delimiting them with properly placed obstacles, thus implementing the spatial constraint feature introduced in \cite{ZhoXu04} as part of the definition of the Virtual Track Group Mobility (VTGM) model, and listed in \cite{AunSee15} as a desirable feature for a mobility model. An example of a mobility pattern obtained by taking advantage of such feature is presented in Figure \ref{fig:OA_spatial_constraint}.
\Figure[t]()[width=0.45\textwidth]{./Figures/Figure_OA_spatial_constraints.eps}
{Example of a mobility pattern for a network of $M=3$ nodes in presence of spatial constraints, defining a set of hallways in a $50x50\,m^2$ room. Mo\textsuperscript{3} settings used to generate the pattern were as follows: $d_{trigger}^{OA}=1\,m$, $T_{OA}=0.01\, s$, $T_{IM}=0.1\, s$, $\gamma_{max}=\pi / 2\, rad/s$, $a_{max}=5\, m/s^2$, $v_{max}=0.5\, m/s$ and $\theta^{OA}=\pi/10$. \label{fig:OA_spatial_constraint}}\\
Finally, the rule can be also used to ensure that nodes remain within the boundaries of the movement area. In most implementations of mobility models this is obtained by adopting a perfect reflection law: nodes hitting a side of the area rebound with same speed and symmetric direction with respect to the normal to the side. The implementation of Mo\textsuperscript{3} made available in \cite{Mo3Rep} supports this approach as well. A smarter solution to this issue can be however adopted by introducing four extremely narrow rectangular obstacles that delimit the movement area. A comparison between the patterns obtained with the two approaches is shown in Figure \ref{fig:smoothborders}, highlighting the smoother patterns obtained when the Obstacle Avoidance rule is used.

\subsection{Upper Bounds Enforcement}
\label{sec:Mo3_UpperBounds}
The Upper Bounds Enforcement rule is derived from the approach proposed for the Boundless model in \cite{Haa97} to limit the linear acceleration in a range $\left[-a_{max},a_{max}\right]$ and angular speed in a range $\left[-\gamma_{max},\gamma_{max}\right]$. At each update, the updated speed vector $\vec{v}_i^{OA}$ for the node $i$ under consideration, obtained after the application of the Individual Mobility, Correlated Mobility, Collision Avoidance and Obstacle Avoidance rules, is compared to the original speed vector $\vec{v_i}^{IN}$ in order to determine the following variation factors:
\begin{equation}
   \left\{
    \begin{aligned}
    \Delta v &=v_i^{OA}-v_i^{IN} \\
    \Delta \theta &=\min \left(\left|\theta_i^{OA}-\theta_i^{IN}\right|,\pi\right)\\
        \Delta \varphi &=\varphi_i^{OA}-\varphi_i^{IN}.
    \end{aligned}\right.
\label{eq:UB_variation_variables}
\end{equation}
These factors will be compared with the maximum variation allowed by the upper bounds on linear acceleration, rotation speed and elevation variation speed, determining the final speed vector $\vec{v_i}^{OUT}$. One has for the speed $v_i^{OUT}$:
\begin{equation}
    v_i^{OUT}=v_i^{IN}+\sign\left(\Delta v \right)\min\left(\left|\Delta v\right|, a_{max}\tau\right),
\label{eq:UB_new_speed}
\end{equation}
where $\tau$ is the time elapsed from the previous mobility update, as already defined in Section \ref{sec:IndividualModels}\footnote{In a discrete implementation of the model, with position updates every $\Delta t$, one has $\tau=\Delta t$. Oppositely, if the position update is triggered on demand, $\tau$ measures the time elapsed from the last position update.}.\\
For the direction, one must consider the boundary between $-\pi$ and $\pi$ in the computation. Let us focus on cases where a violation occurs, corresponding to $\Delta \theta> \gamma_{max}\tau$. For the case $\theta_i^{OA}-\theta_i^{IN}>0$ one can introduce the logical condition:
\begin{equation}
C1=\left[\left(\theta_i^{OA}>0\right) \land  \left( \theta_i^{OA}-\theta_i^{IN} \leq \pi \right)\right] \lor \left(\theta_i^{OA}\leq0\right), 
\end{equation}
that determines $\theta_i^{OUT}$ as follows:
\begin{equation}
\theta_i^{OUT}=
\begin{cases}
\theta_i^{IN} + \gamma_{max}\tau,  & \text{if } C1  \\
\theta_i^{IN} - \gamma_{max}\tau,  & \text{otherwise}.
\end{cases}
\label{eq:UB_new_direction_1}
\end{equation}
For the case $\theta_i^{OA}-\theta_i^{IN}\leq0$, one has the following logical condition:
\begin{equation}
C2=\left[ \theta_i^{IN}>0  \land \left( \theta_i^{OA}-\theta_i^{IN} \geq -\pi \right)\right] \lor \left(\theta_i^{IN}\leq 0\right), 
\end{equation}
that determines $\theta_i^{OUT}$ as follows:
\begin{equation}
\theta_i^{OUT}=
\begin{cases}
\theta_i^{IN} - \gamma_{max}\tau,  & \text{if } C2 \\
\theta_i^{IN} + \gamma_{max}\tau,  & \text{otherwise}.
\end{cases}
\label{eq:UB_new_direction_2}
\end{equation}
In cases where there is no violation, corresponding to $\Delta \theta \leq  \gamma_{max}\tau$, no correction is needed, and thus $\theta_i^{OUT}\equiv\theta_i^{OA}$.\\
In the case of the elevation angle there is no boundary condition to be considered, and the output elevation angle $\varphi_i^{OUT}$ is determined as:
\begin{equation}
    \varphi_i^{OUT}=\varphi_i^{IN}+\sign\left(\Delta \varphi \right)\min\left(\left|\Delta \varphi\right|, \delta_{max}\tau\right).
\label{eq:UB_new_elevation}
\end{equation}
The following observations hold for the Upper Bounds Enforcement rule:
\begin{enumerate}
    \item the rule can be used to introduce a memory effect and enforce bounds in Individual Mobility models that do not provide such features natively;
    \item the rule is mandatory also when the Individual Mobility model has built-in upper bound enforcement, as is the case for the Boundless model, because the Correlated Mobility, Collision Avoidance and Obstacle Avoidance rules may modify the speed vector in a way that would cause violations;
    \item in order to ensure that bounds are never violated, the period of execution of the Upper Bounds Enforcement rule, $T_{UB}$, shall be selected equal to or lower than the minimum update period adopted for the other four rules; 
    \item the enforcement of the bounds in \eqref{eq:UB_new_speed}-\eqref{eq:UB_new_direction_2} may alter the  speed and direction selected as a result of the application of other rules, and in particular of the Collision Avoidance and Obstacle Avoidance rules. In most cases, these alterations will be compensated in subsequent mobility updates, but from time to time they may result in a failure to avoid a collision. Such an event is part of the modeling approach adopted in Mo\textsuperscript{3}: no matter what the mobility scenario under consideration (pedestrian, vehicular, aerial, etc.) collisions do happen in the real world, and Mo\textsuperscript{3} covers this possibility.
\end{enumerate}

\begin{table*}[t!]
\caption{Mo\textsuperscript{3} input parameters. Parameters with the same name used in multiple modules are required to take the same value in order to ensure consistency in the generated mobility patterns.}
\label{tab:Mo3_setup}
\centering
\begin{tabular}{|p{4.2cm}|p{2.5cm}|p{8cm}|}
\hline
\textbf{Module} & \textbf{Parameter}&\textbf{Meaning}\\
\hline
\multirow{5}{*}{Individual Mobility} &  $T_{IM}$& Speed vector update period \\
\cline{2-3}
&  $v_{min}$& Minimum node speed\\
\cline{2-3}
&  $v_{max}$& Maximum node speed\\
\cline{2-3}
&  $a_{max}$& Maximum node linear acceleration\\
\cline{2-3}
&  $\gamma_{max}$& Maximum node rotation speed\\
\cline{2-3}
&  $\delta_{max}$& Maximum elevation angle variation speed\\
\hline
\multirow{6}{*}{Correlated Mobility} &  $T_{CM}$& Speed vector update period\\
\cline{2-3}
& Binding Matrix & Sequence of binding matrices with time of switch (File) \\
\cline{2-3}
&  $D_{c}$& Binding condition threshold\\
\cline{2-3}
 &  $\rho_{min}$& Grouping condition threshold\\
\cline{2-3}
 &  $v_{max}$& Maximum node speed\\
\hline
\multirow{5}{*}{Collision Avoidance} 
 &  $T_{CA}$& Speed vector update period \\
\cline{2-3}
&  $v_{min}$& Minimum node speed\\
\cline{2-3}
 &  $v_{max}$& Maximum node speed\\
\cline{2-3}
&  $d^{CA}_{trigger}$& Node detection threshold\\
\cline{2-3}
&  $d^{CA}_{min}$& Collision risk detection threshold\\
\cline{2-3}
&  $\theta^{CA}$& Direction variation to address frontal collision risks\\
\hline
\multirow{5}{*}{Obstacle Avoidance}  &  $T_{OA}$& Speed vector update period\\
\cline{2-3}
& Obstacles List &List of obstacles with geometrical parameters (File)\\
\cline{2-3}
&  $d^{OA}_{trigger}$& Obstacle detection threshold \\
\cline{2-3}
&  $\theta^{OA}$& Allowed direction margin\\
\hline
\multirow{5}{*}{Upper Bounds Enforcement} &  $T_{UB}$& Speed vector update period\\
\cline{2-3}
&  $v_{min}$& Minimum node speed\\
\cline{2-3}
&  $v_{max}$& Maximum node speed\\
\cline{2-3}
 &  $a_{max}$& Maximum node linear acceleration\\
\cline{2-3}
&  $\gamma_{max}$& Maximum node rotation speed\\
\cline{2-3}
&  $\delta_{max}$& Maximum elevation angle variation speed\\
\hline
\end{tabular}
\end{table*}
\subsection{Model setup}
\label{sec:Mo3_setup}
The setup procedure for Mo\textsuperscript{3} can be summarized as follows:
\begin{enumerate}
    \item define the number $M$ of nodes;
    \item define the coordinate ranges $[x_{min}, x_{max}]$, $[y_{min}, y_{max}]$ and $[z_{min}, z_{max}]$ in the tridimensional case, determining the movement area $A$ (2D case) or space $S$ (3D case);
    \item generate or load $M$ coordinate sets within $S$ that will set the initial position of the nodes. Note that if the Obstacle Avoidance rule is enabled in the bidimensional case, and obstacles are introduced in $A$, the initial positions must be set ensuring that no node is placed inside any region of $A$ occupied by an obstacle;
    \item generate or load $M$ sets of speed, direction and elevation values, defining the initial speed vectors of the $M$ nodes.
\end{enumerate} 
Starting from the initial state resulting from the procedure above, Mo\textsuperscript{3} will periodically update the speed vectors of the nodes, according to the input parameters set for the five modules that compose the model. Table~\ref{tab:Mo3_setup} provides the full list of parameters\footnote{The set of parameters required for the Individual Mobility module might change if an individual mobility model different from the one proposed in Section \ref{sec:Mo3_individual_model} is adopted.}, indicating for each of them which modules they refer to, and whether they are provided in a configuration file. Table~\ref{tab:Mo3_setup} shows that setting up the model is pretty straightforward: Mo\textsuperscript{3} does not require normalizing coefficients and weights as behavioral models do, and all parameters have a clear relation with the desired spatial properties for the mobility patterns. In addition, the mobility patterns generated by RPGM can be emulated by activating only the Individual Mobility, Correlated Mobility and Upper Bounds Enforcement modules, with a comparable set-up complexity. \\
The position of each node will be then updated, typically every $\Delta t$ seconds, by adding to the previous position the displacement in $x$, $y$ and $z$ determined by the node's speed vector.

\subsection{Availability}
\label{sec:Mo3_availability}
The Mo\textsuperscript{3} model was implemented in Matlab. The software includes a main script performing the setup procedure described in Section \ref{sec:Mo3_setup} and a set of supporting functions implementing each of the modules defined in Sections \ref{sec:Mo3_individual_model} to \ref{sec:Mo3_UpperBounds}. The code is available for download under an open source license from a dedicated GitHub repository \cite{Mo3Rep}. The repository also includes an implementation of a subset of the Mo\textsuperscript{3} modules for the OMNeT++ \cite{Var02} discrete event simulator; see Section \ref{sec:PerfEvalPure} for details.
\section{Using Mo\textsuperscript{3} to emulate other correlated mobility models}
\label{sec:Mo3_mimetic_abilities}
The flexibility provided by the combination of the Individual Mobility and Correlated Mobility rules allows Mo\textsuperscript{3} to emulate other correlated mobility models. An exhaustive comparison with the most popular group mobility model, that is RPGM, will be carried out in the next two sections; this section focuses instead on other correlated mobility models that do not fall in the group mobility category as defined in Section~\ref{sec:CorrelatedModels}. Mo\textsuperscript{3} can effectively emulate these models by adopting an asymmetric Binding Matrix, as anticipated in Section~\ref{sec:Mo3_correlated_mobility}. \\
Three notable correlated mobility models, all proposed in \cite{SanMan01}, will be considered in the following: the Pursue, Column and Nomadic Community mobility models.
\subsection{Pursue model}
The Pursue model describes a scenario where a node (the target) moves independently of any other node, and is pursued by the other nodes.\\
This model can be easily emulated by adopting the binding matrix shown in Figure \ref{fig:PursueBM}, where node $1$ is the target and moves independently from the others; correspondingly, row $1$ of the binding matrix is equal to the same row in the identity matrix. All remaining nodes are pursuers, and have thus a binding at position $1$ of their row. The aggressiveness of the pursuers in following the target can be tuned by properly setting the $D_c$ distance in \eqref{eq:Mo3_binding_eq}: the lower $D_c$, the more aggressive the pursuers will be.
\subsection{Column model}
In the Column model the nodes move in line, with some degree of liberty in deviating from regular spacing.\\ This model can be emulated by adopting the binding matrix shown in Figure~\ref{fig:ColumnBM}: node $1$ will set the pace by deciding freely speed and direction, while each node from $2$ to $M$ will follow the node before it: as a result, in the row corresponding to the generic node $j>1$ a binding is present in position $j-1$. By adopting a speed range $\left[v_{min},v_{max}\right]$ leading to a relatively fast movement for node $1$, nodes will tend to organize in a line following the leader. The degree of liberty in deviating from this behavior defined in the Column model can be modeled and tuned by properly choosing the $D_c$ distance: a larger distance will guarantee more freedom for the nodes to deviate from the line formation.
\subsection{Nomadic community model}
The Nomadic community model describes the mobility of a population that alternates moving phases, during which the whole population moves from one area to another, to stationary phases, in which the population members roam freely in an area. As suggested in \cite{SanMan01}, the model is suitable for describing the movement of security patrols or search \& rescue teams, as well as the movement of herds of animals.\\
The model can be emulated in Mo\textsuperscript{3} by adopting a time varying binding matrix. The two phases can be in fact modeled by periodically switching between the two binding matrices shown in Figures \ref{fig:NomadicMovingBM} (moving phases) and \ref{fig:NomadicStationBM} (stationary phases) The binding matrix in Figure \ref{fig:NomadicMovingBM} elects node $1$ as the community leader, defining the path toward the area where the next stationary phase will take place. Note that the Correlated Mobility rule introduced in Section \ref{sec:Mo3_correlated_mobility} will ensure that the switch between the two phases does not cause sudden position changes or other mobility artifacts, as further discussed in Section \ref{sec:PerfEvalMixed}.
\begin{figure*}[!ht]
    \centering
    \subfloat[Pursue]{
        $
        \begin{bmatrix} 
        1&0&0&0&0&0&0\\
        1&1&0&0&0&0&0\\
        1&0&1&0&0&0&0\\
        1&0&0&1&0&0&0\\
        1&0&0&0&1&0&0\\
        1&0&0&0&0&1&0\\
        1&0&0&0&0&0&1
        \end{bmatrix}
$\label{fig:PursueBM}
    }\hspace{0.15in}    
    \subfloat[Column]{
        $
        \begin{bmatrix} 
        1&0&0&0&0&0&0\\
        1&1&0&0&0&0&0\\
        0&1&1&0&0&0&0\\
        0&0&1&1&0&0&0\\
        0&0&0&1&1&0&0\\
        0&0&0&0&1&1&0\\
        0&0&0&0&0&1&1
        \end{bmatrix}
$\label{fig:ColumnBM}
    }\hspace{0.15in}    
    \subfloat[Nomadic community, moving]{
       $
        \begin{bmatrix} 
        1&0&0&0&0&0&0\\
        1&1&1&1&1&1&1\\
        1&1&1&1&1&1&1\\
        1&1&1&1&1&1&1\\
        1&1&1&1&1&1&1\\
        1&1&1&1&1&1&1\\
        1&1&1&1&1&1&1
        \end{bmatrix}
$
        \label{fig:NomadicMovingBM}
    }\hspace{0.15in}
        \subfloat[Nomadic community, stationary]{
        $
\begin{bmatrix}
        1&0&0&0&0&0&0\\
        0&1&0&0&0&0&0\\
        0&0&1&0&0&0&0\\
        0&0&0&1&0&0&0\\
        0&0&0&0&1&0&0\\
        0&0&0&0&0&1&0\\
        0&0&0&0&0&0&1
\end{bmatrix}
        $
        \label{fig:NomadicStationBM}
    }
    \caption{Binding matrices for a network of $M=7$ nodes suitable for emulating the correlated mobility models proposed in \cite{SanMan01}: (a) Pursue model; (b) Column model; (c) Nomadic community - moving phases; (d) Nomadic community - stationary phases.
    \label{fig:BM_mimetic}}
\end{figure*}
\begin{table*}
\caption{Simulation settings for the considered mobility models}
\label{tab:sim_settings}
\centering
\begin{tabular}{|c||c||c||c||c||c|}
\hline
& \multicolumn{2}{c||}{Section \ref{sec:PerfEvalPure}}&\multicolumn{3}{c|}{Section \ref{sec:PerfEvalMixed}}\\
\hline
 & \textbf{Mo\textsuperscript{3}} & \textbf{RPGM}& \textbf{Mo\textsuperscript{3}} & \textbf{RPGM}& \textbf{RVGM}\\
\hline
Groups  & 1 & 1 & 4 & 4 & 4 \\
\hline
Nodes in the group  & N/A & 20 & 4 & 4 & 4 \\
\hline
Binding matrix type & Figure \ref{fig:BM_mimetic}a & N/A & Figure \ref{fig:BM_examples}c & N/A & N/A \\
\hline
$T_{IM}$  & $5\, s$ & N/A & N/A & $5\, s$ & N/A \\
\hline
$T$  & N/A  & $5\, s$ & $5\, s$ & N/A  & N/A\\
\hline
$T_{SU}$  & N/A  &  N/A & N/A  &  N/A & $=\Delta t$ \\
\hline
$v_{max}$  & $1-2\, m/s$ & $1-2\, m/s$ & $5\, m/s$ & $5\, m/s$ & $5\, m/s$ \\
\hline
$v_{min}$  & $0.001\, m/s$ & $0.001\, m/s$ & $0.001\, m/s$ & $0.001\, m/s$ & $0.001\, m/s$ \\
\hline
$a_{max}$ & $5\,m/(s^2)$ & N/A & $5\,m/(s^2)$ & N/A & N/A  \\
\hline
$\gamma_{max}$& $\pi/2\, rad/s$ & N/A & $\pi/2\, rad/s$ & N/A & N/A  \\
\hline
$D_c$ & $7.5\,m$ & N/A & $30\,m$ & N/A & N/A  \\
\hline
$\rho_{min}$ & $1$ & N/A & $0.5$ & N/A & N/A \\
\hline
$T_{CM}$ & $=\Delta t$ & N/A & $=\Delta t$ & N/A & N/A \\
\hline
$T_{UB}$ & $=\Delta t$ & N/A & $=\Delta t$ & N/A & N/A \\
\hline
$d_{max}$ & N/A & \vtop{\hbox{\strut $7.5 \,m$ (RPGM1)}\hbox{\strut $3.75 \,m$ (RPGM2)}} & N/A & $30\,m$ & N/A \\
\hline
$\sigma_v$ & N/A & N/A & N/A & N/A & $1\,m/s$  \\
\hline
$\sigma_{\theta}$ & N/A & N/A & N/A & N/A & $0.26\,rad$ \\
\hline
\end{tabular}
\end{table*}
\section{Performance evaluation}
\label{sec:PerfEvalIntro}
Performance evaluation and comparison of correlated and group mobility models is challenging, due to the lack of mobility traces to be used as a benchmark against the patterns generated by the models. Mobility traces for groups of users/devices are in fact not widely available in the literature, and focus, in general, on very specific scenarios, e.g. large fleets of military vehicles \cite{CheRos10}, civilian vehicles \cite{LiZha13}, or human mobility over daily/weekly epochs \cite{NunVaz16}, \cite{SuzKit18}, and typically provide data collected with periods ranging from one second to several minutes, that make them unsuitable to be used as benchmarks in pure micro-mobility or mixed macro-mobility / micro-mobility scenarios.\\ In order to overcome this limitation, most contributions on mobility modeling compare models in terms of protocol-related performance metrics such as link duration, average capacity, or packet delivery rate, rather than on the basis of the generated patterns. A reference example can be found in \cite{RahAla16}, where the performance of a network using the Ad Hoc on Demand Routing protocol was analyzed in combination with four different mobility models. Performance metrics included packet delivery ratio, end-to-end delay, and routing overhead. We argue however that this approach cannot, in general, provide an insight on the \textit{quality} of mobility models unless a baseline benchmark for the considered performance metrics to compare against is available. As an example, the comparison in \cite{RahAla16} determines that the four models lead to different network performance, but cannot tell which model is better, since no baseline benchmark is provided. For this reason, in this work protocol-related metrics are only used to compare Mo\textsuperscript{3} vs. the RPGM model in a scenario where a baseline benchmark is available. In particular, we selected the pure micro-mobility scenario considered in \cite{ZheHaa17}, where a distributed MIMO algorithm is evaluated in presence of individual mobility, thus providing a baseline benchmark for other mobility models. This comparison is presented in Section \ref{sec:PerfEvalPure}.\\

An approach based on a direct comparison of the generated mobility patterns is adopted next to compare Mo\textsuperscript{3} with the RPGM and RVGM models in a mixed macro-mobility / micro-mobility scenario. The patterns generated by the models are analyzed by assessing whether, and to what extent, they meet predefined upper bounds set on maximum speed and rotation speed. This comparison is presented in Section \ref{sec:PerfEvalMixed}.\\
Table {\ref{tab:sim_settings}} lists the settings of Section {\ref{sec:PerfEvalPure}} vs. Section {\ref{sec:PerfEvalMixed}}, and shows that the main differences between pure micro-mobility and mixed macro-mobility / micro-mobility scenarios are the proximity constraints $D_c$ / $d_{max}$ and the maximum speed with, as expected, a tighter proximity and a smaller scale mobility in the pure micro-mobility scenario.

\subsection{Performance evaluation in a pure micro-mobility scenario}
\label{sec:PerfEvalPure}
The high device density that will characterize 5G and beyond 5G networks will require the introduction of advanced cooperation algorithms in order to turn network density in an advantage, rather than a limitation; an accurate modeling of mobility will be thus fundamental for a reliable evaluation of the performance of such cooperative algorithms. As previously discussed in Section \ref{sec:BenchmarkSelection}, however, the modeling of correlated mobility patterns is typically still based on RPGM, see for example \cite{ZhuGal19}. In this context, this section compares RPGM and Mo\textsuperscript{3} in order to assess their suitability to support the performance analysis of cooperative algorithms, considering the pure micro-mobility scenario analyzed in \cite{ZheHaa17}\footnote{The authors gratefully acknowledge Dr. Zheng and Dr. Haas for sharing the code they developed to carry out the simulations presented in \cite{ZheHaa17}.}.\\
In \cite{ZheHaa17} a distributed MIMO network was considered, in which a transmitter $TX$ selects $L$ relay nodes out of $K>L$ candidates, randomly scattered around the position of $TX$, in order to create a virtual antenna array and send data toward a second virtual antenna array formed by $N$ nodes clustered around a receiver $RX$, placed at $d$ meters from $TX$. The work in \cite{ZheHaa17} focused on the transmitter side, and proposed an algorithm for the selection of the set of $L$ relay nodes, called Reconfigurable Distributed MIMO (RD-MIMO), that favors the nodes with the highest channel gains toward the array at the receiver side. The results presented in \cite{ZheHaa17} showed that RD-MIMO obtains an average achievable rate larger than when all the $K$ nodes are used in the array. The paper also investigated the impact on performance of node mobility, by determining the update period $T_{os}$ for the set of $L$ relays required to compensate for the variations of channel gains introduced by mobility, as a function of the node speed and of the desired trade-off between performance and overhead introduced by the selection procedure. Mobility of nodes was modeled in \cite{ZheHaa17} according to the Random Walk (RW) model, with the additional constraint of not allowing nodes to move outside an area of $s$ by $s$ square meters, centered on $TX$, thus indirectly introducing a correlation between the mobility patterns of the nodes.\\
In the following the analysis carried out in \cite{ZheHaa17} is extended by evaluating the performance of the RD-MIMO algorithm using proper group mobility models, in particular Mo\textsuperscript{3} and RPGM, and comparing it with the performance obtained by the RD-MIMO algorithm using the RW model: the latter can be considered as a reference ground truth, since the RW model does not introduce mobility artifacts, as discussed in Section \ref{sec:IndividualModels}. A good group mobility model should thus lead to a performance of the RD-MIMO algorithm comparable to the one observed with the RW model, while significant discrepancies would suggest the presence of artifacts in the generated mobility patterns.\\
The performance evaluation presented in this section and in Section \ref{sec:PerfEvalMixed} were both carried out using a mobility simulator developed in the OMNeT++ 5.5 simulation environment \cite{Var02}, and implementing the RPGM, RVGM models as well as the subset of Mo\textsuperscript{3} modules required to emulate the above models\footnote{OMNeT++ 5.5 modules implementing the RPGM and RVGM mobility models, the Individual Mobility, Correlated Mobility, Collision Avoidance and Upper Bounds Enforcement Mo\textsuperscript{3} modules, as well as supporting functions to collect the simulation data presented in Sections \ref{sec:PerfEvalPure} and \ref{sec:PerfEvalMixed}, are available for download from the Mo\textsuperscript{3} project GitHub repository \cite{Mo3Rep}.}.
The settings defining the scenario were derived from \cite{ZheHaa17}, and foresee static $TX$ and $RX$ nodes at distance $d=30$ m, with $K=20$ nodes moving within an area centered on $TX$ and $N=8$ relays around $RX$. Also according to \cite{ZheHaa17}, the optimal value $L=12$ was adopted in all simulations.\\
Positions of nodes were updated periodically with a period $\Delta t$. A periodic update was adopted in the analysis for two reasons: a) while RW and Mo\textsuperscript{3} allow to determine the position of nodes at any time by using (\ref{eq:position_inter_update}), RPGM only defines a speed vector for group leaders, and position updates for standard nodes can only be provided with a periodic update; b) most wireless network simulators are actually discrete event simulators, favoring the implementation of mobility models as a sequence of periodic updates rather than through the asynchronous evaluation of a time-continuous function. $\Delta t=0.1\, s$ was selected, in order to allow the analysis for selection updated periods as low as 0.1 s, as also considered in \cite{ZheHaa17}.\\
The constraint on the position of the $K$ candidate relays was implemented in the three mobility models as follows:
\begin{itemize}
\item RW - nodes moved freely in a square area of side $s=15$ m centered on $TX$; nodes reaching an edge of the square were rebounded towards the center with same speed and new direction determined by a perfect reflection;
\item RPGM1 - all $K$ nodes shared a static reference point set on the position of $TX$, and were positioned at each position update at random locations within a distance $d_{max}=s/2$ from the reference point;
\item RPGM2 - all $K$ nodes shared a reference point moving according to the RW model within a circle of radius $s/4$ centered on $TX$, while the $K$ nodes were positioned at each position update at random locations within a distance $d_{max}=s/4$ from the reference point;
\item Mo\textsuperscript{3} - each of the $K$ nodes had $TX$ as its only mate, with $\rho_{min}=1$ and $D_c=s/2$\footnote{The selected $D_c$ value allows nodes to occasionally occupy positions outside the circle of diameter $s$; a strict observance of the constraint on the position of the candidate relays could be enforced by choosing $D_c=s/2-\Delta t v_{max}$, with the undesirable effect of linking a model parameter, $D_c$, to a simulation setting, $\Delta t$.}. All remaining simulation settings for Mo\textsuperscript{3} were as in Table \ref{tab:sim_settings}.
\end{itemize}
\begin{figure*}[!t]
    \centering
    \subfloat[$v_{max}=1$ m/s]{
        \includegraphics[scale=0.5]{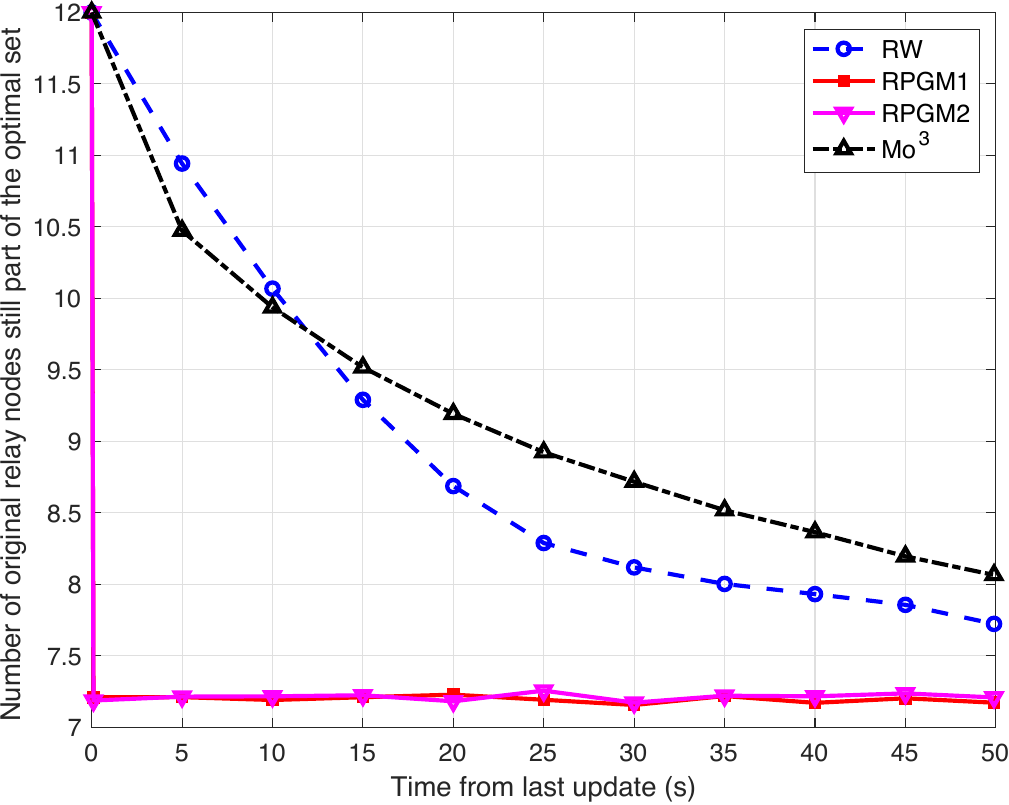}
        \label{fig:NVN_1ms}
    }
    \hfill
    \subfloat[$v_{max}=2$ m/s]{
        \hspace*{-.1in}
        \includegraphics[scale=0.5]{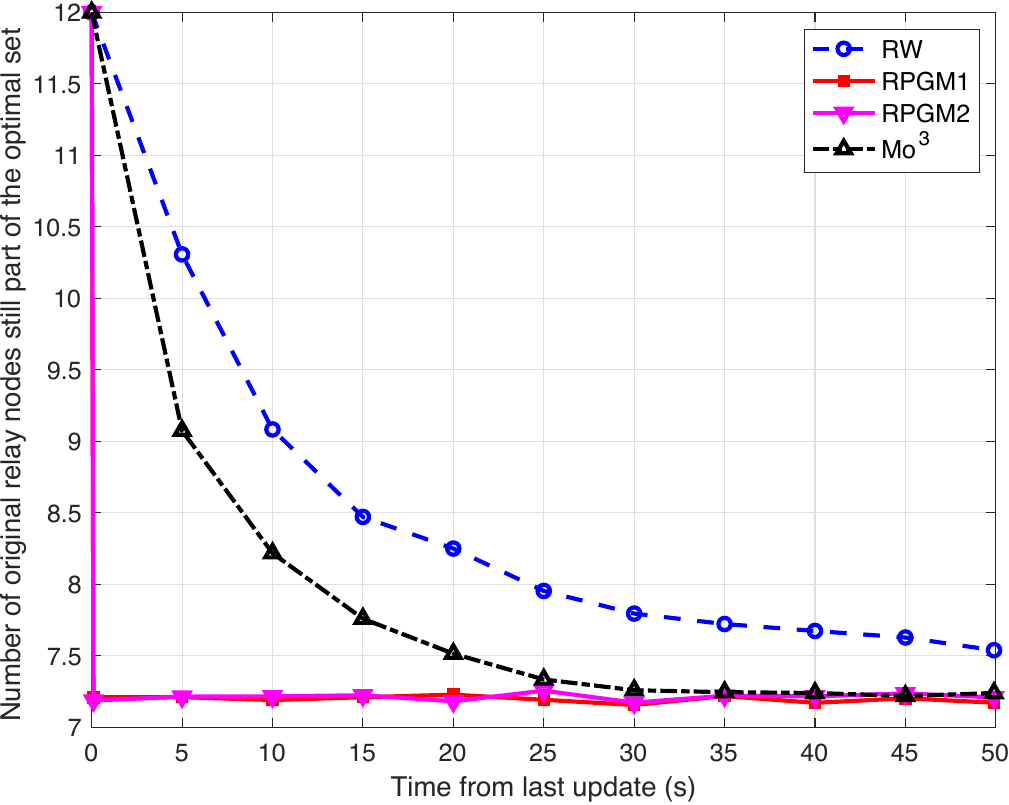}
        \label{fig:NVN_2ms}
    }
    \hfill
    \caption{Number of relays selected in the last relay selection that are still part of the set of best candidates as a function of elapsed time from last relay selection in the network in RD-MIMO adopting RW, RPGM1, RPGM2 and Mo\textsuperscript{3} mobility models, respectively, assuming a maximum speed $v_{max}=1$ m/s (a) and $v_{max}=2$ m/s (b) for each candidate node in RW and Mo\textsuperscript{3}, and for the reference point shared by all candidate nodes in RPGM2. Results for RPGM1 are independent of speed, and are replicated in both figures out of completeness.}
      \label{fig:NVN}
\end{figure*}
\begin{figure*}[!t]
    \centering
    \subfloat[$v_{max}=1$ m/s]{
        \includegraphics[scale=0.5]{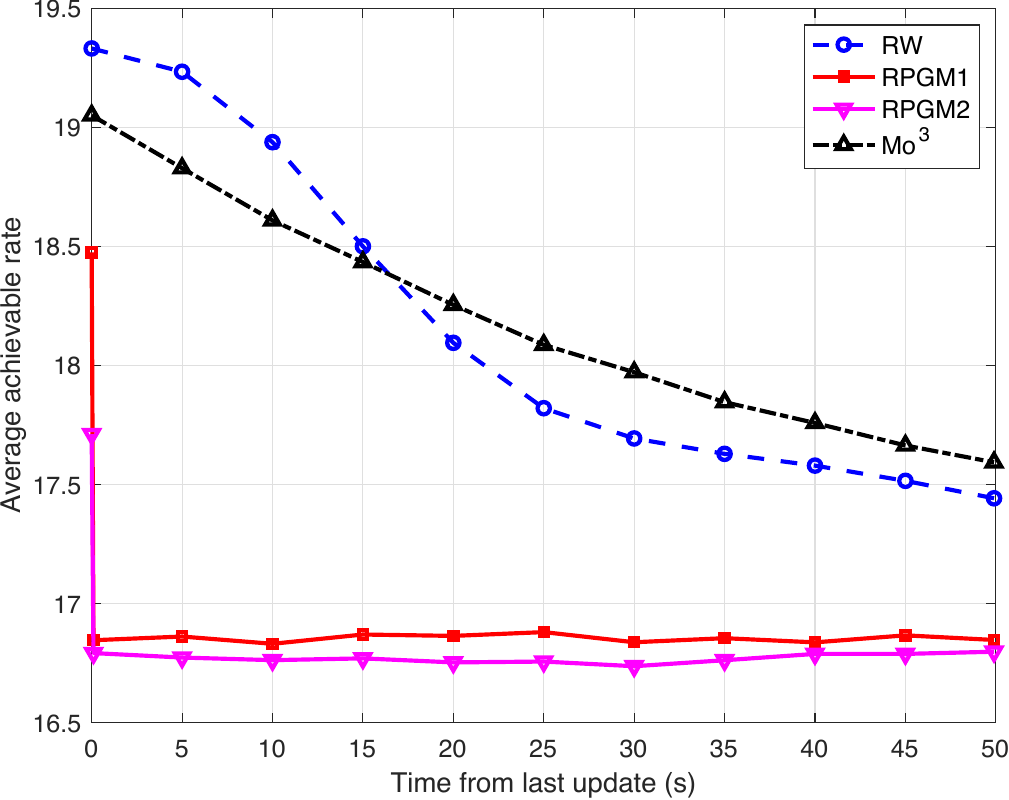}
        \label{fig:AAR_1ms}
    }
    \hfill
    \subfloat[$v_{max}=2$ m/s]{
        \hspace*{-.1in}
        \includegraphics[scale=0.5]{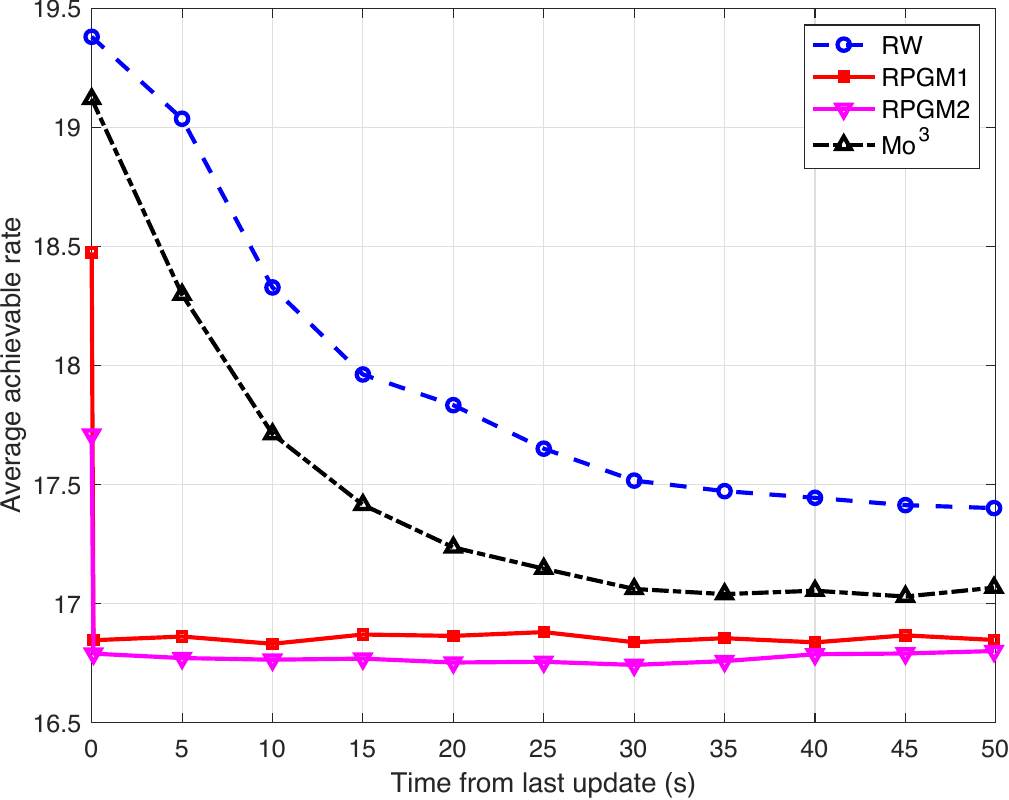}
        \label{fig:AAR_2ms}
    }
    \hfill
    \caption{Average achievable rate as a function of elapsed time from last relay selection in the network in RD-MIMO adopting RW, RPGM1, RPGM2 and Mo\textsuperscript{3} mobility models, respectively, assuming a maximum speed $v_{max}=1$ m/s (a) and $v_{max}=2$ m/s (b) for each candidate node in RW and Mo\textsuperscript{3}, and for the reference point shared by all candidate nodes in RPGM2. Results for RPGM1 are independent of speed, and are replicated in both figures out of completeness.}
      \label{fig:AAR}
\end{figure*}
\begin{figure*}[!t]
    \centering
    \subfloat[RW]{
        \includegraphics[scale=0.4]{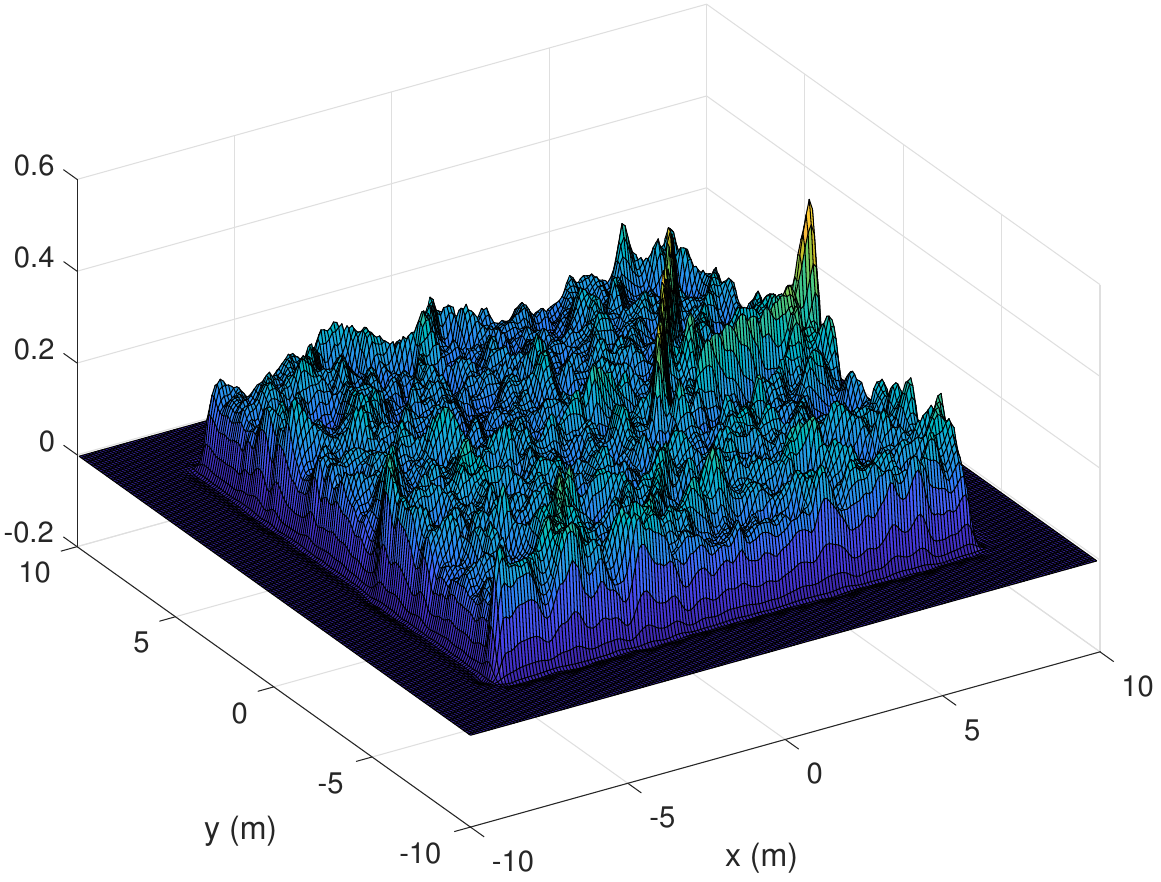}
        \label{fig:SD_RRW}
    }
    \hfill
    \subfloat[RPGM1]{
        \includegraphics[scale=0.4]{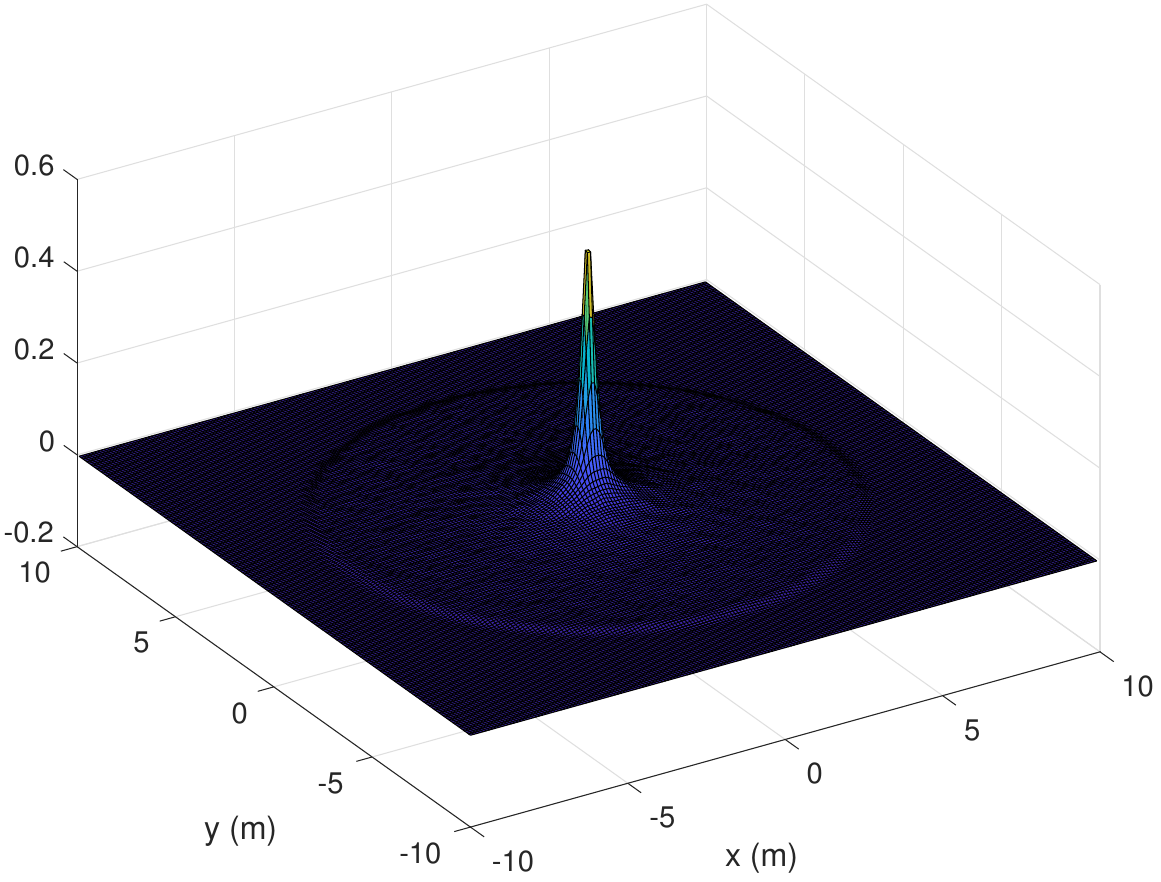}
        \label{fig:SD_RPGM1}
    }
    \hfill
    \subfloat[RPGM2]{
        \includegraphics[scale=0.4]{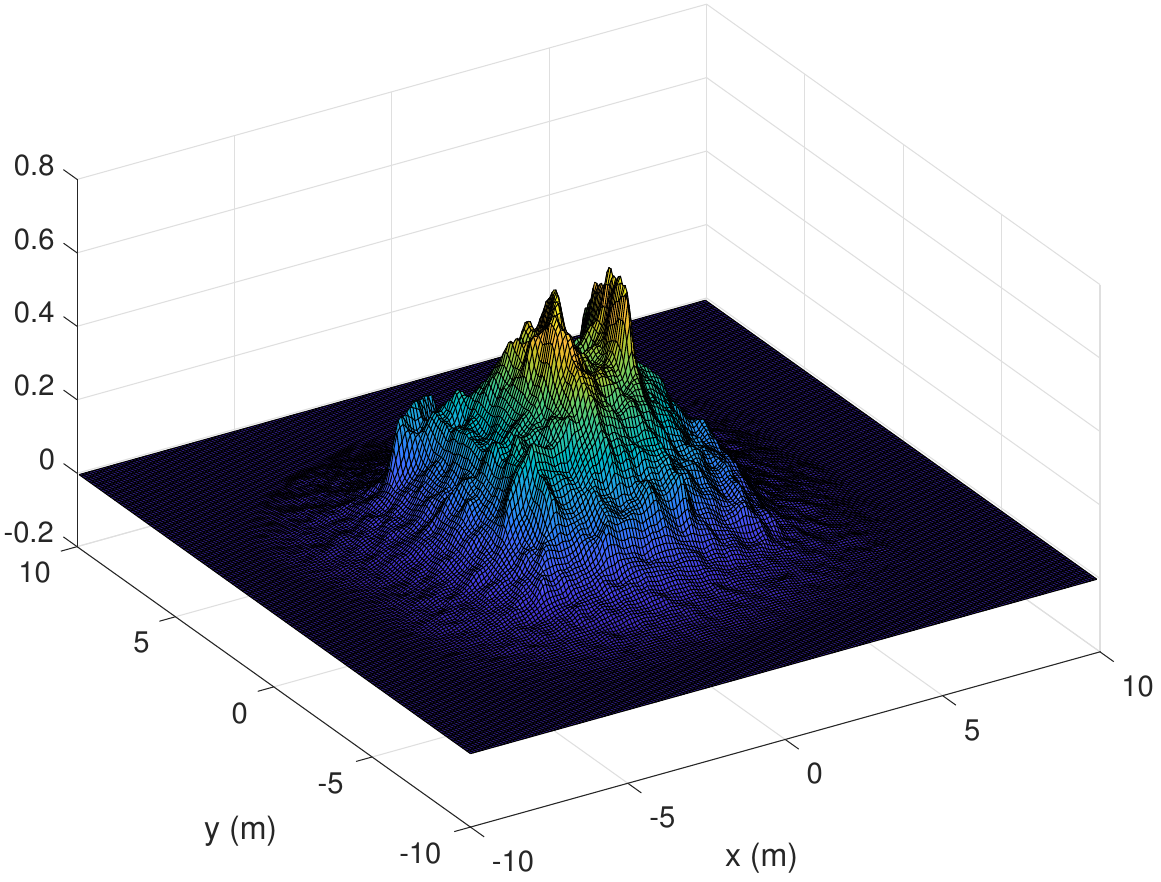}
        \label{fig:SD_RPGM2}
    }
        \subfloat[Mo\textsuperscript{3}]{
        \includegraphics[scale=0.4]{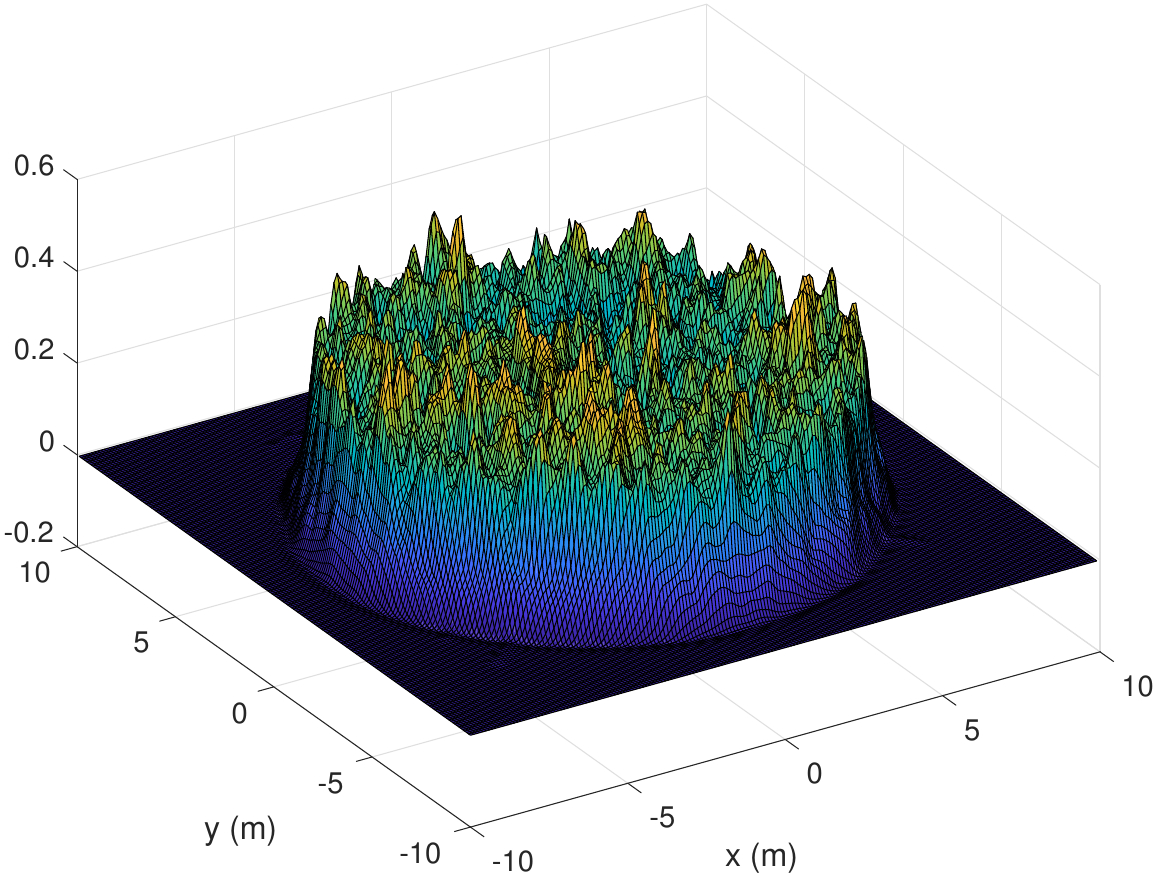}
        \label{fig:SD_Mo3}
    }
    \caption{Spatial probability density function for RW, RPGM1, RPGM2 and Mo\textsuperscript{3} mobility models measured within a square area of size 20x20 square meters centered on the position of $TX$, with a resolution of 0.1 m.
    \label{fig:SD}}
\end{figure*}

Note that the RPGM1 implementation is the most natural solution to introduce the desired mobility constraint in RPGM, but in the considered scenario it would make the RPGM mobility patterns independent from $v_{max}$. The alternative approach labeled as RPGM2 was thus also considered in order to ensure fairness for RPGM in the comparison.\\
Results are presented Figure \ref{fig:NVN} and Figure \ref{fig:AAR}. 
Figure \ref{fig:NVN} shows the number of candidate nodes selected in the optimal set of $L$ relays that are still among the best $L$ relays, as a function of the elapsed time from the last relay selection, referred to as $T_{el}$. Results in Figure \ref{fig:NVN} show that the Mo\textsuperscript{3} mobility model preserves the effect of correlation in the position of nodes over time observed in \cite{ZheHaa17}, and also confirmed in the same Figure, for the RW model. In addition, the two models are also similarly affected by a variation in the maximum speed: as one would intuitively expect, higher mobility leads to a faster disruption of the optimal set of $L$ relays, due to the physical topology changes occurring at a higher pace. Oppositely, both flavors of RPGM fail to preserve any correlation between the positions of nodes, and introduce thus a major artifact in the generated mobility patterns, consisting in the complete loss of spatial correlation between two consecutive positions of the same node. As a result, the number of nodes that are still part of the optimal set drops, immediately after the last selection, to the average size $\overline{\lvert A \cap B \rvert}$ of the intersection of two random subsets $A$ and $B$, both of size $L$, independently extracted out of a set $S=\left\{s_1,\cdots, s_K \right\}$ of size $K$. Under the assumption of uniform, independent extractions, one has: $p_i^A=Prob\left\{s_i\in A\right\}=p_i^B=Prob\left\{s_i\in B\right\}=L/K \; \forall i=1,\cdots, K$, and $\overline{\lvert A \cap B \rvert}$ is thus equal to:
\begin{equation}
\overline{\lvert A \cap B \rvert}=\sum_{i=1}^{K}p_i^A p_i^B=K\frac{L}{K}\frac{L}{K}=\frac{L^2}{K}=144/20=7.2.
\end{equation}
The average achievable rate as a function of $T_{el}$, presented in Figure \ref{fig:AAR}, exhibits the same pattern, with a graceful decrease for RW and Mo\textsuperscript{3} vs. an abrupt drop for both flavors of RPGM.
Results confirm thus the capability of Mo\textsuperscript{3} of correctly modeling the movement of nodes in micro-mobility scenarios, overcoming the inherent limits of RPGM that lead to artifacts affecting the correctness of performance evaluation.\\
Interestingly, Figure \ref{fig:AAR} also shows that the four models present different average achievable rates even at $T_{el}=0$, when the optimal set is considered. This result can be explained by observing that the achievable rate depends on the spatial distribution of the candidate nodes: mobility models that lead to a distribution of candidate relays farther away from $TX$ will in general allow for the selection of nodes closer to $RX$, and thus lead to a higher average rate. This observation is confirmed by the analysis of the spatial distribution of the four models, presented in Figure \ref{fig:SD}, showing the estimate of the probability density function for the position of candidate nodes in an area of $20x20 \, m^2$ centered on the position of $TX$.
Figure \ref{fig:SD} highlights that RW and Mo\textsuperscript{3} lead to a more uniform distribution of nodes in the area, corresponding to a higher probability of finding relays closer to $RX$. Both RPGM flavors lead to a distribution biased toward the center of the area and thus, as an average, to relays closer to $TX$ and farther from $RX$. Correspondingly, the average achievable rate at $T_{el}=0$ is higher for RW and Mo\textsuperscript{3} than for the RPGM flavors. This is a second artifact introduced by the RPGM model, since a model should not lead to uneven spatial distributions of nodes, unless appositely designed to do so to describe a specific mobility scenario. Regarding the difference between the two RPGM implementations, a close examination of Figure \ref{fig:SD_RPGM1} vs. Figure \ref{fig:SD_RPGM2} shows that RPGM1 leads to a higher probability of nodes being on the edge of the circular area compared to RPGM2, and, correspondingly, to a slightly higher average achievable rate, as shown in Figure \ref{fig:AAR}.\\
Figure \ref{fig:AAR} also shows a slight gap in the average achievable rate between RW and Mo\textsuperscript{3}; this can be explained by the different shape of the area around $TX$ occupied by nodes in the two models: the additional surface available in the case of RW occasionally allows to achieve optimal configurations that are impossible to achieve when using Mo\textsuperscript{3}.
\subsection{Performance evaluation in a mixed macro-mobility / micro-mobility scenario}
\label{sec:PerfEvalMixed}
The Mo\textsuperscript{3} model was also compared with the RPGM and RVGM group mobility models in a mixed macro-mobility / micro-mobility scenario as defined in Section \ref{sec:introduction}, consisting in a typical ad-hoc network scenario, that is obstacle-free movement in a large movement area, combined with tight group requirements. As anticipated in the same Section, due to the lack of real world traces to be used as ground truth, a criterion should be defined to compare the different models. The proposed criterion and the corresponding performance indicators are introduced in Section \ref{sec:CompCriterion}, followed by the simulation settings in Section \ref{sec:simulation_settings} and by the results of the analysis, presented in Section \ref{sec:perfEval}.
\subsubsection{Comparison criterion and performance indicators}
\label{sec:CompCriterion}
The definition of the comparison criterion moved from the following observation: all mobility models allow the introduction of bounds on one or more mobility parameters. As a bare minimum, models allow to define an upper bound on linear speed, $v\leq v_{max}$, and in some cases, as in the Boundless model and in Mo\textsuperscript{3}, also on linear acceleration, $\left|a\right| \leq a_{max}$, and on angular speed, $\left|\gamma\right| \leq \gamma_{max}$.\\
Once the bounds are defined at model setup, mobility patterns generated by the model are, by definition, expected to meet these bounds at all times.\\
The comparison criterion proposed in this work aims at assessing if, and to what degree, models actually meet this reasonable expectation in a mixed macro-mobility / micro-mobility scenario. The corresponding performance indicators adopted to evaluate the models are the following:
\begin{itemize}
\item probability of violating the upper bound on linear speed, $P_{v}^{out}$;
\item probability of violating the upper bound on angular speed, $P_{\gamma}^{out}$;
\item average linear speed $v_{mean}$.
\end{itemize}
Note that RPGM and RVGM do not provide a way to limit rotation speed: for such models it would be thus unfair to consider $P_{\gamma}^{out}>0$ as a failure, since they do not promise to enforce an upper bound on $\gamma$. It is nevertheless interesting to measure $P_{\gamma}^{out}$ for RPGM and RVGM as well, as it may provide insights on the models, as well as quantify how well they would cope with scenarios requiring a bound on $\gamma$.\\
The performance indicators were evaluated by collecting samples at each position update, performed periodically with period $\Delta t$.\\
The performance indicators were measured based on the actual movement of a node $i$ during the period $\Delta t$ associated to the $k$ update, from $\left\{x_i\left((k-1)\Delta t\right),y_i\left((k-1)\Delta t\right)\right\}\equiv \left\{x_i,y_i\right\}$ to $\left\{x_i(\left(k \Delta t\right), y_i\left(k \Delta t\right)\right\}\equiv \left\{x_i',y_i'\right\}$, that corresponds to the following linear and angular speeds:
\begin{equation}
\left\{
\begin{aligned}
v_i\left(k\Delta_t\right)&=\frac{\sqrt{\left(x_i'-x_i\right)^2+\left(y_i'-y_i\right)^2}}{\Delta t}\\
\gamma_i\left(k\Delta_t\right)&=\left|\frac{\arctantwo(y_i', x_i')-\arctantwo(y_i, x_i)}{\Delta t}\right|.
\end{aligned}\right.
\end{equation}
The occurrence of a violation of the linear speed bound for node $i$ at update $k$ can be then associated to the following binary variable:
\begin{equation}
v_{i,k}^{out}=
\begin{cases}
1,  & \text{if }v_i\left(k\Delta_t\right)>v_{max}\\
0,  & \text{otherwise},
\end{cases}
\end{equation}
and similarly one has for the angular speed bound:
\begin{equation}
\gamma_{i,k}^{out}=
\begin{cases}
1,  & \text{if }\left|\gamma_i\left(k\Delta_t\right)\right|>\gamma_{max}\\
0,  & \text{otherwise}.
\end{cases}
\end{equation}
The three performance indicators can be thus expressed as:
\begin{equation}
\label{eq:perfIndicators}
\left\{
\begin{aligned}
P_{v}^{out}&=\lim_{K\rightarrow \infty}\frac{1}{K}\frac{1}{M}\sum_{k=1}^{K}\sum_{i=1}^M v_{i,k}^{out}\\
P_{\gamma}^{out}&=\lim_{K\rightarrow \infty}\frac{1}{K}\frac{1}{M}\sum_{k=1}^{K}\sum_{i=1}^M \gamma_{i,k}^{out}\\
v_{mean}&=\lim_{K\rightarrow \infty}\frac{1}{K}\frac{1}{M}\sum_{k=1}^{K}\sum_{i=1}^M v_i\left(k\Delta_t\right).
\end{aligned}\right.
\end{equation}
The truncation of the sums in \eqref{eq:perfIndicators} to a finite $K$, that corresponds to a finite observation time $T_{o}=K\Delta t$, will lead to estimated values for the three indicators, that can be expected to be accurate as long as $T_{o}$ is sufficiently long.
\subsubsection{Simulation settings}
\label{sec:simulation_settings}
The simulation scenario considered a network of $M=16$ nodes divided into 4 groups of 4 nodes. The movement area $A$ was $5000x5000\, m^2$, and each simulation run lasted $T_o=10000$ seconds. The upper bounds on linear and angular speed to be met by models were $v_{max} = 5\, m/s$ and $\gamma_{max} = \pi/2\, rad/s$. The following implementation choices were made, in order to ensure a fair comparison:
\begin{itemize}
    \item  \emph{Mo\textsuperscript{3}}: a binding matrix as in Figure~\ref{fig:GrpMobilityEx3} was adopted, but with four blocks of size $4x4$; $\rho_{min}$ was set so that the grouping condition defined in \eqref{eq:Mo3_condition} was only satisfied for a node if all of its mates were part of its connected set. The Collision Avoidance and Obstacle Avoidance modules were disabled, since these features are not available in RPGM and RVGM;
\item \emph{RPGM}: a group leader was selected in each group. The reference path for each group leader was generated using the Random Walk model in place of the Random Waypoint model originally proposed in \cite{HonGer99}, in order to avoid the side effects of the latter model, discussed in Section \ref{sec:IndividualModels};
\item \emph{RVGM}: a group leader was selected in each group. The possibility to enforce bounds on minimum and maximum speed was introduced by adopting for both group leaders and standard nodes a truncated Gaussian distribution taking values in $[v_{min},v_{max}]$, rather than the Gaussian distribution originally proposed in \cite{WanLi02}.
\end{itemize}
The indicators introduced in Section \ref{sec:CompCriterion} were evaluated as a function of the system parameter most directly related to mobility, that is the mobility update period $\Delta t$, taking values in the set $S_{\Delta t}=\left\{0.1,0.25,0.5,1,1.25,5\right\}\,s$. In the case of the Mo\textsuperscript{3} and RPGM models, that share a mechanism to enforce a maximum distance between group members, the analysis also included the impact of the model parameter that controls this mechanism, that is the $D_c$ parameter for Mo\textsuperscript{3} and the $d_{max}$ parameter for RPGM, both varied over the set $S_d=\left\{15,75,135,195\right\}\,m$. The values in $S_d$ were selected so to cover scenarios ranging from tight to loose correlation between patterns of nodes in the same group.\\ 
Model-specific settings presented in Table \ref{tab:sim_settings} were adopted during simulations, unless otherwise stated.

\subsubsection{Results}
\label{sec:perfEval}
\begin{figure}[t]
  \centering
\includegraphics[width=3.2in]{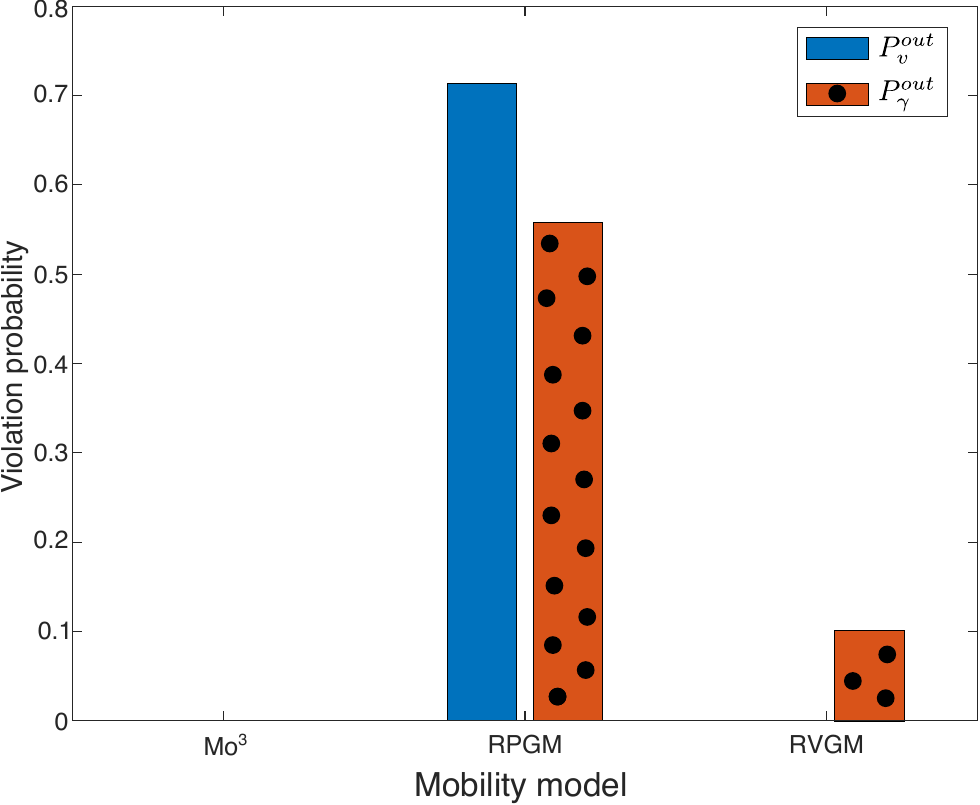}
 \caption{$P_{v}^{out}$ and $P_{\gamma}^{out}$ for the Mo\textsuperscript{3}, RPGM and RVGM mobility models, configured according to Table \ref{tab:sim_settings}, with $\Delta t=1\,s$.}
 \label{fig:accuracy_violations}
\end{figure}

\begin{figure}[t]
  \centering
\includegraphics[width=3.2in]{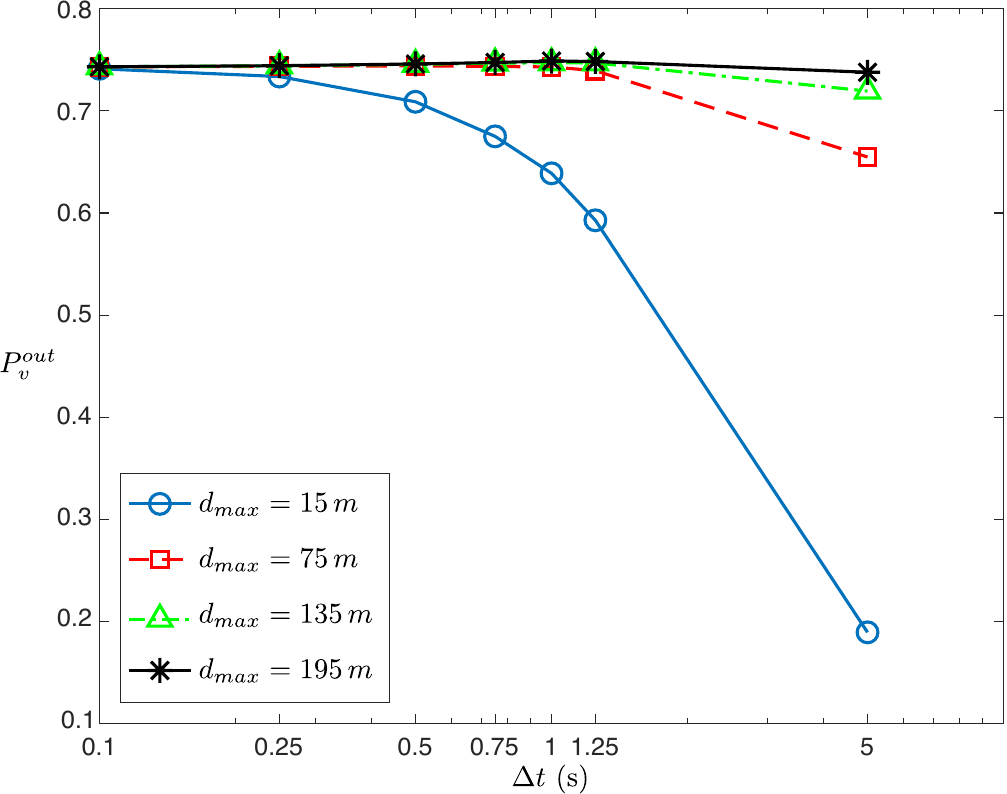}
 \caption{$P_{v}^{out}$ measured for RPGM as a function of $\Delta t \in S_{\Delta t}$, for all values of $d_{max} \in S_{d}$; all parameters of RPGM except for $d_{max}$ were configured according to Table \ref{tab:sim_settings}.}
 \label{fig:RPGM_speed_violations}
\end{figure}

\begin{figure}[t]
  \centering
\includegraphics[width=3.2in]{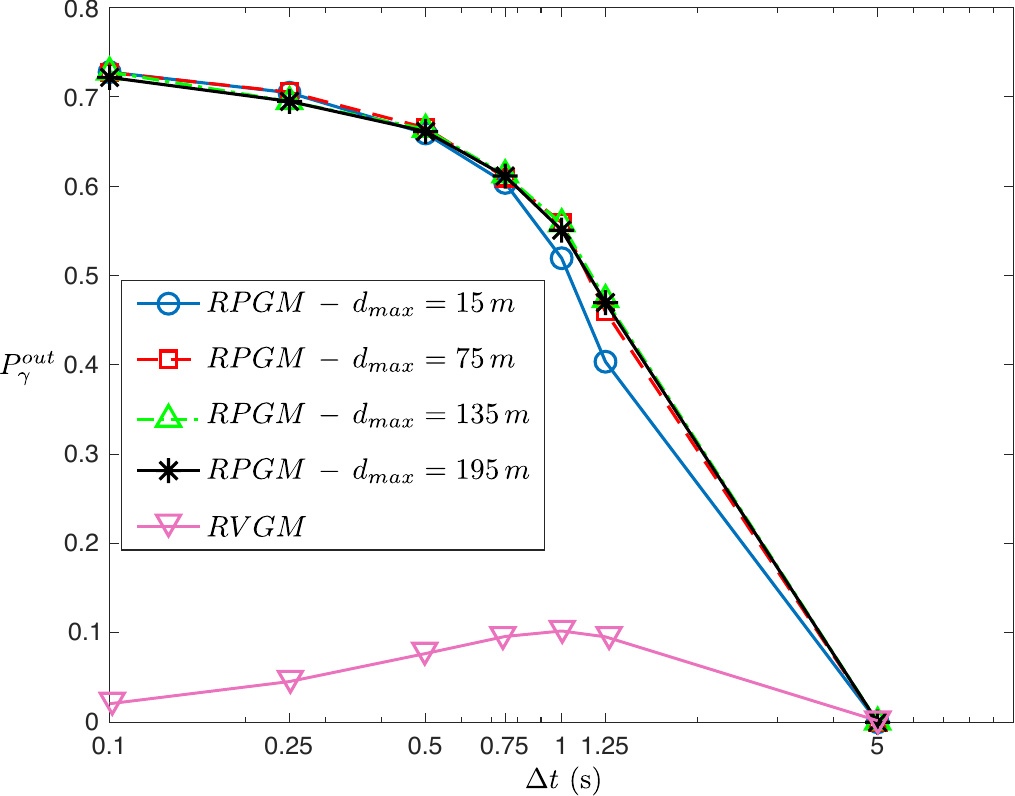}
 \caption{$P_{\gamma}^{out}$ measured for RPGM and RVGM as a function of $\Delta t \in S_{\Delta t}$ and for all values of $d_{max} \in S_{d}$ for RPGM; all parameters of RPGM except for $d_{max}$ were configured according to Table \ref{tab:sim_settings}.}
 \label{fig:RPGM_RVGM_rotation_violations}
\end{figure}

\begin{figure}[t]
  \centering
\includegraphics[width=3.2in]{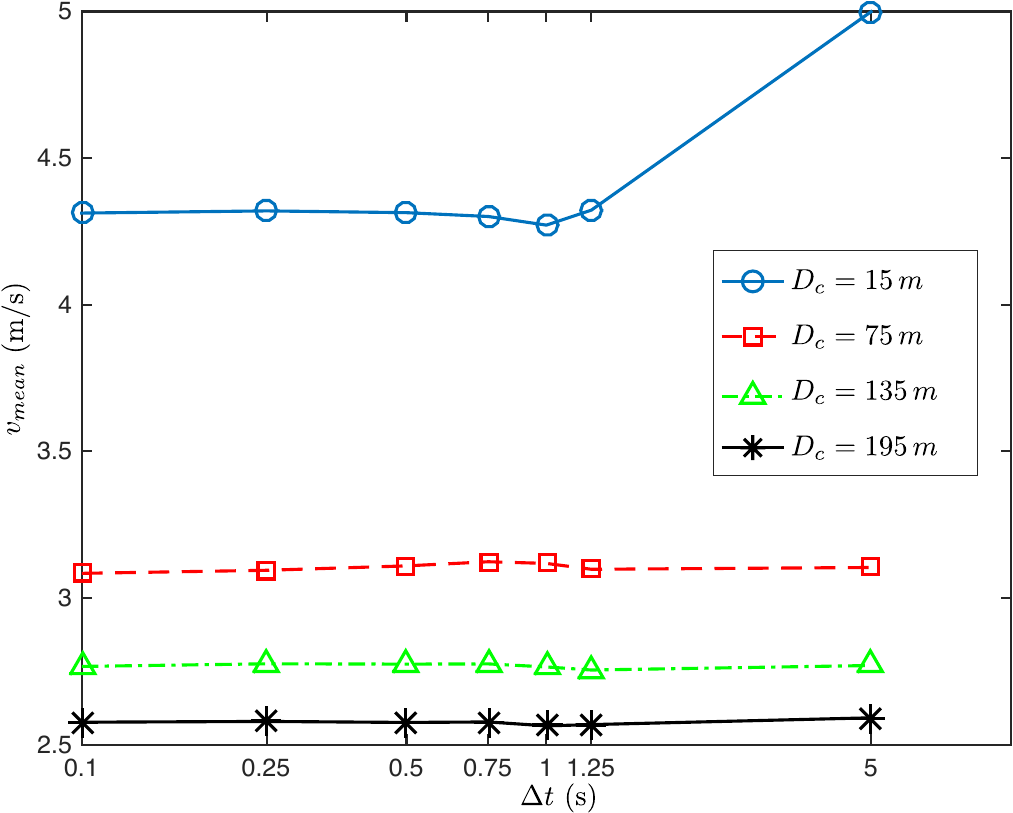}
 \caption{$v_{mean}$ measured for Mo\textsuperscript{3} as a function of $\Delta t \in S_{\Delta t}$, for all values of $D_{c} \in S_{d}$; all parameters of Mo\textsuperscript{3} except for $D_{c}$ were configured according to Table \ref{tab:sim_settings}.}
 \label{fig:Mo3_average_speed}
\end{figure}

\begin{figure}[t]
  \centering
\includegraphics[width=3.2in]{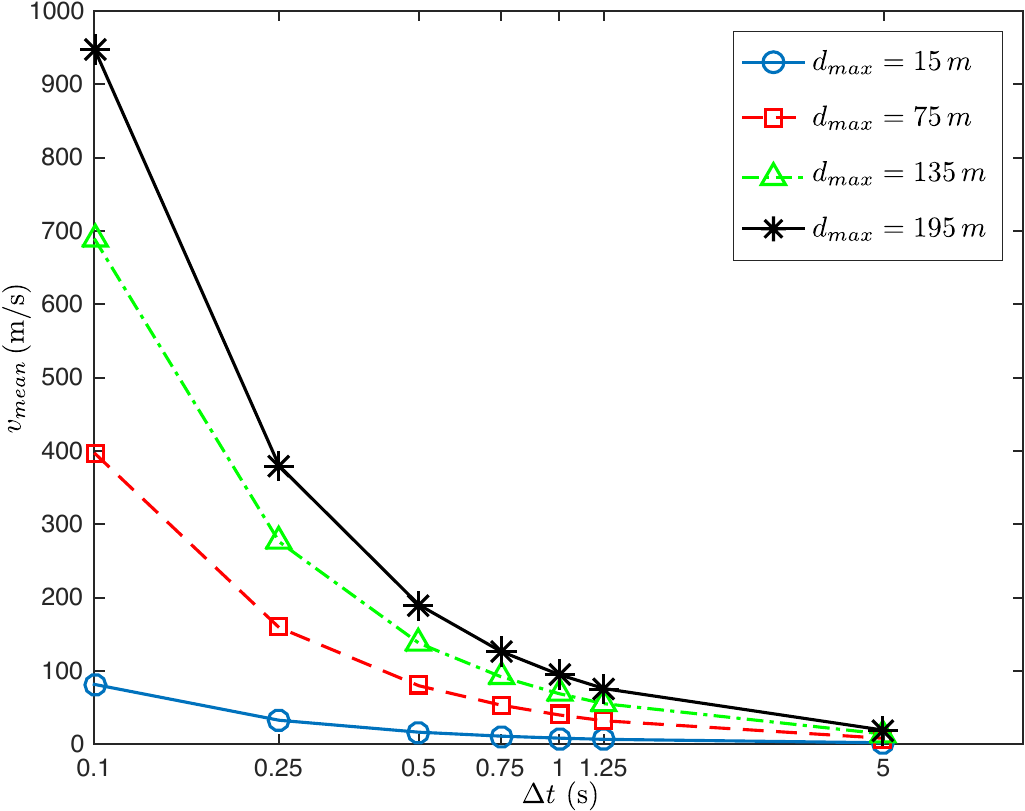}
 \caption{$v_{mean}$ measured for RPGM as a function of $\Delta t \in S_{\Delta t}$, for all values of $d_{max} \in S_{d}$; all parameters of RPGM except for $d_{max}$ were configured according to Table \ref{tab:sim_settings}.}
 \label{fig:RPGM_average_speed}
\end{figure}

\begin{figure}[t]
  \centering
\includegraphics[width=3.2in]{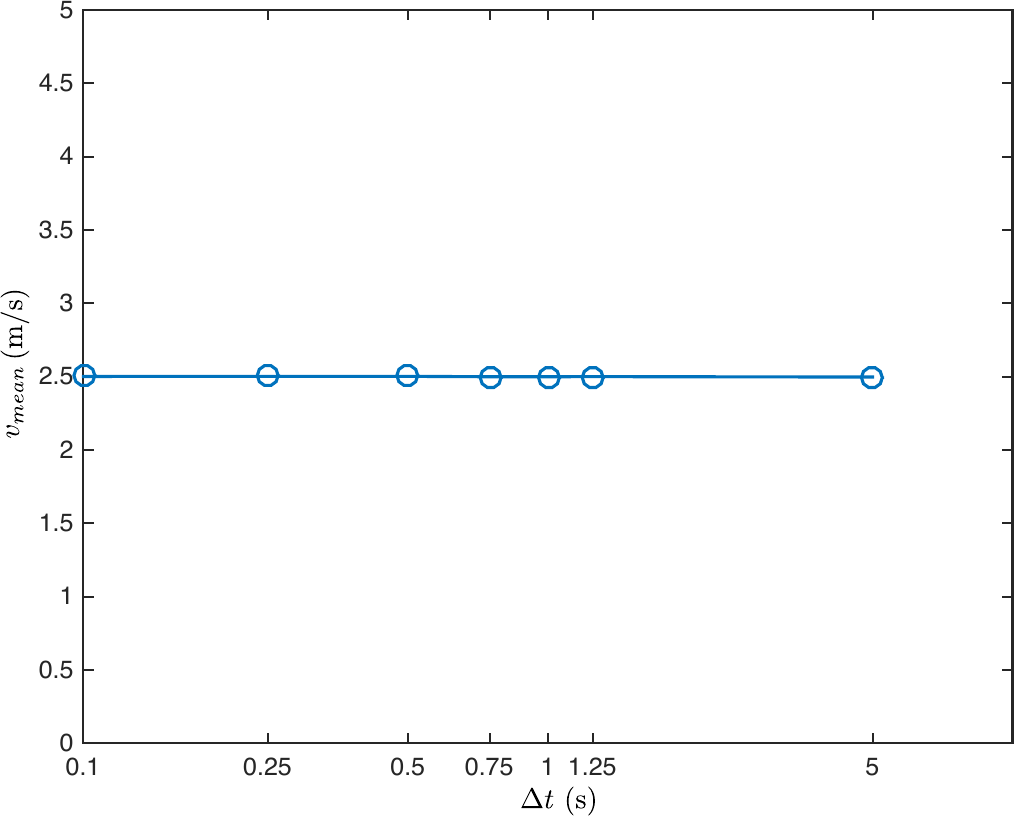}
 \caption{$v_{mean}$ measured for RVGM as a function of $\Delta t \in S_{\Delta t}$; all parameters of RVGM were configured according to Table \ref{tab:sim_settings}.}
 \label{fig:RVGM_average_speed}
\end{figure}

Based on the comparison criterion defined in Section \ref{sec:CompCriterion}, the research questions we sought to answer are:
\begin{enumerate}
    \item are models capable at all to meet upper bounds?
    \item are they robust to different settings of system and model parameters?
    \item how do they cope with time-varying correlated mobility scenarios?
\end{enumerate}
The answer to the first question can be found in Figure \ref{fig:accuracy_violations}, showing $P_{v}^{out}$ and $P_{\gamma}^{out}$ for the three models with the settings in Table \ref{tab:sim_settings}, defining a typical macro-mobility scenario with reasonably tight group bindings, and $\Delta t=1$ s.
Figure \ref{fig:accuracy_violations} shows that both Mo\textsuperscript{3} and RVGM are able to meet at all times the limit on linear speed, leading to $P_{v}^{out}=0$. The RPGM model, on the other hand, shows a high probability to violate the bound on maximum speed, with $P_{v}^{out}>0.7$.\\ 
As for rotation speed, Mo\textsuperscript{3} is again able to meet at all times the limit on upper rotation speed, with no recorded violations. As expected, this is not the case for RPGM ($P_{\gamma}^{out}=0.5$) and RVGM ($P_{\gamma}^{out}=0.1$), since they do not provide a mechanism to limit rotation speed.\\
The answer to the second question can be found by analyzing the impact of $\Delta t$ on the performance indicators for the three models. In the case of Mo\textsuperscript{3}, the Upper Bounds Enforcement module ensures that the upper bound on speed is always met: as a consequence, $P_{v}^{out}=0$ was observed for all combinations of $\Delta t$ and $D_c$. The same is true for RVGM, thanks to the adoption of a truncated Gaussian distribution for the absolute value of speed. This is not the case for RPGM, as shown in Figure \ref{fig:RPGM_speed_violations}, presenting $P_{v}^{out}$ as a function of $\Delta t \in S_{\Delta t}$, for all values of $d_{max} \in S_{d}$.
Results presented in Figure \ref{fig:RPGM_speed_violations} clearly highlight that the RPGM model fails to meet the upper bound on node speed. For low values of $\Delta t$, in particular, nodes that are not group leaders (that is 75$\%$ of the total number of nodes in the considered scenario) almost never meet the bound on maximum speed, leading to an overall $P_{v}^{out} \approx 0.75$. 
The behavior of RPGM is directly related to the way positions of standard nodes are determined. For small $\Delta t$ values the positions of nodes are updated very often; since at every position update positions of standard nodes are randomly generated within $d_{max}$ meters of group leader position, a small $\Delta t$ leads to a high probability of violating the speed upper bound. The longest distance a standard node can cover in RPGM at each update is in fact equal to:
\begin{equation}
d=2d_{max}+v_{leader}\Delta t,
\label{eq:RPGM_max_distance}
\end{equation}
where $v_{leader}$ is the current speed of the group leader. The corresponding maximum speed, for low values of $\Delta t$, can be approximated by:
\begin{equation}
v=\frac{d}{\Delta t}=2\frac{d_{max}}{\Delta t}+v_{leader}\cong 2\frac{d_{max}}{\Delta t}.
\label{eq:RPGM_max_speed}
\end{equation}
For example, when $d_{max}=10\, m$ and $\Delta t = 0.1\, s$, one has $v \le 200\,m/s$, independently of the maximum speed allowed for group leaders.\\ 
Figure \ref{fig:RPGM_speed_violations} highlights that the adoption of a large $\Delta t$ mitigates the issue; this result comes, however, at the price of a lower accuracy in mobility modeling since, as already said, in RPGM the position of standard nodes is only known at position update times.\\ 
The analysis of $P_{\gamma}^{out}$ reveals a similar trend: in this case as well the Mo\textsuperscript{3} model always meets the bound for all combinations of $D_c$ and $\Delta t$, while is this is not the case for RPGM and RVGM, as expected. Figure \ref{fig:RPGM_RVGM_rotation_violations} presents $P_{\gamma}^{out}$ as a function of $\Delta t$ for RPGM, for all values of $d_{max}$ in $S_d$, and for RVGM. For both RPGM and RVGM $P_{\gamma}^{out}$ drops eventually to zero when $\Delta t$ is large enough to allow a full rotation between two updates without causing a rotation speed violation, as shown in Figure \ref{fig:RPGM_RVGM_rotation_violations} for $\Delta t = 5\, s$. However, for lower $\Delta t$ values the two models behave rather differently: RVGM is far less prone to cause violations ($P_{\gamma}^{out} \leq 0.1$ in all cases) than RPGM ($P_{\gamma}^{out} > 0.4$ in all cases).
An interesting trend can be observed for RVGM: $P_{\gamma}^{out}$ increases with $\Delta t$ for small values of $\Delta t$, while the opposite trend is observed for higher values of $\Delta t$. This can be explained by observing that, as $\Delta t$ increases, two opposite phenomena coexist:
\begin{enumerate}
\item the number of position updates decreases; since in RVGM rotation speed violations can only happen just after the selection of a new reference speed vector, the number of rotation violations is mainly depending on $T$, which is kept fixed in the simulations. As a consequence, although a detailed analysis of simulation results shows that the number of violations decreases with $\Delta t$, they become more relevant in percentage because the total number of position updates decreases at a faster rate;
\item the maximum rotation allowed for a node between two updates, given by $\gamma_{max} \Delta t$, increases, thus decreasing the probability for a direction update to cause a rotation speed violation.
\end{enumerate}
For low $\Delta t$, phenomenon 1) prevails, leading to an overall increase in $P_{\gamma}^{out}$; as $\Delta t$ increases beyond 1,  phenomenon 2) becomes predominant and $P_{\gamma}^{out}$ decreases with $\Delta t$.\\
Moving to the third indicator defined in Section \ref{sec:CompCriterion}, that is $v_{mean}$, Figure \ref{fig:Mo3_average_speed} presents $v_{mean}$ as a function of $\Delta t$ for Mo\textsuperscript{3} for all values of $D_c\in S_d$. Figure \ref{fig:Mo3_average_speed} highlights two phenomena: first, $v_{mean}$ increases as $D_c$ decreases, since a tighter binding condition leads nodes to spend a larger amount of time in \emph{Forced} state. Secondly, $v_{mean}$ is independent of $\Delta t$ for most $D_c$ values. The only exception is the combination of $D_c= 15\,m $ and $\Delta t=5\,s$, where the average speed gets very close to the allowed maximum speed $v_{max}=5\, m/s$. This result can be explained by observing that in all simulations $T_{CM}=\Delta t$: when node positions are not updated very often, also the grouping condition is checked less often, leading to more frequent cases where $\rho<\rho_{min}$, forcing the nodes to catch up at maximum speed. This can be easily avoided by setting $T_{CM}$ at a constant value, independent of $\Delta t$.\\
Results for RPGM, presented in Figure \ref{fig:RPGM_average_speed}, show that, for this model, the average speed strongly depends on $\Delta t$, due to the effect described by equation (\ref{eq:RPGM_max_speed}): in particular, the average speed is extremely unrealistic for low $\Delta t$ that is, incidentally, the setting required for modeling the mobility of standard nodes with high accuracy.\\
Finally, Figure \ref{fig:RVGM_average_speed} shows that RVGM is very robust to variations of $\Delta t$, since the speed selection process is not influenced by the update period.\\

\begin{figure}[t]
  \centering
\includegraphics[width=3.2in]{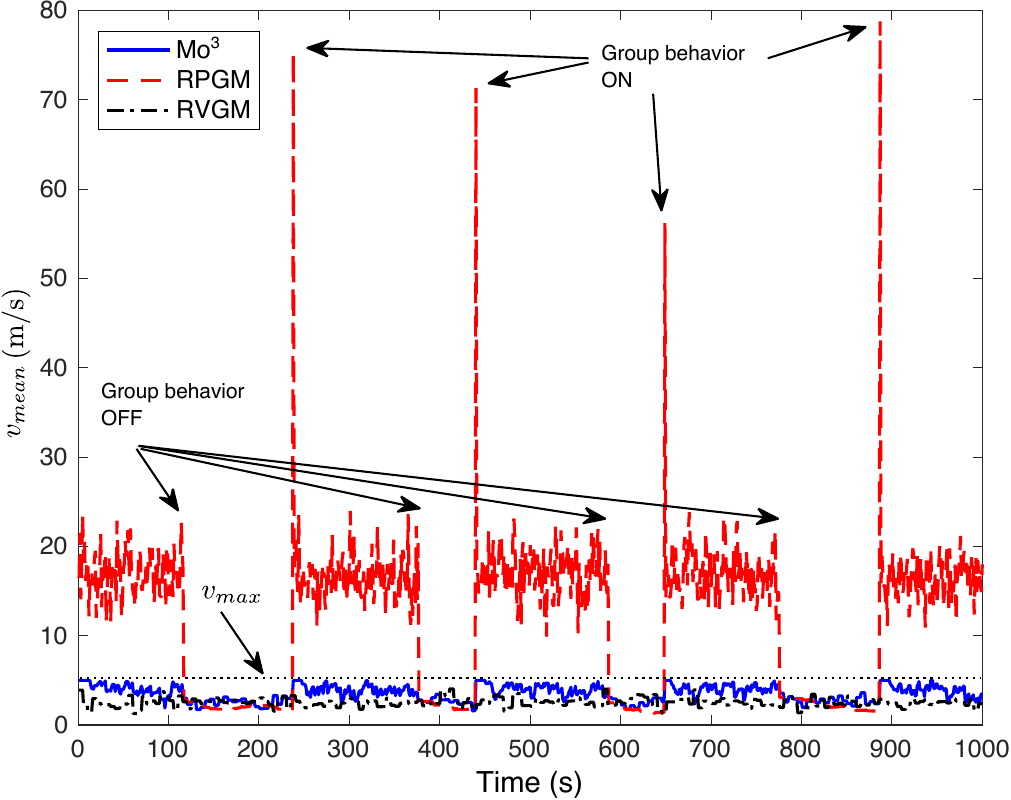}
 \caption{$v_{mean}$ for Mo\textsuperscript{3}, RPGM and RVGM as a function of time in a time-varying correlated mobility scenario; the upper bound $v_{max}$ is also shown as a reference. Group behavior was switched on and off during the observation, with average duration of on/off periods equal to $T_{switch}=100\,s$.  All remaining settings were as in Table \ref{tab:sim_settings}.}
 \label{fig:average_speed_switch}
\end{figure}

\begin{figure}[t]
  \centering
\includegraphics[width=3.2in]{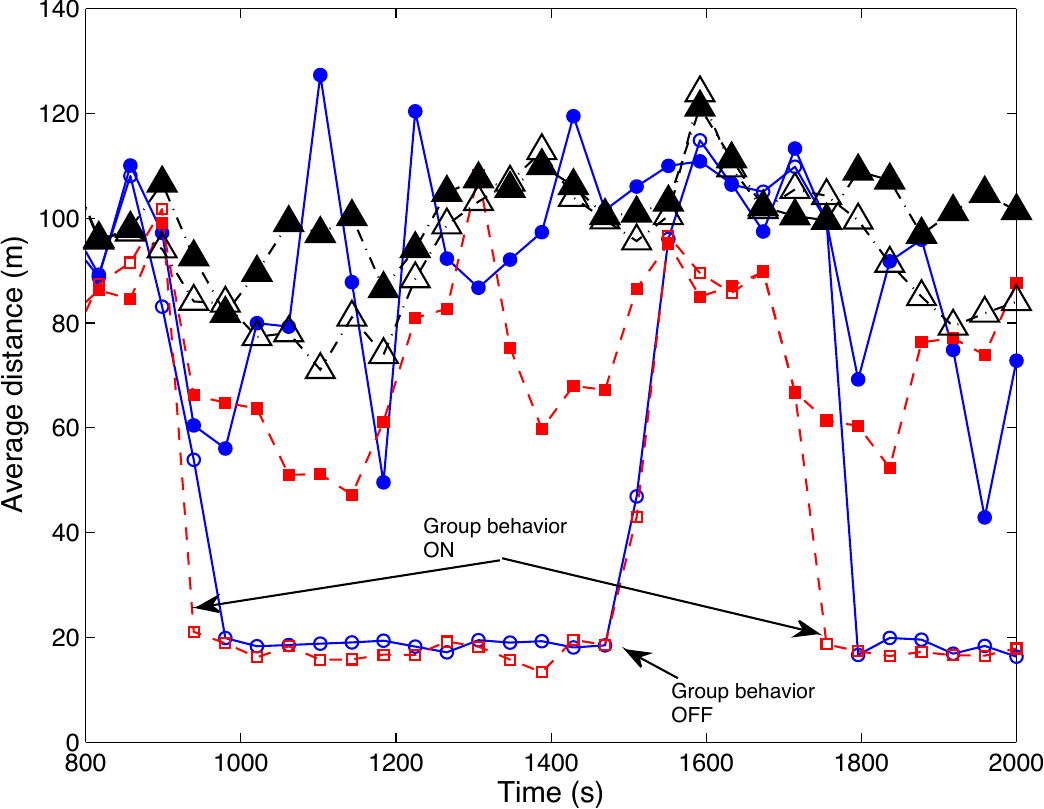}
 \caption{Average distance for Mo\textsuperscript{3}, RPGM and RVGM models as a function of time in a dynamic group mobility scenario. Group behavior was alternatively switched on and off with average duration of on/off periods equal to $T_{switch}=100\,s$. All remaining settings were as in Table \ref{tab:sim_settings} (Mo\textsuperscript{3}: circles, RPGM: squares, RVGM: triangles. Empty markers: intra-group average distance; filled markers: network-wide average distance).}
 \label{fig:average_distance_switch}
\end{figure}

It is interesting to observe that the settings $D_c=195\,m$ for Mo\textsuperscript{3} and $d_{max}=195\,m$ for RPGM lead to a configuration in which both intra-group and inter-group mobility occur on a large scale, corresponding thus to a pure macro-mobility scenario according to the classification introduced in Section {\ref{sec:introduction}}. In such a scenario, nodes may significantly depart from group mates, and group structure preservation mechanisms are not frequently activated; as a result, node mobility patterns should be mostly determined by the underlying individual mobility model. This behavior is indeed observed in Mo\textsuperscript{3}, as highlighted by the results for $v_{mean}$ presented in Figure {\ref{fig:Mo3_average_speed}}: $v_{mean}$ is in fact very close to 2.5 m/s, that is the expected average speed determined by the Individual Mobility settings listed in Table {\ref{tab:sim_settings}}. Oppositely, Figure {\ref{fig:RPGM_speed_violations}} shows that RPGM in this configuration has the same probability of violating the upper bound on linear speed as with smaller $d_{max}$ values, and actually leads to the most blatant violations in terms of average speed when used in combination with a low $\Delta t$, as shown in Figure {\ref{fig:RPGM_average_speed}}, confirming the difficulty for RPGM to properly describe  mobility in 5G and B5G scenarios requiring frequent position updates.\\  
Moving to the third question posed at the beginning of this section, a time-varying correlated mobility was modeled by introducing a random switching mechanism from group to individual mobility and viceversa; the duration of each group/individual mobility period was obtained as the outcome of a uniform random variable with average value $T_{switch}=100\,s$, taking values in $[50,150]\,s$. The switch between group and individual mobility for the three models was implemented as follows:
\begin{itemize}
\item for Mo\textsuperscript{3}, the binding matrix switched from the one in Figure \ref{fig:GrpMobilityEx2} to the identity matrix shown in Figure \ref{fig:IndMobilityEx};
\item in RPGM and RVGM all nodes were considered as group leaders, and thus started moving independently of other nodes in the network, starting from the last known position.
\end{itemize}
Figure \ref{fig:average_speed_switch} shows $v_{mean}$ as a function of time for all models, in a typical simulation run under these settings. The figure also shows as a reference the upper bound on speed, $v_{max}$. Results highlight the failure of RPGM to cope with time-varying correlated mobility settings: the average speed for this model presents high spikes when nodes switch back to group mobility, due to the abrupt displacement of standard nodes from their previous position to a random position within $d_{max}$ meters from their group leader. Figure \ref{fig:average_speed_switch} also confirms that the reason for speed upper bound violations in RPGM is the grouping mechanism: in periods where the group behavior is off, the underlying Random Walk individual mobility model is adopted for all nodes, and $v_{mean}$ immediately falls well below the $v_{max}$ threshold.\\
Mo\textsuperscript{3}, on the other hand, does not show any anomalous behavior during transitions. Following a transition from individual to group mobility, in particular, nodes check the grouping condition and, if required, switch from \emph{Free} to \emph{Forced} state and update their speed vector accordingly, without any discontinuity in their position. During the periods characterized by individual mobility, nodes part of a same group will drift away moving according to their individual mobility model; as a consequence, every time the group behavior is switched back on, a short period ensues in which most nodes move at speed $v_{max}$ to reestablish the group and satisfy the grouping condition. Correspondingly, Figure \ref{fig:average_speed_switch} shows that $v_{mean}$ is close to $v_{max}$ in each of such periods, as expected.\\
Figure \ref{fig:average_speed_switch} shows that the RVGM model suffers no discontinuities in average speed on transitions as well, again thanks to the adoption of a truncated Gaussian distribution.\\
Figure \ref{fig:average_distance_switch} completes the analysis by presenting both the average distance between nodes in the same group, and the average distance between all nodes as a function of time for the three models. The results highlight a strong difference of RPGM and Mo\textsuperscript{3} vs. RVGM. As already discussed in Section \ref{sec:ReferenceBasedModels}, RVGM is in fact unable to preserve spatial proximity between group members, in particular after long periods of individual mobility. During these periods, nodes belonging to the same group spread across the movement area, since their speeds and directions become independent. As a consequence, RVGM is only suitable for very specific mobility scenarios, where group members are not required to meet any bound on intra-group average distance. Figure \ref{fig:average_distance_switch} shows indeed that RVGM does not lead to different intra-group vs. inter-group average distances, while Mo\textsuperscript{3} and RPGM present similar properties from a topological point of view, and are characterized by a markedly shorter intra-group vs. inter-group average distance. However, the results shown earlier in this section and in Section \ref{sec:PerfEvalPure} demonstrate that the behavior of the two models in terms of the patterns generated for nodes in the same group is extremely different, as RPGM fails to maintain any spatial correlation in the mobility patterns of standard nodes. The adoption of RPGM in a scenario requiring accurate micro-mobility scale movements would thus lead to misleading results, as shown in Section \ref{sec:PerfEvalPure}.

\section{Conclusion}
\label{sec:Conclusion}
A new rule-based mobility model, called Mo\textsuperscript{3}, was proposed. Mo\textsuperscript{3} defines 5 rules, Individual Mobility, Correlated Mobility, Collision Avoidance, Obstacle Avoidance and Upper Bounds Enforcement, designed to accurately model the mobility of a node on both the micro-mobility and macro-mobility scale, as typically required in short distance wireless communications in 5G and beyond 5G networks.\\
Mo\textsuperscript{3} was compared against existing correlated/group mobility models using a multi-fold approach. First, features made available by Mo\textsuperscript{3} were compared with those provided in a wide set of existing group mobility models; the comparison highlighted that Mo\textsuperscript{3} provides a set of features on par or beyond existing models. Secondly, the accuracy of Mo\textsuperscript{3} vs. the RPGM model in generating patterns in a pure micro-mobility scenario was analyzed by comparing their impact on the performance of a communication network. Third, Mo\textsuperscript{3} was compared with RPGM and RVGM models in a mixed macro-mobility / micro-mobility scenario, using a newly defined set of performance indicators, related to the capability of a model to generate patterns that meet mandatory bounds on mobility parameters.\\
Results showed that Mo\textsuperscript{3} overcomes the limitations of existing group mobility models, that is lack of flexibility and accuracy for reference-based models, and difficulty in configuring and tuning the models for behavioral models. Mo\textsuperscript{3}  can in fact reliably and accurately describe mobility patterns in scenarios ranging from loose / no group mobility to tight group mobility, while taking into account the interaction with other nodes and with obstacles present in the mobility area. Furthermore, Mo\textsuperscript{3} is robust to variations in the position update period typically used in discrete event simulators, in particular to its reduction to sub-second duration, as required to accurately track nodes in micro-mobility scenarios. Mo\textsuperscript{3} is therefore a suitable candidate for modeling correlated and individual mobility in future network scenarios, and in particular in short distance wireless network scenarios, that will characterize 5G and beyond 5G systems.\\
A key characteristic of Mo\textsuperscript{3} is its native support for tridimensional mobility for a subset of the rules, in particular Individual Mobility, Correlated Mobility and Upper Bounds Enforcement. This feature enables Mo\textsuperscript{3} to model any scenario involving correlated movement of nodes in a tridimensional landscape, such as a swarm of UAVs, but also a flock of birds, or a bank of fish. Future work will focus on the extension to the tridimensional case of the two remaining rules, Collision Avoidance and Obstacle Avoidance. In the case of Collision Avoidance this will require the extension of the concept of a collision risk zone from a circle of radius $d_{min}^{CA}$, as shown in Figure {\ref{fig:CA_example}}, to a sphere with the same radius. For the Obstacle Avoidance rule, the most immediate extension will consist in considering rectangular cuboids and ellipsoids in place of rectangles and ellipses as obstacle shapes, and extending the approach described in Appendix {\ref{sec:OAdetails}} to find tangent lines to such obstacles.\\
Future work will furthermore focus on the acquisition and use of captured mobility traces in 5G networks in order to compare them with mobility patterns generated with the Mo\textsuperscript{3} model, and on the extension of Mo\textsuperscript{3}, by providing additional basic shapes for obstacle avoidance both in the bidimensional and tridimensional cases. \appendices
\begin{table*}
\caption{Analysis of correlated mobility models in terms of target mobility scenarios and impact, measured by the number of citations received in the Scopus database overall and in the last 5 years.}
\label{tab:relevance_and_impact}
\centering
\begin{tabular}{|c|c|c|c|c|c|}
\hline
  \textbf{Model} & \textbf{Year} & \textbf{Mobility scenarios}&\textbf{Overall citations}&\textbf{Citations in 2017-2021}\\
\hline
RPGM \cite{HonGer99}& 1999 & General purpose & 989 & 149 \\
\hline
RVGM \cite{WanLi02}& 2002  & General purpose & 155 & 17 \\
\hline
GFMM \cite{WilHua09} & 2009 & General purpose & 17 & 5 \\
\hline
MGCM \cite{WuYu06} & 2006 & Military pedestrian & 5 & 2 \\
\hline
RRGM \cite{NgZha05} & 2005  & Security / Search \& rescue & 26 & 4 \\
\hline
VTGM \cite{ZhoXu04} & 2004  & Military vehicular& 107 & 15 \\
\hline
CMM \cite{MusMas06} & 2006 & Human social interactions & 254 & 20 \\
\hline
ECMM \cite{VasYan12} & 2012 & Human social interactions & 32 & 19 \\
\hline
SGMM \cite{BlaLow04} & 2004 & Human social interactions & 52 & 9 \\
\hline
BMM\cite{LegBor06} &2006 & General purpose & 20 & 3\\
\hline
BMM-GC \cite{BeuMiw13} & 2013 & Urban mobility & 3 & 1\\
\hline
SIMPS \cite{BorLeg09} & 2009 & Human social interactions & 59 & 11\\
\hline
\end{tabular}
\end{table*}
\section{Impact of existing mobility models}
\label{sec:CitationAnalysis}
Table {\ref{tab:relevance_and_impact}} presents the models analyzed in Sections {\ref{sec:introduction}} and {\ref{sec:CorrelatedModels}} and compares them in terms of the mobility scenarios they address, the general impact in terms of overall citations in the Scopus database \cite{scopus} and their impact in the last five years, measured by the citations in the period 2017-2021. Table {\ref{tab:relevance_and_impact}} confirms that RPGM is by far the most used group mobility model. Other models with good impact on the research community are CMM, that however targets mobility scenarios focusing on social interactions, RVGM, also a general purpose model, and VTGM, designed for military/vehicular networks with specific spatial constraints.

\section{Collision avoidance}
\label{sec:CAdetails}
In this appendix a detailed description of the algorithm proposed for the Collision Avoidance rule introduced in Section \ref{sec:Mo3_CollisionAvoidance} is provided.
\subsection{Path crossing identification} 
\label{sec:CA_path_crossing} 
The path crossing identification relies on the representation of trajectories as rays in the bi-dimensional plane. Keeping the notation introduced in Section \ref{sec:Mo3_CollisionAvoidance} for speed and direction, and indicating with $\left\{x_i, y_i\right\}$ the coordinates of the node $i$ executing the algorithm, the trajectory followed by $i$ can in fact be represented as a ray centered on its current position, $\left\{x_i^0, y_i^0\right\}$, with direction given by $\theta_i$ using the following notation for the two Cartesian components:
\begin{equation}
\left\{
\begin{aligned}
&x_i=x_i^0+cos\left(\theta_i\right)r\\
&y_i=y_i^0+sin\left(\theta_i\right)r
    \end{aligned}
\quad r\geq 0.\right.
\end{equation}
Similarly, for another generic node $k$ one has
\begin{equation}
\left\{
\begin{aligned}
&x_k=x_k^0+cos\left(\theta_k\right)s\\
&y_k=y_k^0+sin\left(\theta_k\right)s
    \end{aligned}
\quad s\geq 0.\right.
\end{equation}

The future trajectories of $i$ and $k$ will cross if a solution exists to the following system of equations in $r$, $s$:
\begin{equation}
\label{eq:PCI_system}
\left\{
\begin{aligned}
&x_i=x_k\\
&y_i=y_k
    \end{aligned}\right.
\Leftrightarrow
\left\{
\begin{aligned}
&x_i^0+cos\left(\theta_i\right)r=x_k^0+cos\left(\theta_k\right)s\\
&y_i^0+sin\left(\theta_i\right)r=y_k^0+sin\left(\theta_k\right)s
    \end{aligned}\right.
\end{equation}
with $r\geq 0$ and $s\geq 0$\footnote{Solutions obtained for either $r<0$ or $s< 0$ would correspond to cases where the future trajectory of one node crosses the past trajectory of the other. These cases are not considered by the Collision Avoidance rule, since an actual collision is impossible even if the nodes were to be at distance lower than $d_{min}^{CA}$ at crossing time.}. Solving for $r$ and $s$ leads to:

\begin{equation}
\label{eq:PCI_solution}
\left\{
\begin{aligned}
&r=\frac{\left[\left(y_k^0-y_i^0\right)\cos\left(\theta_k\right)-\left(x_k^0-x_i^0\right)\sin\left(\theta_k\right)\right]}{D}\\
&s=\frac{\left[\left(y_k^0-y_i^0\right)\cos\left(\theta_i\right)-\left(x_k^0-x_i^0\right)\sin\left(\theta_i\right)\right]}{D}
    \end{aligned}\right.
\end{equation}
where $D=\cos\left(\theta_k\right)\sin\left(\theta_i\right)-\sin\left(\theta_k\right)\cos\left(\theta_i\right)$, and $D\neq 0$ is assumed.\\ Note that $D=0$ can be obtained only if  $\theta_i=\theta_k$ or if $\left|\theta_i-\theta_k\right|=\pi$. $\theta_i=\theta_k$ corresponds to two parallel rays that never cross unless $\left\{x_i^0, y_i^0\right\} \equiv \left\{x_k^0, y_k^0\right\}$, and requires thus no correction. On the other hand, $\left|\theta_i-\theta_k\right|=\pi$ corresponds to $i$ and $k$ moving on the same line in opposite directions, eventually leading to a frontal collision. In this case, since no $v_i'\neq v_i$ would avoid the collision, a small deviation $\theta^{CA}$ is added to $\theta_i$, and the algorithm is applied again from scratch.\\
Each node $k$ such that $d_{ik} \leq d_{trigger}^{CA}$ will be analyzed by solving the system in \eqref{eq:PCI_system}; if $r\geq0$ and $s\geq0$, $k$ will be added to the $\left\{PCN\right\}$ set, and the corresponding crossing point $\left\{x_C^{i,k},y_C^{i,k}\right\}$ will be stored for further evaluation in the next step of the algorithm.
\subsection{Speed bounds identification} 
\label{sec:CA_speed_bound} 
The goal of this step is to determine a set of acceptable ranges for the new speed $v_i'$, one for each node $k$ in $\left\{PCN\right\}$, so to guarantee that when $i$ arrives at $\left\{x_C^{i,k},y_C^{i,k}\right\}$, $k$ will be at least $d_{min}^{CA}$ meters away.\\
First, the times at which $i$ and $k$ will reach the crossing point $\left\{x_C^{i,k},y_C^{i,k}\right\}$, referred to as  $T_C^{i}$ and $T_C^{k}$, are determined:
\begin{equation}
\label{eq:TimeToCollision}
\left\{
\begin{aligned}
&T_C^{i}=\frac{d_{iC}}{v_i}\\
&T_C^{k}=\frac{d_{kC}}{v_k},
    \end{aligned}\right.
\end{equation}
where $d_{iC}$ and $d_{kC}$ indicate the distance between the crossing point and the current position of $i$ and $k$, respectively.\\
If $T_C^{i}< T_C^{k}$, at its current speed $i$ will reach $\left\{x_C^{i,k},y_C^{i,k}\right\}$ before $k$.  In this case, a lower bound is defined on $v_i'$, so to ensure that, when $i$ reaches the crossing point, $k$ will still be at least $d_{min}^{CA}$ away from this point on its trajectory. The bound is defined as follows:
\begin{equation}
\label{eq:LowerBound}
v_i' > v_L^k= v_k \frac{d_{min}^{CA}+d_{iC}}{d_{kC}}.
\end{equation}
Note that $T_C^{i}< T_C^{k}$ already implies $v_i > v_k \left(d_{iC}/d_{kC}\right)$, so that the bound defined in (\ref{eq:LowerBound}) could be already satisfied by the current value of $v_i$. Furthermore, since any speed above $v_L^k$ will satisfy the bound, one can also introduce the upper bound $v_i'<v_U^k=v_{max}$.\\
Oppositely, if $T_C^{i} > T_C^{k}$, at the current speed $v_i$ $i$ will reach $\left\{x_C^{i,k},y_C^{i,k}\right\}$ after $k$; in this case an upper bound is defined on $v_i'$, so to ensure that, when $k$ reaches the crossing point, $i$ will still be at least $d_{min}^{CA}$ away. The bound is thus equal to:
\begin{equation}
\label{eq:UpperBound}
v_i' < v_U^k =v_k \frac{d_{iC}}{d_{min}^{CA}+d_{kC}}
\end{equation}
Again, $T_C^{i} > T_C^{k}$ already implies $v_i < v_k \left(d_{iC}/d_{kC}\right)$, that is a weaker upper bound on $v_i$; furthermore, any speed below $v_U^k$ will satisfy the bound, leading to the lower bound $v_i'>v_L^k=v_{min}$.\\
Finally, if $T_C^{i} = T_C^{k}$, at their current speeds $i$ and $k$ will reach $\left\{x_C^{i,k},y_C^{i,k}\right\}$ at the same time. In this case the bound on $v_i'$ will be defined according to \eqref{eq:LowerBound} if $i<k$, and according to \eqref{eq:UpperBound} otherwise.\\
The analysis of all nodes in the $\left\{PCN\right\}$ set will thus lead to the system of inequalities on $v_i'$ given by:
\begin{equation}
\label{eq:system}
 v_L^k \leq v_i' < v_U^k \quad \text{for} \, k=1,\cdots, \left| \left\{PCN\right\}\right|
\end{equation}
where, for each $k$, either $v_L^k=v_{min}$ or $v_U^k=v_{max}$.
\subsection{Collision avoidance} 
The set of constraints on the new speed $v_i'$ determined at the previous step is analyzed in order to determine whether the corresponding system of inequalities can be solved. If solutions exist, they will be in the range $[v_{L}^{i},\; v_U^{i}]$, given by:
\begin{equation}
\left\{
\begin{aligned}
    v_{L}^{i}&=\max\left\{v_L^1,\cdots,v_L^N\right\}\\
    v_{U}^{i}&=\min\left\{v_U^1,\cdots,v_U^N\right\}
\end{aligned}
\right.,
\label{eq:CollAvoidanceSolution}
\end{equation}
where $N=\left|\left\{PCN\right\}\right|$.\\
The analysis will lead to one of three possible outcomes:
\begin{enumerate}
    \item solutions exist, and $v_i \in [v_{L}^{i},\; v_U^{i}]$: no collision risk is identified, and no action is taken, leading to $v_i'\equiv v_i$;
    \item solutions exist, but $v_i \notin [v_{L}^{i},\; v_U^{i}]$: a collision risk is identified, and $v_i'$ is set to the acceptable value closest to $v_i$, in order to minimize the variation with respect to the current $v_i$. This corresponds to
    \begin{equation}
    v_i'=\left\{
    \begin{aligned}
    \min\left\{v_{L}^{i},v_{max}\right\}&\quad \text{if}\; v_i< v_{L}^{i}\\
    \max\left\{v_{U}^{i},v_{min}\right\}& \quad \text{if} \; v_i> v_{U}^{i}
    \end{aligned}
    \right.,
\label{eq:CollAvoidanceChoice}
\end{equation}
\item no solution exists: $v_i'$ is set so to satisfy the largest possible number of inequalities, and inequalities that are not satisfied are left to be addressed at the next application of the Collision Avoidance rule.
\end{enumerate} 

\section{Obstacle avoidance}
\label{sec:OAdetails}
This Appendix provides details on the algorithm proposed for the Obstacle Avoidance rule introduced in Section \ref{sec:Mo3_ObstacleAvoidance}. Since throughout this section only one obstacle is considered, the $k$ subscript used in Section \ref{sec:Mo3_ObstacleAvoidance} is dropped in order to simplify notation; the angles that define the range of forbidden directions  are thus indicated in the following as $\theta_{i}^{min}$ and $\theta_{i}^{max}$.\\
Section \ref{sec:OAdetails_ellipse} describes the procedure adopted in Mo\textsuperscript{3} to determine whether a ellipse-shaped obstacle is within $d_{trigger}^{OA}$ from a node $i$ in position $\left\{x_i,y_i\right\}$ and, if this is the case, to compute the angles $\theta_{i}^{min}$ and  $\theta_{i}^{max}$; Section \ref{sec:OAdetails_rectangle} does the same for a rectangular obstacle. 
\subsection{Ellipse}
\label{sec:OAdetails_ellipse}
\subsubsection{Distance}
\label{sec:OAdetails_ellipse_distance}
Let us consider an ellipse centered in $\left\{x_{obs},y_{obs}\right\}$, with semi axes $a$ and $b$.\\
The problem of finding the minimum distance $d_{ie}$ between $\left\{x_i,\,y_i\right\}$ and the ellipse has no straightforward mathematical solution, and requires the solution of an equation by the Newton method. Mo\textsuperscript{3} adopts thus an approximated approach, by evaluating the distance $\hat{d}_{ie}$ between $\left\{x_i,\,y_i\right\}$ and the closest intersection point between the ellipse and a line passing through $\left\{x_i,\,y_i\right\}$ and $\left\{x_{obs},\,y_{obs}\right\}$, as shown in Figure \ref{fig:OA_details_ellipse_distance}.\\
The coordinates of the point $\left\{x_p,\,y_p\right\}$ can be obtained by defining a ray originating in $\left\{x_i,\,y_i\right\}$ with direction $\theta_{obs}$, also shown in Figure \ref{fig:OA_details_ellipse_distance}. The value of $\theta_{obs}$ is given by:
\begin{equation}
    \theta_{obs}=\arctantwo\left(y_{obs}-y_i,x_{obs}-x_i\right).
\label{eq:OA_distance_ray_direction}
\end{equation}
The ray will be thus defined as:
\begin{equation}
\left\{
\begin{aligned}
&x=x_i+cos\left(\theta_{obs}\right)r\\
&y=y_i+sin\left(\theta_{obs}\right)r
    \end{aligned}
\quad r\geq 0,\right.
\label{eq:OA_distance_ellipse_ray}
\end{equation}
and the coordinates of the intersection $\left\{x_p,\,y_p\right\}$ can be obtained by solving for $r$ the following system:
\begin{equation}
\left\{
\begin{aligned}
\frac{\left(x-x_{obs}\right)^2}{a^2}+\frac{\left(y-y_{obs}\right)^2}{b^2}=1\\
x=x_i+cos\left(\theta_{obs}\right)r\\
y=y_i+sin\left(\theta_{obs}\right)r.
    \end{aligned}\right.
\end{equation}
The system has two solutions, $r_1$ and $r_2$, with $r_1<r_2$, corresponding to the two intersection points with coordinates $\left\{x_p,\,y_p\right\}$ and $\left\{x_p',\,y_p'\right\}$, shown in Figure \ref{fig:OA_details_ellipse_distance}; the coordinates of $\left\{x_p,\,y_p\right\}$ can be obtained by substituting $r_1$ in \eqref{eq:OA_distance_ellipse_ray}.
\Figure[t]()[width=0.45\textwidth]{./Figures/Figure_OA_example_ellipse_distance.eps}
{Evaluation of the approximated minimum distance $\hat{d}_{ie}$ between a node $i$ in $\left\{x_i,\,y_i\right\}$ and an ellipse centered in $\left\{x_{obs},\,y_{obs}\right\}$, defined as the distance between $\left\{x_i,\,y_i\right\}$ and the intersection point $\left\{x_p,\,y_p\right\}$. The direction of a ray originating in $\left\{x_i,\,y_i\right\}$ and pointing at $\left\{x_{obs},\,y_{obs}\right\}$ is also shown. \label{fig:OA_details_ellipse_distance}}
\subsubsection{Forbidden range}
\label{sec:OAdetails_ellipse_range}
The two directions $\theta_{i}^{min}$ and $\theta_{i}^{max}$ correspond to the two lines tangent to the ellipse and passing in $\left\{x_i,y_i\right\}$. The coordinates of the two tangent points can be found by solving the following system of equations in $x,\,y$, imposing the same slope for line and ellipse:
\begin{equation}
   \left\{
    \begin{aligned}
    \frac{\left(x-x_{obs}\right)^2}{a^2}+\frac{\left(y-y_{obs}\right)^2}{b^2}=1\\
    \left(y-y_i\right)=-\frac{b^2\left(x-x_{obs}\right)}{a^2\left(y-y_{obs}\right)}\left(x-x_i\right),
    \end{aligned}
    \right.
\label{eq:TangentPointsSystem}
\end{equation}
leading to the solutions in $x$ in the usual form:
\begin{equation}
    x_{1,2}=\frac{-B \pm \sqrt{B^2-4AC}}{2A},
\label{eq:TangentPointsSolution}
\end{equation}
where:
\begin{equation}
   \left\{
    \begin{aligned}
    A&=c_2c_1^2+c_4\\
    B&=2c_2 c_1 c_0+ c_1 c_3 + c_5\\
    C&=c_2 c_0^2+c_3 c_0 +c_6,
    \end{aligned}\right.
\label{eq:TangentPointsSolutionCoeffs}
\end{equation}
and finally:
\begin{equation}
   \left\{
    \begin{aligned}
    c_0&=\frac{a^2 \left(y_i y_{obs}-y_{obs}^2\right)+b^2 \left(x_i x_{obs}-x_{obs}^2\right)+a^2 b^2}{a^2\left(y_i-y_{obs}\right)}\\
    c_1&=-\frac{b^2\left(x_i-x_{obs}\right)}{a^2\left(y_i-y_{obs}\right)}\\
    c_2&=a^2\\
    c_3&=-2 a^2 y_{obs}\\
    c_4&=b^2\\
    c_5&=-2b^2 x_{obs}\\
    c_6&=a^2 y_e^2+b^2 x_{obs}^2-a^2 b^2.
    \end{aligned}\right.
\label{eq:TangentPointsSolutionCoeffsDetail}
\end{equation}
The corresponding $y_{1,2}$ coordinates can be obtained by substituting $x_{1,2}$ in either equation in \eqref{eq:TangentPointsSystem}. The two directions are then obtained as:

\begin{equation}
   \left\{
    \begin{aligned}
    \theta_{i}^{1}&=\arctantwo\left(y_1-y_i,x_1-x_i\right)\\
    \theta_{i}^{2}&=\arctantwo\left(y_2-y_i,x_2-x_i\right).\\
    \end{aligned}\right.
\label{eq:TangentDirections}
\end{equation}
Depending on the relative position of the node with respect to the ellipse, both cases $\theta_{i}^{1}\geq\theta_{i}^{2}$ and $\theta_{i}^{1}<\theta_{i}^{2}$ are possible; the association of the two directions to $\theta_{i}^{min}$ and $\theta_{i}^{max}$ will be thus done so to ensure that $\theta_{i}^{min}<\theta_{i}^{max}$.\\
Note that the relative position of $i$ with respect to the ellipse also determines the forbidden range $F_\theta$ corresponding to the two limit angles $\theta_{i}^{min}$ and $\theta_{i}^{max}$, according to the following rule:

\begin{equation}
F_\theta=
    \begin{cases}
\left[\theta_{i}^{min},\theta_{i}^{max}\right],  \text{if }\left(\theta_{i}^{min}\cdot\theta_{i}^{max}>0\right)\lor \left(x_i<x_{obs}\right)\\
\left[-\pi,\theta_{i}^{min}\right] \cup \left[\theta_{i}^{max}, \pi\right], \text{otherwise}.
\end{cases}
\label{eq:forbiddenRangeCases}
\end{equation}

\subsection{Rectangle}
\label{sec:OAdetails_rectangle}
\subsubsection{Distance}
\label{sec:OAdetails_rectangle_distance}
Let us consider in this case a rectangle centered in $\left\{x_{obs},y_{obs}\right\}$, with horizontal side of length $2a$ and vertical side of length $2b$. The coordinates of the four corners of the rectangle, starting from the bottom left corner and proceeding clockwise, are thus: $\left\{x_{obs}^{min},\,y_{obs}^{min}\right\}$, $\left\{x_{obs}^{min},\,y_{obs}^{max}\right\}$, $\left\{x_{obs}^{max},\,y_{obs}^{max}\right\}$ and $\left\{x_{obs}^{max},\,y_{obs}^{min}\right\}$, where:
\begin{equation}
   \left\{
    \begin{aligned}
    x_{obs}^{min}&=x_{obs}-a\\
   x_{obs}^{max}&=x_{obs}+a\\
   y_{obs}^{min}&=y_{obs}-b\\
   y_{obs}^{max}&=y_{obs}+b.
    \end{aligned}\right.
\label{eq:CornersCoordinates}
\end{equation}
The above coordinates are used to determine the minimum distance $d_{ir}$ between a node $i$ in $\left\{x_i,\,y_i\right\}$ an the rectangle as follows. The movement area is divided in 8 sectors around the rectangle: 4 corner sectors (Bottomleft, Topleft, Topright and Bottomright) and 4 side sectors (Left, Top, Right and Bottom), as shown in Figure~\ref{fig:OA_details_rectangle_distance}. If the point $\left\{x_i,\,y_i\right\}$ falls in a corner sector, the minimum distance is the distance between $\left\{x_i,\,y_i\right\}$ and the corresponding corner; if it falls in a side sector, the minimum distance is determined as the distance between $\left\{x_i,\,y_i\right\}$ and the point $\left\{x_p,\,y_p\right\}$ on the corresponding side crossed by a line passing through $\left\{x_i,\,y_i\right\}$ and orthogonal to the side. As an example, Figure~\ref{fig:OA_details_rectangle_distance} shows a point $\left\{x_i,\,y_i\right\}$ in the Left side quadrant; the point of the rectangle at minimum distance from $i$ is $\left\{x_p,\,y_p\right\}\equiv \left\{x_{obs}^{min},\,y_i\right\}$.
\Figure[t]()[width=0.45\textwidth]{./Figures/Figure_OA_example_rectangle_distance.eps}
{Evaluation of the minimum distance $d_{ir}$ and of the forbidden range $\left[\theta_{i}^{min},\,\theta_{i}^{max}\right]$ for a node $i$ in $\left\{x_i,\,y_i\right\}$ and a rectangle centered in $\left\{x_{obs},\,y_{obs}\right\}$ with sides $2a$ and $2b$.\label{fig:OA_details_rectangle_distance}}
\subsubsection{Forbidden range}
\label{sec:OAdetails_rectangle_range}
The evaluation of the angles $\theta_{i}^{min}$ and $\theta_{i}^{max}$ determining the forbidden range takes advantage of the quadrants defined in the previous subsection. In all cases, each of the two directions delimiting the range corresponds to the angle defined by a line passing through $\left\{x_i,\,y_i\right\}$ and one of the corners of the rectangle; the sector $\left\{x_i,\,y_i\right\}$ falls in determines which corners will be considered:
\begin{itemize}
    \item if $\left\{x_i,\,y_i\right\}$ falls in a side sector, the corners determining the two angles will be those at the two ends of the corresponding side. This case is shown in Figure~\ref{fig:OA_details_rectangle_distance} for a node $i$ in the Left quadrant, where one has:
\begin{equation}
   \left\{
    \begin{aligned}
    \theta_{i}^{min}&=\arctantwo\left(y_{obs}^{min}-y_i,x_{obs}^{min}-x_i\right)\\
    \theta_{i}^{max}&=\arctantwo\left(y_{obs}^{max}-y_i,x_{obs}^{min}-x_i\right);\\
    \end{aligned}\right.
\label{eq:TangentDirections_rectangle_side}
\end{equation}
\item if $\left\{x_i,\,y_i\right\}$ falls in a corner sector, the corners determining the two angles will be those adjacent to the corner delimiting the sector. As an example, for a node $i$ in the Topleft sector, the two corners would be  $\left\{x_{obs}^{min},\,y_{obs}^{min}\right\}$ and $\left\{x_{obs}^{max},\,y_{obs}^{max}\right\}$, leading to:
\begin{equation}
   \left\{
    \begin{aligned}
    \theta_{i}^{min}&=\arctantwo\left(y_{obs}^{min}-y_i,x_{obs}^{min}-x_i\right)\\
    \theta_{i}^{max}&=\arctantwo\left(y_{obs}^{max}-y_i,x_{obs}^{max}-x_i\right).\\
    \end{aligned}\right.
\label{eq:TangentDirections_rectangle_corner}
\end{equation}
\end{itemize}
Once $\theta_{i}^{min}$ and $\theta_{i}^{max}$ have been determined, the corresponding forbidden range $F_\theta$ can be obtained by applying the rule defined in \eqref{eq:forbiddenRangeCases}.
\bibliographystyle{IEEEtran}
\bibliography{Latest_LDN_bibs}
\begin{IEEEbiography}[{\includegraphics[width=1in,height=1.25in,clip,keepaspectratio]{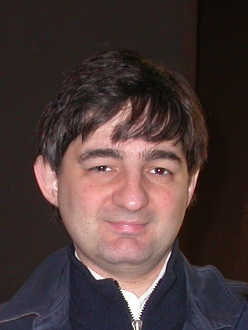}}]{Luca De Nardis} (M'02)  is an Assistant Professor with the Department of Information Engineering, Electronics and Telecommunications at Sapienza University of Rome, Italy. He received the Ph.D. degree from Sapienza University of Rome in 2005. In 2007, he was a Postdoctoral Fellow with the University
of California, Berkeley. He authored or co-authored
over 120 international peer-reviewed publications.
His research interests focus on cognitive communications, medium access control, routing protocols, and wireless positioning systems.
\end{IEEEbiography}

\begin{IEEEbiography}[{\includegraphics[width=1in,height=1.25in,clip,keepaspectratio]{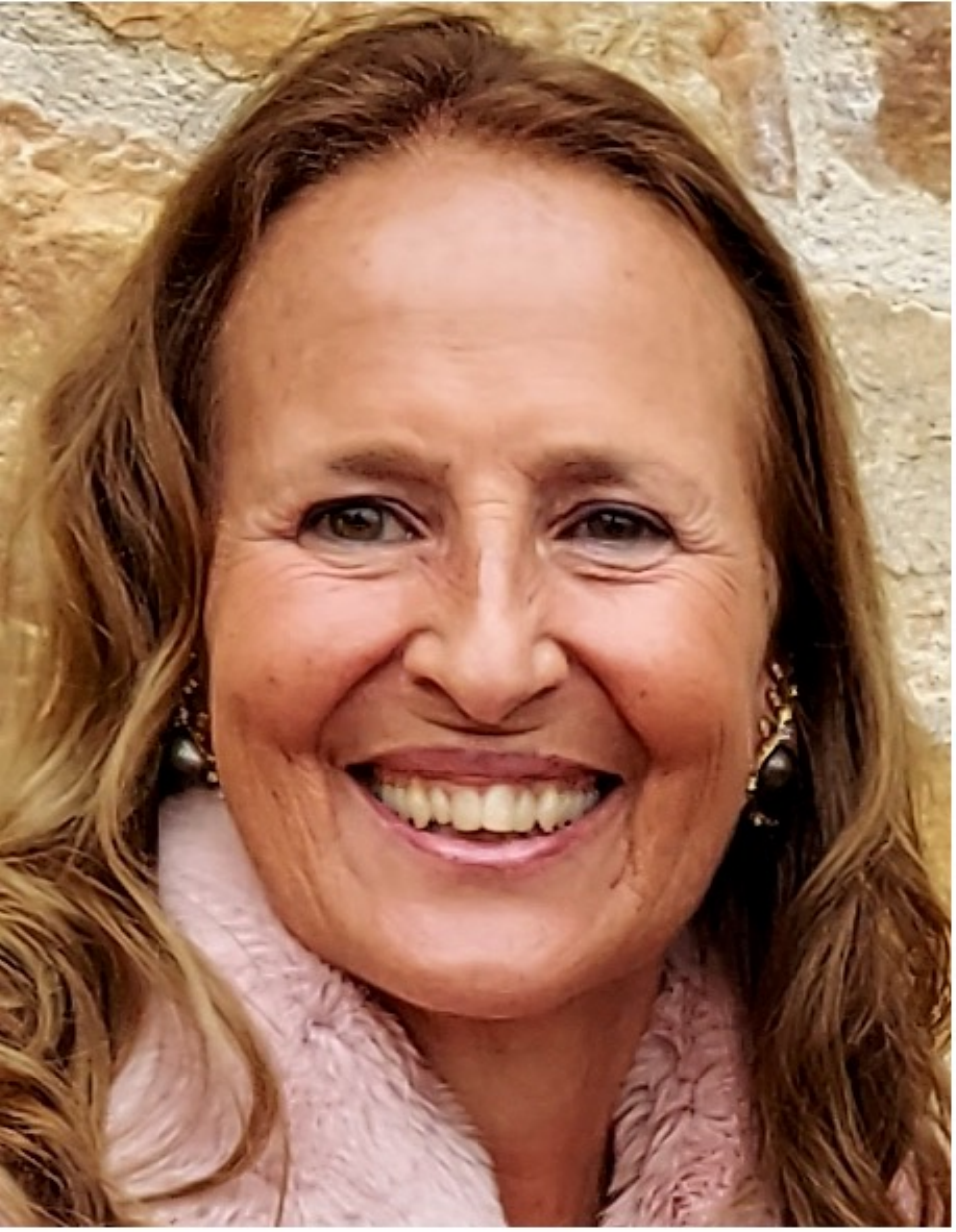}}]{Maria-Gabriella Di Benedetto}(F'16) received the Ph.D. degree from the Sapienza
University of Rome, Italy, in 1987. In 1991, she joined the Faculty of Engineering, Sapienza University of Rome, where she is currently a Full Professor of telecommunications. She held various visiting positions with the Massachusetts Institute of Technology (MIT), University of California, Berkeley, and the University of Paris XI. She is currently a Research Affiliate at the Research Laboratory of Electronics (RLE) of MIT. She is a fellow of the Radcliffe Institute for Advanced Study, Harvard University, Cambridge, MA, USA. In 1994, she received the Mac Kay Professorship Award from the University of California, Berkeley. Her research interests include wireless communication systems, impulse radio communications, and speech. 
\end{IEEEbiography}

\EOD

\end{document}